\documentclass[reprint, prb,unsortedaddress]{revtex4-1}

\usepackage{graphicx, bm, amsmath, amsfonts,amssymb,amsthm}

\usepackage{float}
\usepackage{caption}
\usepackage{subcaption}
\usepackage{color}
\usepackage[titletoc,title]{appendix}

\newcommand{\al}{\alpha}
\newcommand{\ep}{\epsilon}

\newcommand{\spn}{\text{span}}

\newcommand{\hil}{\mathcal{H}}
\newcommand{\ham}{\mathbf{H}}
\newcommand{\Z}{\mathbb{Z}}
\newcommand{\ket}[1]{|#1\rangle}

\newcommand{\Tr}{\text{Tr }}

\newcommand{\mz}{\mathbb{Z}}

\newcommand{\tF}{\tilde{F}}

\graphicspath{{figures/}}

\begin{document}

\title{Symmetry fractionalization and twist defects}
\author{Nicolas Tarantino}
\affiliation{Department of Physics and Astronomy, Stony Brook University, Stony Brook, NY 11794-3800, USA}
\author{Netanel H. Lindner}
\affiliation{Physics Department, Technion, 320003 Haifa, Israel}
\author{Lukasz Fidkowski}
\affiliation{Department of Physics and Astronomy, Stony Brook University, Stony Brook, NY 11794-3800, USA}

\begin{abstract}
Topological order in two dimensions can be described in terms of deconfined quasiparticle excitations - anyons - and their braiding statistics.  However, it has recently been realized that this data does not completely describe the situation in the presence of an unbroken global symmetry.  In this case, there can be multiple distinct quantum phases with the same anyons and statistics, but with different patterns of symmetry fractionalization - termed symmetry enriched topological (SET) order.  When the global symmetry group $G$, which we take to be discrete, does not change topological superselection sectors - {\emph{i.e.}} does not change one type of anyon into a different type of anyon - one can imagine a local version of the action of $G$ around each anyon.  This leads to projective representations and a group cohomology description of symmetry fractionalization, with $H^2(G,{\cal A})$ being the relevant group.  In this paper, we treat the general case of a symmetry group $G$ possibly permuting anyon types.  We show that despite the lack of a local action of $G$, one can still make sense of a so-called twisted group cohomology description of symmetry fractionalization, and show how this data is encoded in the associativity of fusion rules of the extrinsic `twist' defects of the symmetry.  Furthermore, building on work of Hermele \cite{Hermele_string_flux}, we construct a wide class of exactly solved models which exhibit this twisted symmetry fractionalization, and connect them to our formal framework.
\end{abstract}

\maketitle

\tableofcontents

\section{Introduction}
In the last 30 years it has been realized that there exist quantum phases of matter that - unlike ordinary crystals, magnets, or superconductors - cannot be understood in terms of symmetry breaking and local order parameters.  The main example is the fractional quantum Hall effect, which instead exhibits a subtle non-local order manifested in exotic properties like emergent excitations with exotic statistics (anyons), protected gapless edge modes, and ground state degeneracy on surfaces of non-trivial topology.  In general, one can use the braiding statistics of the anyons to give a coarse classification of gapped Hamiltonians; this is called intrinsic topological order.  Although intrinsic topological order is independent of any symmetry considerations, it has recently been realized that the presence of a global symmetry can further refine the coarse classification given by intrinsic topological order.  In particular, there can exist several `symmetry protected' quantum phases (SPTs) realizing trivial intrinsic topological order \cite{TurnerBerg, Fidkowski1d, Chen1d, Chen2d}, and multiple `symmetry enriched' phases (SETs) corresponding to the same intrinsic topological order, with the latter being the focus of this paper.  Our main result concerns an unconventional type of symmetry action where acting with the global symmetry on some excitations turns them into new excitations which cannot be obtained from the original ones via the action of a local operator.  We refer to such excitations as being in different `topological superselection sectors', or being anyons of different type; thus our symmetry non-trivially permutes the topological superselection sectors.  We give a general prescription for understanding symmetry fractionalization in this case, show how it fits into the classification of SETs, and construct a wide class of exactly solved examples illustrating our results.

One way to distinguish among different symmetry enriched phases is based on how the symmetry fractionalizes on the anyons \cite{Essin2013, Mesaros2013}.  Let us for the moment review the better understood case where the symmetry group $G$ does not permute any of the topological superselection sectors.  Then, because anyons are excitations that cannot be created locally, they may carry fractional symmetry quantum numbers.  For example, if $G$ is spin rotation symmetry $SO(3)$, certain anyons might carry half integral spins.  This fractionalization of the symmetry on a given anyon $b$ is captured by a collection of Berry phases $\{\omega_b(g,h)\}$, where $g,h$ are group elements in $G$:

\begin{equation} \label{eq:def_Berry}
U_g^{(b)} U_h^{(b)}=\omega_b(g,h) U_{gh}^{(b)},
\end{equation}
where $U_g^{(b)}$ is the `local' action of $g$ on the anyon $b$, to be defined in more detail below.  In mathematical language, for each anyon $b$, the set of Berry's phases $\{\omega_b(g,h)\}$ defines a so-called group cohomology class in $H^2(G,U(1))$.  The assignment of fractional symmetry quantum numbers to anyons must also be consistent with the anyon fusion rules, which leads to the compatibility conditions
\begin{equation}
\omega_b(g,h) = \omega_c(g,h) \omega_d(g,h)
\end{equation}
whenever the anyon $b$ is an allowed fusion product of $c$ and $d$.

Another way to study symmetry fractionalization is using twist defects of the symmetry $G$ \cite{Maissam_twist, Teo1, Lindner_defects}, which are just extrinsic modifications of the Hamiltonian that insert a flux of $G$ in a particular location.  For example, an extrinsic defect of the $\Z_2$ spin-flip symmetry in a nearest neighbor Ising model is constructed by reversing the sign of $J \sigma^z_i \sigma^z_{j}$ terms on edges $( i j )$ bisected by a branch cut extending from the defect core out to infinity; see appendix \ref{long_appendix} for a precise definition.  Such extrinsic defects have topological superselection sectors, which can be changed by fusing and splitting anyons from the defect.  Suppose we fix a choice of such superselection sector, denoted $\tau_g$, for each $g$ defect.  Then generically defect fusion rules will close only modulo an anyon ambiguity $a_{g,h}$:
\begin{equation} \label{eq:basic_defect_fusion}
\tau_g \times \tau_h = \tau_{gh} \times a_{g,h}
\end{equation}
Non-trivial symmetry fractionalization is then reflected in the fact that there is no choice of $\tau_g$ which makes all of the $a_{g,h}=1$.  These $a_{g,h}$ (which we will see later can all be chosen to be abelian anyons) are directly related to the Berry phases $\omega_b(g,h)$ defined above.  Indeed, braiding the defect $\tau_g$ around the anyon $b$ gives a local action $U_g^{(b)}$ of $g$ on $b$ discussed above, so that, using eq. \ref{eq:basic_defect_fusion}, the phase difference $\omega_b(g,h)$ in eq. \ref{eq:def_Berry} is just the full braiding phase of $a_{g,h}$ around $b$.  Because of this connection we introduce new notation, and denote $a_{g,h}$ by $\omega(g,h)$.  Thus $\omega(g,h)$, without a subscript, is an anyon-valued function of pairs of group elements, and all of the Berry phases $\omega_b(g,h)$ can be recovered from it:
\begin{equation}
\omega_b(g,h)=\langle \omega(g,h), b \rangle,
\end{equation}
with the angular brackets denoting the full braiding phase.  In mathematical language, $\omega(g,h)$ defines a group cohomology class valued in the abelian anyons, i.e. an element of $H^2(G,{\cal A}_{\rm{abelian}})$; see appendix \ref{long_appendix} for more details.

The discussion so far applies only to the special case of $G$ acting trivially on the quasiparticle topological superselection sectors.  To what extent does it generalize to a situation where the symmetry might non-trivially permute the topological superselection sectors?  Symmetries with such non-trivial permutation action have been dubbed `anyonic symmetries' \cite{Hughes1, Teo_Hughes}.  For example, it is possible for a certain $\mz_2$ symmetry to turn an electric `$e$' excitation into a magnetic `$m$' excitation in the $\mz_2$ toric code \cite{Bombin,Kitaev_Kong}.  In this general permuting (or `twisted') setting, it is difficult to make sense of fractional symmetry quantum numbers assigned to anyons, since even the notion of a local action of $G$ on anyons does not make sense: {\emph{e.g.}} there is no local operator that turns an `$e$' anyon into an `$m$' anyon in the toric code.  Another complication is that in the permuting case, extrinsic twist defects are generically non-abelian \cite{Lu2013, Maissam_twist, You_Wen, Hughes1, Oleg1, Mesaros1}.  This makes it more difficult to study their fusion rules and extract from them any information about the symmetry enriched phase.

In this paper, we study this general permuting situation.  Our first approach is to build concrete exactly solved Hamiltonians which realize symmetries that permute anyons.  This is inspired by work of Hermele \cite{Hermele_string_flux}, who built such models for $\Z_n$ gauge theories with non-permuting symmetries, and found a class of distinct SET Hamiltonians naturally parametrized by a function $\omega(g,h)$ in $H^2(G,\Z_n)$.  The physical interpretation of this $\omega(g,h)$ is exactly what was discussed above, with the $\Z_n$ in $H^2(G,\Z_n)$ interpreted as the subgroup of fluxes $(0,k)$, $k=0,1,\ldots,n-1$, among the set of all anyons, which are are just the charge $j$, flux $k$, composites $(j,k)$, $j,k=0,1,\ldots,n-1$.  In this special context of a $\Z_n$ gauge theory we will abuse notation slightly and identify this $\Z_n$ subgroup of fluxes with the multiplicative group of $n$th roots of unity $e^{2\pi i k/n}$, $k=0,1,\ldots,n-1$, so that we can equivalently think of $\omega(g,h)$ as being $U(1)$-valued.  This just amounts to identifying $\omega(g,h)$ with the braiding phase $\omega_{(1,0)}(g,h)$ of the fundamental $\Z_n$ charge $(1,0)$ around $\omega(g,h)$, which contains all the information about $\omega(g,h)$ in this special $\Z_n$ gauge theory case.  We will make this identification throughout sections \ref{our_SET} and \ref{gauge_analyze} of our paper, which deal only with $\Z_n$ gauge theories, and where it will thus not cause confusion.

The model Hamiltonians of reference \onlinecite{Hermele_string_flux} are explicitly designed to produce a Berry's phase of $\omega(g,h)$ for the fundamental $\Z_n$ charge under the $G$ symmetry action.  Now, in our anyon-permuting situation, we find that we can construct a similar class of $G$ symmetric Hamiltonians - again with the topological order of a $\Z_n$ gauge theory - with only a slight modification of the constraints on $\omega(g,h)$.  These modified constraints turn out to define a mathematically well known generalization of group cohomology, called twisted group cohomology, $H^2_{\rm{twisted}}(G,{\cal A}_{\rm{abelian}})$.  The symmetry in these models ends up permuting the gauge charges, and also permuting the gauge fluxes in the same way.

Although generalizing the models of reference \onlinecite{Hermele_string_flux} to the permuting case is rather straightforward, the physical interpretation of $\omega(g,h)$ is now less clear.  First of all, as discussed above, the naive interpretation of $\omega(g,h)$ in terms of symmetry fractionalization on the $\Z_n$ charges is unavailable to us in this permuting setting.  One can still study extrinsic twist defects of the symmetry however, and hope that $\omega(g,h)$ shows up in their fusion rules, as in equation \ref{eq:basic_defect_fusion} in the non-permuting case.  However, it turns out this is not always the case: there exist gauge inequivalent choices of $\omega(g,h)$ in our models which nevertheless give rise to the same defect fusion rules, at the level of superselection sectors.\footnote{These fusion rules for defects in a permuting theory will generically be non-abelian.}  The corresponding Hamiltonians then cannot be distinguished by the fusion rules of the defects, at least at the level of superselection sectors.  Nevertheless, these Hamiltonians do define distinct SET phases, as we check by fully gauging $G$ in our models and examining the statistics of the resulting quasiparticle excitations, which turn out to be different in the two cases.  Indeed, the gauged models have the topological order of an $E$ gauge theory, where $E$ is a `twisted' product of $G$ and $\Z_n$, with the twist determined by $\omega(g,h)$; distinct $\omega(g,h)$ give rise to distinct $E$.

A more complete picture of how defect fusion data relate to SET order can be obtained by studying defect fusion not only at the level of superselection sectors, but also at the level of `F-matrices', i.e. associativity relations for the fusion of defects and anyons.  At this level, it turns out that gauge inequivalent choices of $\omega(g,h)$ do indeed give rise to inequivalent collections of defect fusion and associativity data.  In particular, even when the defect fusion rules are the same at the level of superselection sectors for two such theories with inequivalent $\omega(g,h)$, the two collections of F-matrices will be distinct and gauge inequivalent.  In order to see this, we move beyond our specific class of lattice model examples, and develop a general framework for studying arbitrary SETs with permuting symmetries.  The basic assumption in this formal algebraic approach is that extrinsic defects can be braided and fused with each other and with the anyons.  Just as in the case of ordinary anyons, whose fusion and braiding structures - namely unitary modular tensor categories (UMTCs) - are highly constrained, the algebraic structures in the present setting involving extrinsic defects, so-called `braided $G$-crossed categories' \cite{Maissam_twist, Teo1}, are also highly constrained.  Note that these are not the same structures, because extrinsic defects do not behave exactly like anyons: instead, they have branch cuts which are visible to the other excitations.  For example, braiding around a defect can change anyon type, something that is not allowed in a UMTC.

Classifying all braided $G$-crossed categories is at least as difficult as classifying UMTCs, since the latter are a subset of the former.  However, in trying to distinguish SETs, we are really interested in the simpler problem of classifying all braided $G$-crossed categories with a given fixed anyon content and symmetry group $G$.  This classification problem has been solved in reference \onlinecite{ENO} and the resulting mathematical machinery has been applied to classify SETs in references \onlinecite{Maissam_twist, Teo1}.  Using this general classification, one finds an invariant which distinguishes braided $G$-crossed categories with the same permutation action of $G$ which is valued in $H^2_{\rm{twisted}}(G,{\cal A}_{\rm{abelian}})$.  This invariant reduces to the ordinary fractionalization class in the non-permuting case, where it is seen in the defect fusion rules already at the level of superselection sectors.  In the more general permuting case, though, it can generically only be obtained from knowledge of both fusion rules and F-matrices involving 2 defects.  We will review the stepwise construction of braided $G$-crossed categories, following reference \onlinecite{ENO} and using an intuitive graphical calculus, and see explicitly how the invariant shows up in fusion and F-matrix data.

To connect this formal approach to our class of model Hamiltonians, we study a specific example: a $\Z_4$ gauge theory with a $\Z_2$ symmetry acting by $(i,j)\rightarrow(4-i,4-j)$ on the charge/flux composites.  There are two distinct lattice Hamiltonians of the type we consider with this symmetry action, corresponding to two inequivalent sets of Berry phases $\omega_{+}(g,h)$ and $\omega_{-}(g,h)$, and they are exactly of the type discussed above: their defect fusion rules are identical at the level of superselection sectors, but they correspond to distinct SETs, because they gauge to different topologically ordered theories.  Therefore, they should differ in their F-matrix data, for F-matrices involving two defects and an anyon.  It is difficult to extract such F-matrix data from the lattice Hamiltonians directly, but fortunately, because of the strong algebraic constraints within the braided $G$-crossed category, this F-matrix data is also reflected in defect braiding data.  Specifically, we will see that, for this example, the F-matrix data should be encoded in certain anyon-defect braiding processes, and we explicitly confirm that this is the case for our lattice models.

The remainder of the paper is structured as follows.  In section \ref{our_SET} we construct our exactly solved lattice SET models.  As in reference \onlinecite{Hermele_string_flux}, they are given by coupling $|G|$ copies of a $\Z_n$ gauge theory (here $|G|$ is the number of elements in the group $G$), although in our case the symmetry action non-trivially permutes the $\Z_n$-charges among themselves, and similarly for the $\Z_n$-fluxes.  In section \ref{gauge_analyze} we explicitly gauge $G$ in these models, study the topological superselection sectors of their defects, and show that the topological order of the gauged theory is the quantum double of the non-central extension of $G$ by $\Z_n$ determined by $\omega(g,h)$, generalizing the non-permuting result of \cite{Hermele_string_flux}.  In particular, whenever these non-abelian gauge theories are distinct, so are the underlying SETs, showing that this construction does indeed produce non-trivial SETs.  Of course, these SETs are far from the most general ones possible - in particular, since after gauging we obtain discrete gauge theories, all of our defects have integral quantum dimension.  Nevertheless, they still form a wide class of explicit realizations of the various phases allowed by the recent general classification of SETs in two dimensions.  In particular, we discuss in detail the simplest example of $\Z_4$ gauge theories with symmetry $G=\mz_2$ acting by $k \rightarrow -k$ for $k\in \Z_4$, where there are two symmetry enriched phases, which give non-abelian ${\mathbb D}_4$ (dihedral group of symmetries of the square) and ${\mathbb Q}_8$ (quaternion group) gauge theories respectively upon gauging $G$.  Finally, in section \ref{general_theory} we develop the general theory of defect fusion rules and their deformations, applicable both in the non-permuting and permuting cases.  We use a graphical formalism to introduce the mathematical description of defect superselection sectors, and describe defect fusion rules within this formalism.  Mathematical results of reference \onlinecite{ENO} then allow us to enumerate the gauge equivalence classes of such defect products, and show that they are in one to one correspondence with $H^2_{\rm{twisted}}(G,{\cal A})$.  We then again study the $\Z_4$ gauge theory example mentioned above, and treat it within the context of this general theory.  Finally, we summarize and discuss new directions in section \ref{summary}.

%\section{Physical picture of symmetry localization and defect fusion}
%\label{physical_picture}
%In this section we give a physical picture of symmetry localization and defect fusion, using lots of pictures and not much math...

%%%%%%%%%%%%%%%%%%%%%%%%%%%%%%%

\section{Exactly solved lattice Hamiltonian}
\label{our_SET}

In this section we write down a family of exactly solved lattice models of $G$-symmetric Hamiltonians, with $G$ permuting the anyons.  The goal here is simply to describe the Hilbert space, operators, and symmetry action in as explicit a way as possible, and motivate the form of the Hamiltonian in equation \ref{eq:SETham}.  In later sections we analyze the models described by this Hamiltonian in detail, and see that they correspond to distinct SETs.

We will take $G$ to be abelian for convenience, though we believe our results generalize to non-abelian $G$.  Although we work with the topological order of an abelian $\Z_n$ gauge theory, our results readily generalize to arbitrary abelian groups.  We also treat the special case $G=\Z_2$, $n=4$ in detail.

\subsection[Hilbert space]{Hilbert space}
 
Our model is a $\Z_n$ gauge theory living on a certain oriented, quasi-$2d$ lattice.  Following reference \onlinecite{Hermele_string_flux}, we start with a truly 2d oriented lattice, which can be taken to be a square lattice in the $xy$ plane for all of the examples we consider, and stack $|G|$ identical copies of it.  This stacking allows us to identify corresponding vertices and links in each copy.  In particular, consider the set of $|G|$ vertices that all have the same $x,y$ coordinate.  For any ordered pair $(v,w)$ of such vertices, we add an oriented link connecting $v$ to $w$.  For clarity, we refer to these $|G|(|G|-1)$ links as \emph{vertical}, as opposed to the links within layers, which will be called \emph{horizontal}.  The orientation of horizontal links is the same across all layers.

The set of $|G|$ corresponding vertices together with the $|G|(|G|-1)$ links connecting them will also be referred to as a {\emph{supervertex}} (reference \onlinecite{Hermele_string_flux} calls this a Cayley graph).  Likewise, the set of $|G|$ links which project to the same $2d$ link will be referred to as a {\emph{superlink}}.

\begin{figure}[htbp]
\begin{center}
\includegraphics[width=0.4\textwidth]{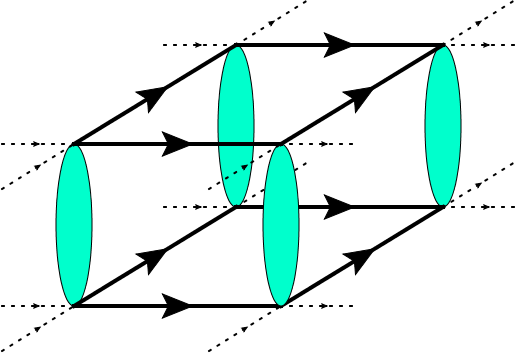}
\caption{Lattice on which our model is defined.  The links carry $\Z_n$ labels.}
\label{lattice}
\end{center}
\end{figure}

The Hilbert space is taken to be spanned by $\Z_n$ labellings of the links of our lattice.  From now on we will identify $\Z_n$ with the set of $n$'th roots of unity, i.e. complex numbers of the form $e^{2\pi i j/n}, j=0,\ldots,n-1$.  Formally, we define an $n$ dimensional link Hilbert space $\hil_l$ whose basis states are in one to one correspondence with such roots of unity:
\begin{align}
\hil_l& = \spn\{~\ket{\eta_l}~| \eta_l \in \{ e^{2\pi i j/n}, j\in \Z \}\}
\end{align}
and take the total Hilbert space $\hil$ to be the tensor product of these link Hilbert spaces (including both horizontal and vertical links):
\begin{align}
\hil &= \bigotimes_l \hil_l.
\end{align}
On each link Hilbert space we define the usual `phase' and `charge' measuring operators $a_l$ and $e_l$:
\begin{align}\label{eq:el}
e_l \ket{\eta_l} &= \ket{ e^{2\pi i/n}\eta_l} \\ 
\label{eq:al}
a_l\ket{ \eta_l} &=\eta_l\ket{ \eta_l}
\end{align}
By tensoring with the identity on all other links, we can think of $a_l$ and $e_l$ as being defined on the total Hilbert space $\hil$.  Note that two such operators acting on different links $l$ and $l'$ commute.

The Hamiltonians we will work with contain terms which act on certain groupings of links, associated to vertices and plaquettes, and before we can write them down we need to establish some effective notation.  First of all, as we mentioned above, our quasi-2d lattice is oriented, which means that there is a preferred choice of direction for each link.  This orientation is efficiently encoded in a function $s_v(l)$, where $l$ is any link and $v$ is one of the two endpoint vertices of this link:
\begin{align}\label{sv_orient}
s_v(l) &=
\begin{cases}
1 &\text{ if } l \text{ points toward } v\\
-1 &\text{ if } l \text{ points away from } v.
\end{cases}
\end{align}
We will assume that our orientation is consistent across the $|G|$ layers, i.e. $s_v(l)=s_{v'}(l')$ whenever $v,v'$ are in the same supervertex, and horizontal links $l,l'$ are in the same superlink.

Additionally, we now also assign an orientation to all plaquettes $p$ (i.e. plaquettes involving any combination of horizontal and vertical links).  This orientation is just a choice of direction, either clockwise or counterclockwise, along the links that border $p$.  For each such link $l$ bordering a plaquette $p$, this choice of direction could be the same or opposite to the one defined by Eq. \ref{sv_orient}.  This distinction is encoded in a function $s_p(l)$, where $l$ is a link bordering the plaquette $p$:
\begin{align}
\label{eq:spl}s_p(l) &=
\begin{cases}
1 &\text{ if } l \text{ oriented with } p\\
-1 &\text{ if } l \text{ oriented against } p
\end{cases}
\end{align}
We will assign this plaquette orientation consistently across the layers, in that if $p$ and $p'$ are plaquettes made up entirely of horizontal links that project to the same plaquette in the $xy$ plane, we assign them the same orientation.  This just means that if $l$ and $l'$ are corresponding links of $p$ and $p'$ respectively (so that $l,l'$ are in the same superlink), then $s_{p}(l) = s_{p'}(l')$.  Besides this constraint, the plaquette orientations are chosen arbitrarily.

Now, in reference \onlinecite{Hermele_string_flux}, $G$ acts by permuting layers, and since such a permutation induces a one to one mapping of the underlying oriented quasi-2d lattice to itself, an example of a $G$-invariant Hamiltonian is:
\begin{align}
\label{eq:notwist}
\ham &= -\sum_v  A_v - \sum_p B_p + h.c.\\
A_v &= \prod_{l\sim v} e^{s_v(l)}_l\\
B_p &=  \prod_{l\in p} a^{s_p(l)}_l.
\end{align}
where the notation $l \sim v$ refers to all links $l$ that begin or end at $v$, and $l \in p$ refers to all the links that border a given plaquette $p$.  While eq. \ref{eq:notwist} is adequate in the case where the symmetry does not permute the gauge theory quasiparticles (anyons), we will need a slightly different construction for a symmetry action which does permute the anyons.

\subsection{Symmetry action and Hamiltonian}

In our model, $G$ will act both by permuting the links and changing the $\Z_n$ labels on the links.  The permutation of the links induced by $g\in G$ is the same as that in reference \onlinecite{Hermele_string_flux}: given a vertex $v$ in layer $h$, we let $gv$ denote the vertex in layer $gh$ which is in the same supervertex as $v$.  Then, for a link $l = \langle v\, v' \rangle$, we define $gl=\langle gv \,gv' \rangle$.  The change in the $\Z_n$ label that goes together with this link permutation - which is the new feature of our model, and is referred to as a twisting - is encoded in an integer valued function $\rho(g)$, with each $\rho(g)$ relatively prime to $n$ (i.e. having no common factors with $n$).  $\rho$ is required to satisfy $\rho(gh)=\rho(g)\rho(h)$ (note that this is multiplication of integers modulo $n$) and allows us to define a permutation action of $G$ on $\mz_n$, namely $\eta \rightarrow \eta^{\rho(g)}$.  More explicitly, if $\eta=e^{2\pi i k/n}$, then this action just takes $k\rightarrow \rho(g) k \,\,\rm{mod}\,\,n$.  An example that we will focus on in the rest of the paper is $G=\mz_2$ and $n=4$, and $\rho(g)=-1$ for the non-trivial generator $g$ of $\mz_2$.

Using $\rho$, we define the global action of $G$ on the Hilbert space as follows.  With a slight abuse of notation, we denote the unitary action of $g$ by $U_g$, regardless of what Hilbert space is being acting upon.  For the link degrees of freedom we let:
\begin{align}\label{eq:def_Uf}
U_g \ket{\eta_l} &=\ket{\eta_{gl}^{\rho(g)}}
\end{align}
This induces the action on operators:
\begin{align}\label{eq:Ufl_operators}
U_g a_l U_g^{-1} &=a_{gl}^{\rho(g^{-1})} \\
U_g e_l U_g^{-1} &= e_{gl}^{\rho(g)}
\end{align}
We can immediately infer that the action on vertex and plaquette terms defined in eq. \ref{eq:notwist} is
\begin{align}
U_g B_p U_g^{-1} &=B_{gp}^{\rho(g^{-1})} \\
U_g A_v U_g^{-1} &= A_{gv}^{\rho(g)},
\end{align}
where $gp$ is the plaquette made up of the links $gl$, for all $l\in p$.  Note that for $\rho(g)\neq 1$, the Hamiltonian defined in eq. \ref{eq:notwist} is not invariant under this global action of $g$.  Instead, we consider the following more general Hamiltonian:
\begin{align}
\label{eq:twist}
\ham_{SET}  &= -\sum_{m=1}^n \left(\sum_v  A^{m}_v + \sum_p \left( \omega_p^{-1} B_p\right)^{m}\right).
\end{align}
Here $\omega_p$ are phases - in fact, $n$th roots of unity - associated with each plaquette $p$, which satisfy

\begin{align} \label{eq:omega_cond}
\omega_{gp} = \omega_p^{\rho(g)}
\end{align}
We can verify that $\ham_{SET}$ is $G$-invariant:
\begin{align}
U_g \ham_{SET} &U_g^{-1}=\notag\\
&\phantom{{}=}  -\sum_{m=1}^n \left(\sum_v  A^{m\rho(g)}_{gv} + \sum_p \left( \omega_p^{-1} B^{\rho(g^{-1})}_{gp}\right)^{m}\right)\notag\\
&=  -\sum_{m=1}^n \left(\sum_v  A^{m\rho(g)}_{gv} + \sum_p \left( \omega_p^{-\rho(g)} B_{gp}\right)^{m}\right)\notag\\
\label{eq:gaction}
&= \ham_{SET},\text{ if }\omega_{gp} = \omega_p^{\rho(g)}
\end{align}
Thus, for $\omega_p$ which satisfy $\omega_{gp} = \omega_p^{\rho(g)}$, eq. \ref{eq:twist} describes a Hamiltonian that is invariant under the twisted $G$ action.

Let $\ket{\Psi}$ be a state of $\hil$ corresponding to a specific $\Z_n$ labeling of links.  Recalling that the spectrum of $B_p$ consists of the roots of unity $e^{2\pi i k/n}$, we see that
\begin{align}\label{eq:proj_def}
\sum_{m=1}^n \left( \omega_p^{-1} B_p\right)^{m}\ket{\Psi}=
\begin{cases}
n\ket{\Psi} &\text{ if } B_p\ket{\Psi} = \omega_p\ket{\Psi}\\
0 &\text{ otherwise}
\end{cases}
\end{align}
Thus the operator defined on the left side of eq. \ref{eq:proj_def} is equal to $n$ times a projector.  We now describe a notation that will let us concisely express such operators; we emphasize that this formulation is nothing more than a notational convenience.  First, recall that the \emph{regular} representation of a group $H$ is an $|H|$ dimensional vector space with basis $\{ |h'\rangle | h'\in H\}$, where $h\in H$ acts by
\begin{equation} \label{eq:regular_rep}
h: |h'\rangle \rightarrow |hh'\rangle
\end{equation}
Thus $h\in H$ is represented by an $|H|$ by $|H|$ matrix $M(h)$, where each column and row is labeled by a group element $g,k \in H$, and the matrix elements are
\begin{align}
M(h)_{g,k}= \delta_{g,hk}
\end{align}
A feature of these matrices is that $\Tr M(h) =|H|\delta_{h,1}$. Now, recall that the operator $a_l$ acts by the phase $\eta_l$ on $\ket{\eta_l}$. In our new notation, acting with the operator $a_l$ yields the matrix $M(\eta)$, where now $H=\Z_n$. In the case where $n=4$ we have
\begin{align}
M\left(e^{\frac{i\pi}{2}}\right)=
\begin{pmatrix}
	0 & 0	& 0	& 1\\
	1 & 0	& 0	& 0\\
	0 & 1	& 0	& 0\\
	0 & 0	& 1	& 0
\end{pmatrix}
\end{align}
To construct the plaquette terms, we take a trace of the matrix produced by a closed loop of $a_l$'s.  For example, consider a triangular plaquette $p$ with links $1,2,3$, and $\omega_p=1$, and suppose $\ket{\Psi}$ is an eigenvalue $\eta_j$ eigenvector of each $a_j$.  Then:
\begin{align}
\left(\Tr a_1 a_2 a_3\right) \ket{\Psi}&=\left( \Tr M(\eta_1)M(\eta_2)M(\eta_3)\right)\ket{\Psi}\\
&=
\begin{cases}
n\ket{\Psi} &\text{ if } \eta_1\eta_2\eta_3 = \mathbf{1}\\
0 &\text{ otherwise}
\end{cases}
\end{align}
Thus, the operator in \ref{eq:proj_def} can be rewritten in this new notation as:
\begin{align}
\sum_{m=1}^n \left( \omega_p^{-1} B_p\right)^{m} \rightarrow \Tr  \omega_p^{-1} B_p
\end{align}
yielding a notationally convenient way of writing $n$ times the projector onto the eigenvalue $1$ subspace of $\omega_p^{-1} B_p$. Here it is understood that the complex number $\omega_p$ is substituted with its regular representation matrix.  Notice that the trace on the right hand side of this equation is over the auxiliary regular representation, and not the many body Hilbert space; both sides are operators in the many body Hilbert space.

With this new notation, our Hamiltonian takes the form
\begin{align} \label{eq:hSET}
\ham_{SET} = -\sum_v\left( \sum_{m\in \Z_n} A^m_v \right) -\sum_p \Tr \omega^{-1}_p B_p
\end{align}
One benefit of our new notation is that it makes it easy to generalize $\ham_{SET}$ to the case of an arbitrary abelian gauge group $H$, rather than just $\mz_n$.  Indeed, to do this one just needs to let $\rho$ be a map from $G$ to $\text{Aut}(H)$, the group of automorphisms of $H$.  Nevertheless, we will stick to a $\mz_n$ gauge group in the remainder of this paper.  

\subsection{Supervertices and superlinks}

The next step is to discuss the phases $\omega_p$, different choices of which will give rise to different twisted symmetry enriched phases.  To facilitate this discussion, it is useful to first group the degrees of freedom in our model in a slightly more convenient way.  First, recall that a supervertex $V$ is a collection of $|G|$ vertices which all project onto the same point in the $xy$ plane, i.e. are vertically aligned.  We now tensor the Hilbert spaces living on these links into a single supervertex Hilbert space $\hil_V$ spanned by $\{ \ket{ \eta_{V}(g,h)} \}$, where $\eta_{V}(g,h)$ is the $\Z_n$ label of the link $l_{g,gh}$ connecting layer $g$ to layer $gh$ (here $h$ is necessarily different from the identity in $G$).  We then denote by $a_V(g,h), e_V(g,h)$ the action of the operators $a_{l_{g,gh}}, e_{l_{g,gh}}$, tensored with the identity on the remaining links, in the Hilbert space $\hil_V$.

Similarly, we define a superlink $L$ to be the collection of $|G|$ horizontal links whose projections to the $xy$ plane are all the same, and likewise define a superlink Hilbert space $\hil_L$ to be the tensor product of the associated $|G|$ link Hilbert spaces.  We denote by $a_L(g), e_L(g)$ the action of the operators $a_{l_g}, e_{l_g}$ on the link $l_g \in L$ in layer $g$, tensored with the identity on the remaining $|G|-1$ links.

Our total Hilbert space $\hil$ is thus a tensor product of the supervertex and superlink Hilbert spaces:

\begin{align} \label{eq:newh}
\hil= \left(\bigotimes_V \hil_V \right) \otimes \left(\bigotimes_L \hil_L \right)
\end{align}

\subsection{Conditions on the fluxes $\omega_p$ \label{ssect:H2}}

We now discuss the choice of $U(1)$ phases $\omega_p$, which we also refer to as $\mz_n$ fluxes, since they are restricted to take values in the $n$'th roots of unity.  First of all, throughout this paper we will deal exclusively with the situation where the only non-trivial $\omega_p$ (i.e. $\omega_p \neq 1$) occur for plaquettes $p$ that sit entirely within a single supervertex, or, in other words, contain no horizontal links.  Let us focus on a specific supervertex $V$.  Each vertical plaquette $p$ within it is labeled by a triple $(f,g,h)$, where $g,h \neq 1$, indicating that it involves the links connecting layers $f$, $fg$ and $fgh$.  The corresponding term in the Hamiltonian (eq. \ref{eq:hSET}) is:

\begin{align} \label{eq:supervertex_plaq}
\Tr  \omega^{-1}_p B_p &\equiv \Tr  \omega^{-1}_p  B_{V}(f,g,h) \notag\\
&= \Tr  \omega^{-1}_p a_{V}(f,gh)^{-1}a_{V}(fg,h)a_{V}(f,g)
\end{align}
While most plaquettes will be 3-edged, we are allowed to set $h = g^{-1}$, producing a degenerate, 2-edged plaquette. This situation can still be captured by eq. \ref{eq:supervertex_plaq} by defining $a_V(f,e) \equiv \mathbf{I}$ (the identity operator).  

Using eq. \ref{eq:omega_cond} we see that all of the $\omega_p$ are uniquely determined by the $\omega_p$ for $p=(e,g,h)$, i.e. for plaquettes $p$ which start at the identity element of $G$.  Letting $\omega(g,h)$ denote $\omega_p$ for $p=(e,g,h)$, we then have that for a general plaquette $p'=(f,g,h)$, $\omega_{p'}=\omega(g,h)^{\rho(f)}$.

\begin{figure}[htbp]
\begin{center}
\includegraphics[width=0.4\textwidth]{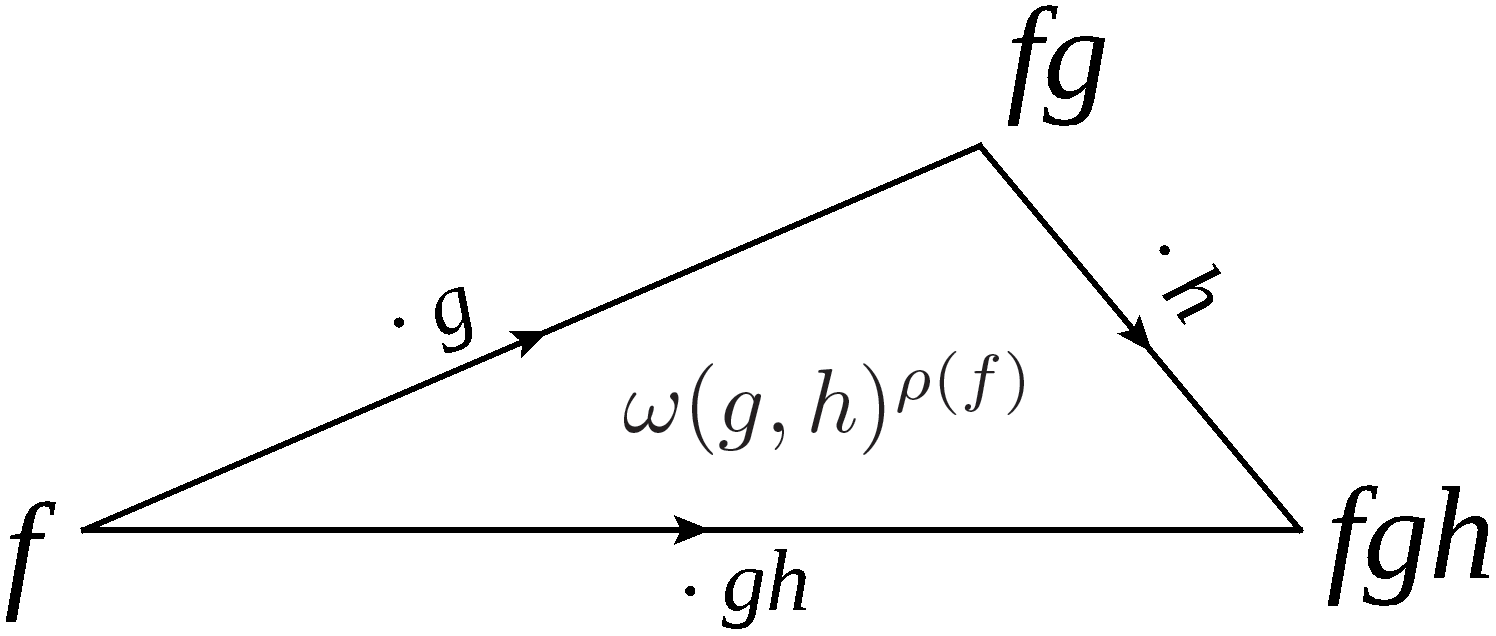}
\caption{A diagram showing the flux piercing the plaquette $(f,g,h)$. Note that it spans layers $f$, $fg$ and $fgh$}
\label{tetrahedron}
\end{center}
\end{figure}

\begin{figure}[htbp]
\begin{center}
\includegraphics[width=0.4\textwidth]{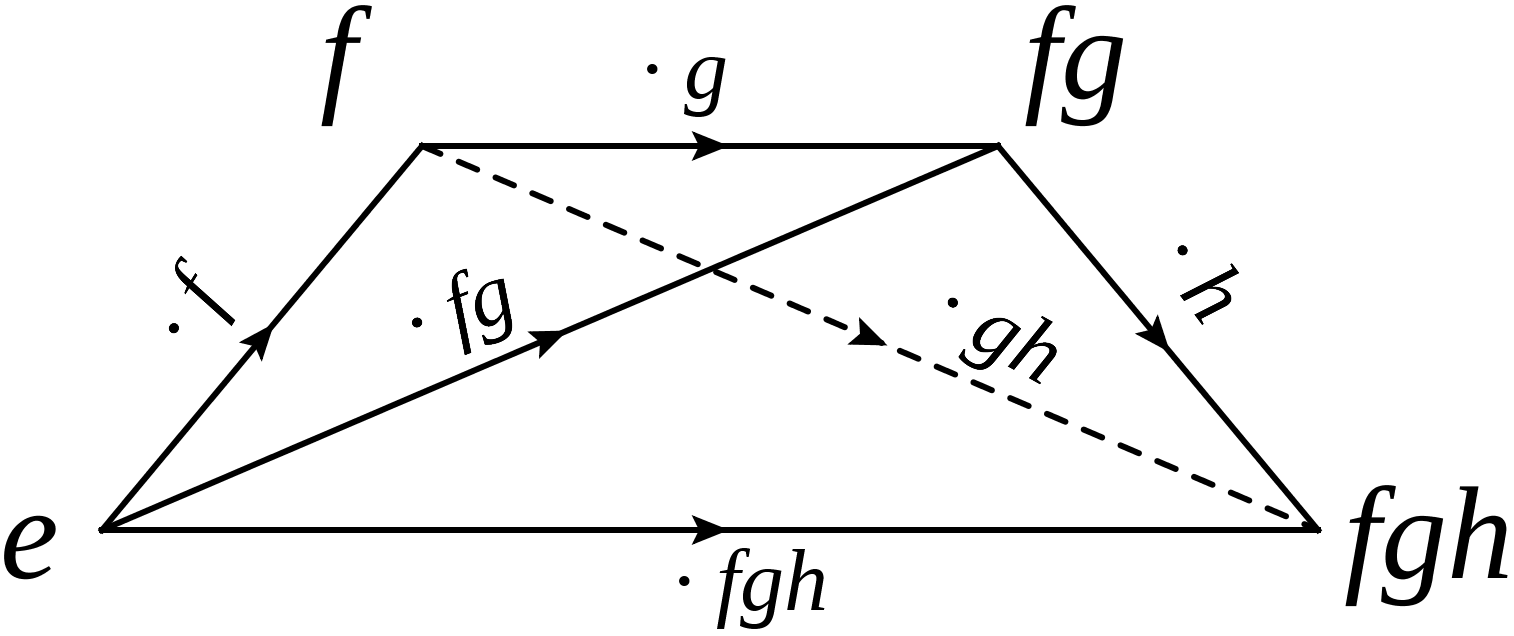}
\caption{A tetrahedron spanning the layers $\{e,f,fg,fgh \}$. The total $\Z_n$ flux emanating out of this tetrahedron must be trivial (eq. \ref{eq:cocycle1}) in order to avoid a degenerate set of frustrated ground states.}
\label{tetrahedron}
\end{center}
\end{figure}

Now, we would like to work with models which are unfrustrated, {\emph i.e.} whose ground states are lowest energy eigenstates of all of the plaquette terms.  Consider the tetrahedron formed by the layers $\{e,f,fg,fgh\}$, which contains 4 plaquettes: $(e,f,g)$, $(f,g,h)$, $(e,fg,h)$, and $(e,f,gh)$.  A necessary and sufficient condition for the model to be unfrustrated is that the $\Z_n$ flux emanating out of any such tetrahedron be zero, {\emph i.e.}
\begin{align}\label{eq:cocycle1}
\omega(g,h)^{\rho(f)}\omega(f,gh) = \omega(f,g)\omega(fg,h).
\end{align}
Indeed, if the ground state is unfrustrated, there must be some labeling $\{\eta_V(g,h)\}$ of the links in the supervertex $V$ (namely one that corresponds to a configuration that enters the unfrustrated ground state with non-zero amplitude) such that
\begin{align}
\eta_V(f,g)\eta_V(fg,h)\eta^{-1}_V(f,gh)=\omega(g,h)^{\rho(f)},\forall f,g,h \in G
\end{align}
Expressing $\omega$ in terms of the $\eta_V$ using this equation, we see that it satisfies eq. \ref{eq:cocycle1}.  Conversely, given a choice of $\omega$'s that satisfy eq. \ref{eq:cocycle1}, we can simply set $\eta_V(g,h) = \omega(g,h)$.  This link labeling satisfies all of the plaquette terms within the supervertex $V$.  Later, we will see that it can be extended to an unfrustrated ground state of all of the vertex and plaquette terms in our model - indeed, we will explicitly solve the model for any choice of fluxes satisfying eq. \ref{eq:cocycle1}.

Certain different choices of $\omega(f,g)$ actually define Hamiltonians which can be made equivalent by redefining link variables:
\begin{equation}
\eta_{V}(g,h) \rightarrow \mu(h)^{\rho(g)} \eta_{V}(g,h)
\end{equation}
where $\mu:G \rightarrow \Z_n$ is an arbitrary function.  This redefinition then takes
\begin{align}\label{eq:cobound}
\omega(g,h) \rightarrow \omega'(g,h) = \omega(g,h) \mu(h)^{\rho(g)}\mu(g)\mu(gh)^{-1}.
\end{align}
Note that the new $\omega'(g,h)$ also satisfy equation \ref{eq:cocycle1}.  In group cohomology terms this means that equivalence classes of non-frustrated Hamiltonians of the above form are parametrized by \emph{twisted} group cohomology classes in $H_{\rho}^2(G,\Z_n)$.\footnote{To interpret $\omega(g,h)$ as a group cocycle, we also need to define its values when either $g=1$ or $h=1$, which are cases that do not correspond to any plaquettes in our model.  We simply set these equal to $1$.} 

For simple enough $G$, it is easy to compute these cohomology groups explicitly.  For example, take $G=\Z_2 = \{1,-1\}$, $n=4$.  Then there is only a single plaquette, bounded by the links $1 \rightarrow -1, -1 \rightarrow 1$, which is pierced by a $\Z_4$ flux $\omega(-1,-1)$, subject to a twisting $\rho(-1) = - 1$. Equation \ref{eq:cocycle1} produces a non-trivial constraint only for $f,g,h = -1$:
\begin{align}
\omega(-1,-1)^{\rho(-1)} &= \omega(-1,-1)
\end{align}
which implies $\omega(-1,-1)=\pm 1$.  As we will see, these two choices of $\omega$, which we call $\omega_{\pm}$, will produce two inequivalent SET Hamiltonians, which in turn yield two distinct non-abelian gauge theories once we gauge the $\Z_2$ symmetry.

\subsection{Form of the SET Hamiltonian}

Let us write out the final form of the SET Hamiltonian in a form convenient for gauging $G$, which we do in the next section.  Recall (eq. \ref{eq:newh}) that our total Hilbert space $\hil$ is a tensor product of supervertex and superlink Hilbert spaces $\hil_V$ and $\hil_L$.  Because all of the links in a superlink are oriented the same way we can define the orientation factor $s_V(L)=s_v(l) = \pm 1$ where $v,l$ are any vertex and adjoining link in $V$ and $L$ respectively.  Now, plaquettes containing only horizontal links can similarly grouped into \emph{superplaquettes} $P$.   Again, because the plaquette orientations have been chosen so that all plaquettes $p$ in a superplaquette $P$ are oriented the same way, we can define $s_P(L)=s_p(l)$ for any $p$ in $P$ and $l$ in $L$ bordering $p$.  We then have the form of the Hamiltonian:

%  We organized vertical links into supervertices $V$ in \ref{ssect:H2}.  We can do the same for horizontal links, resulting in a collection of \emph{superlinks} $L$ which carry states $\ket{\eta_{L}(g)}$, where $\eta_{L}:G \rightarrow \Z_n$.  We assume that we have chosen the graph orientation so that all links in a superlink are oriented the same way; this allows us to define $s_V(L)=s_v(l)$ where $v,l$ are any vertex and link in $V$ and $L$ respectively.  We also introduce new notation for the operator algebra in a superlink: the operators that act on a layer $g$ of a superlink $L$ are:
%\begin{align}
%&a_{L}(g)\ket{\eta_{L}(g)}=\eta_{L}(g)\ket{ \eta_{L}(g)}\\
%&e_{L}(g)\ket{ \eta_{L}(g')}=\ket{ e^{\frac{2\pi i}{n}\delta_{g,g'}}\eta_{L}(g')}
%\end{align}

\begin{align}
\ham_{SET} &= -\sum_{g \in G}\left(\sum_{V}\left( \sum_{m \in \Z_n} A^m_{V}(g) \right) +\sum_{P} \Tr  B_{P}(g)\right)\notag\\
\label{eq:SETham}
&\phantom{{}=}-\sum_{\rm{vertical}\,p} \Tr \omega_p^{-1} C_p
\end{align}
where $V$ and $P$ range over supervertices and superplaquettes respectively, and the various terms in the above sum are defined as follows.  We've rewritten our vertex and plaquette operators as
\begin{align} 
A_V(g) &= \prod_{h \in G} e_V( g, h)^{-1}e_V(gh, h^{-1})\notag\\
&\phantom{{}=}\times \prod_{L\sim V} e_L(g)^{s_V(L)},\label{eq:Avdef}
\end{align}
which denotes the term acting on the vertex on layer $g$ of the supervertex $V$, and
\begin{equation}
B_P(g) =  \prod_{L\in P} a^{s_P(L)}_L(g)
\end{equation}
denotes the  term acting on the plaquette on layer $g$ of the superplaquette $P$.
Finally, for any plaquette $p$ which is not composed solely of horizontal links, which we refer to as `vertical' in eq. \ref{eq:SETham} above,
\begin{equation}
C_p= \prod_{l\in p} a^{s_p(l)}_l.
\end{equation}
Note that there are two distinct kinds of vertical plaquettes: ones contained entirely in a single supervertex, and ones involving a superlink and the adjoining two supervertices.  Only for the ones contained entirely in a single supervertex can we have $\omega_p \neq 1$.

\begin{figure}[htbp]
\begin{center}
\includegraphics[width=0.4\textwidth]{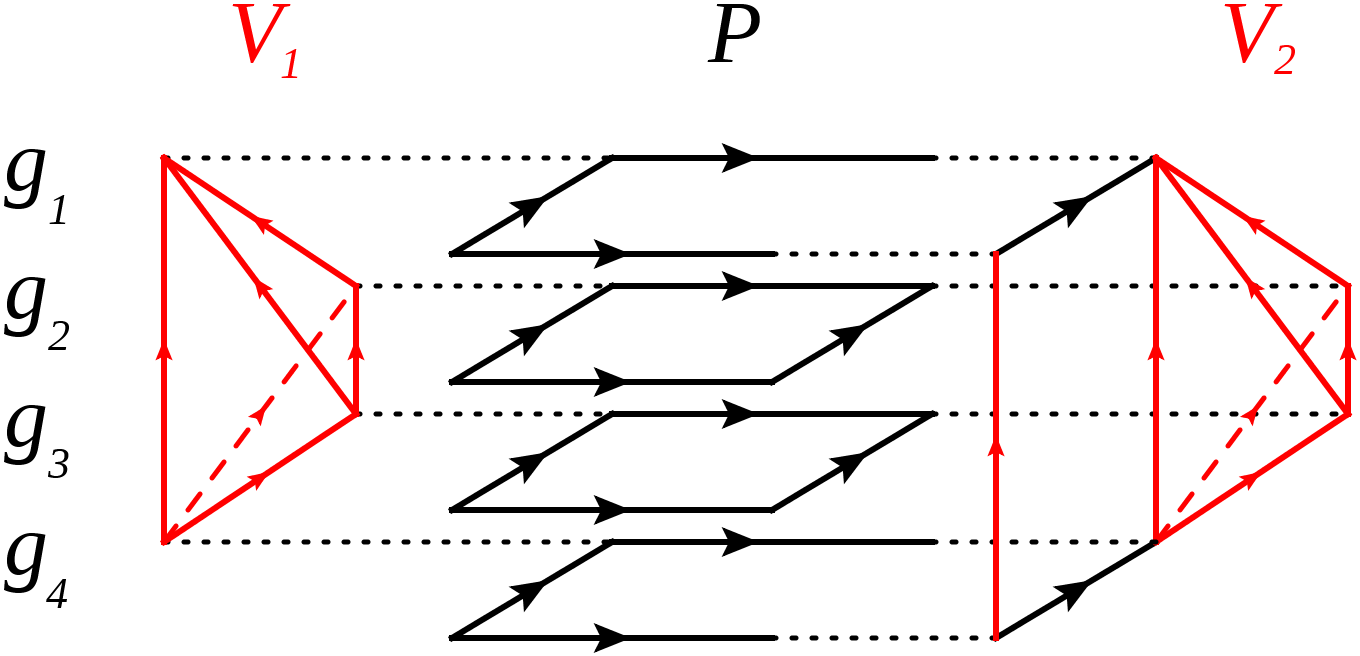}
\caption{Vertical links are colored red and horizontal links are colored black.  Collections of vertical links connecting points with the same x-y coordinate constitute supervertices, labeled $V_1$ and $V_2$.  Plaquettes made entirely out of horizontal links that project to the same plaquette form a superplaquette, labeled $P$ in the figure.  Plaquettes containing at least one vertical link are deemed vertical.  The dotted lines identify vertices and help offset the figure.  Certain links are dashed in order to add perspective.}
\label{lattice}
\end{center}
\end{figure}

\section{Distinguishing SET phases \label{sect:SETphases}}
\label{gauge_analyze}

In the case of symmetries that do not permute anyons, the fluxes $\omega(g,h)$ appearing in the Hamiltonian in equation \ref{eq:SETham} can be physically interpreted as Berry phases for the symmetry action on the fundamental $\Z_n$ charge \cite{Hermele_string_flux}.  However, such an interpretation does not generalize readily to the anyon permuting case, and this motivates us to couple the model to a $G$ gauge field and examine the resulting `gauged' theory.  There are two different versions of such a gauged theory: one can either make the $G$ gauge field a fully dynamical degree of freedom, or one can treat it as a background probe field.  

In the case of a dynamical gauge field, it turns out that different SET Hamiltonians - i.e. different choices of $\omega(g,h)$ - can be distinguished by the statistics of the excitations in the gauged theory.  Demonstrating this fact will take up the bulk of this section.  Indeed, to fully understand the gauged theory with dynamical gauge field $G$, we first perform a minimal coupling procedure of the Hamiltonian in eq. \ref{eq:SETham} to such a dynamical $G$ gauge field, and then perform a series of transformations, analogous to those in reference \onlinecite{Hermele_string_flux}, to simplify the form of the resulting gauged Hamiltonian, without altering the low energy physics.  Although these transformations are technically complicated, there is a simple intuitive picture for what is going on: essentially, the gauge field $G$ inserted along superlinks should be viewed as allowing permutations between the different layers.  When the symmetry is gauged, any potential physical distinction between the different layers is therefore lost, and hence the physical states in the gauge theory live on an ordinary 2d lattice, as opposed to a $|G|$-fold stacked quasi-2d lattice.  Indeed, we find explicitly that the simplified gauged theory is just a discrete gauge theory of a group $E$ on an ordinary 2d lattice.  Here $E$ is a group which has $\Z_n$ as a normal subgroup, $E/\Z_n \cong G$, and $\rho(g)\eta = {\tilde{g}^{-1}} \eta {\tilde{g}}$, where $\tilde{g}$ is any lift of $g$ in $E$.  $E$ is called the group extension of $G$ by $\Z_n$ determined by the permutation $\rho$ and an element of $[\omega]$ of $H^2_{\rho}(G,\Z_n)$.  Each $\omega$ corresponds precisely to one such group extension (See Appendix~\ref{ap:extend}).

The second version of a $G$ gauged theory is one where the $G$ gauge field is treated as a background field - we will refer to this as the non-dynamical case.  Here the fluxes of $G$ are not dynamical excitations, but rather extrinsic defects, requiring a branch cut in the Hamiltonian.  One reason one might want to examine this case is that it is conceptually simpler than that of the fully dynamical $G$ gauge field, as it requires no extra degrees of freedom.  Another reason is that recent work (reference \onlinecite{Maissam2014}) has studied the general mathematical structure encoded in such extrinsic defects, called a braided $G$-crossed category.  We will also discuss general aspects of such braided $G$-crossed categories later in the paper, but in this section we will just study them in the context of the class of lattice models we have just introduced.  Having already coupled these models to a dynamical $G$ gauge field, it turns out that the analysis of the non-dynamical case is easy: in the final form of our dynamical gauged model as an $E$ gauge theory on a 2d lattice, it just amounts to including only vertex terms corresponding to $\Z_n \subset E$, and setting the coefficients of the vertex terms corresponding to other elements of $E$ to $0$.

One may wonder how two of our SET models corresponding to two distinct choices of $[\omega] \in H^2_{\rho}(G,\Z_n)$ can be distinguished when coupled only to a non-dynamical $G$ gauge field.  In particular, can different choices of $[\omega]$ lead to different extrinsic defect types?  It turns out that the answer is no: the properties of a single extrinsic $g$-defect are uniquely determined by $g$ and the permutation $\rho$, and independent of $[\omega]$.  Specifically, the set of topological superselection sectors bound to a single extrinsic $g$-defect, the fusion rules of these superselection sectors with external anyons (that is $\mz_n$ gauge charges and fluxes), and the F-matrix associativity constraints involving two anyons and a defect are all uniquely determined by $g$ and $\rho$ (the F-matrices are unique only up to the appropriate gauge degree of freedom).  This collection of data is known as an invertible bimodule category, and will be discussed in section \ref{general_theory}.  Thus, in order to distinguish SETs with different $[\omega]$, we have to go beyond the case of a single extrinsic defect, and look at pairs of such extrinsic defects and their fusion rules.  Indeed, in the non-anyon-permuting case of reference \onlinecite{Hermele_string_flux}, different choices of $[\omega]$ lead to different fusion rules for extrinsic $g$ defects, at the level of topological superselection sectors.

In our anyon-permuting case, the situation is more subtle.  It is still true that different choices of $[\omega]$ lead to different fusion rules for extrinsic $g$ defects, but not at the level of topological superselection sectors.  In other words, the fusion rules (which tell us which topological superselection sectors in defect $gh$ can end up as the fusion product of specific sectors in defects $g$ and $h$) might be the same for different choices of $[\omega]$.  In this case, the distinction between two such different choices of $[\omega]$ can only be seen in the F-matrices involving two defects and an anyon.  More precisely, since the F-matrices are not gauge invariant, the distinction can only be seen in the gauge equivalence classes of such F-matrices.  We explain this precisely in section \ref{ssec:twisted} below.

We will discuss the general theory of defect fusion in section \ref{general_theory}.  At the end of the present section, however, we will analyze the specific example of a $\Z_4$ gauge theory with $G = \Z_2 = \{1,-1 \}$, with $\rho(-1) = -1$.  We will see that $H^2_\rho(G,\Z_4)=\Z_2$, so there are two inequivalent choices of $\omega(f,g)$, which we denote $\omega_+$ and $\omega_-$.  The group extension $E$ corresponding to the trivial co-cycle $\omega_+$ is $\mathbb{D}_8$ (the dihedral group on 4 points), while that corresponding to the non-trivial one $\omega_-$ is the quaternion group $\mathbb{Q}_8$.  We will find that, regardless of whether we choose $\omega_+$ or $\omega_-$, the extrinsic $\Z_2$ defects in the two SETs have the same superselection sectors and the same fusion rules for these sectors: as discussed above, these are independent of $\omega$.  Thus, to tell the difference between the two theories, we must probe more subtle data.  Indeed, we can either look at the quasiparticle statistics in the dynamical $G$ gauged theory, i.e. the $\mathbb{D}_8$ and $\mathbb{Q}_8$ gauge theories, and see that they are different, or, as discussed above, we can detect the difference in the F-matrices corresponding to the defect fusion rules.

\subsection{Gauging prescription}

The goal of this somewhat technical appendix is to derive eqs. \ref{eq:gauge_ham} and \ref{eq:gauge_dyn}, which describe our SET Hamiltonian in eq. \ref{eq:SETham} coupled to a $G$ gauge field.  First we introduce $G$ gauge field degrees of freedom, which are just $|G|$ dimensional Hilbert spaces which we insert between any supervertex and an $outgoing$ superlink (see figure \ref{lattice}).  The $|G|$ degrees of freedom are incorporated by enlarging each superlink Hilbert space ${\cal H}_{L} \rightarrow {\cal H}_{L} \otimes {\mathbb C}^{|G|}$, with states in this larger Hilbert space carrying an extra $G$ gauge field label $g_{L}$:
\begin{align}
\ket{ \{\eta_{L}(g)\}_g} \rightarrow  \ket{ \{\eta_{L}(g)\}_g, g_{L}}
\end{align}

\begin{figure}[htbp]
\begin{center}
\includegraphics[width=0.35\textwidth]{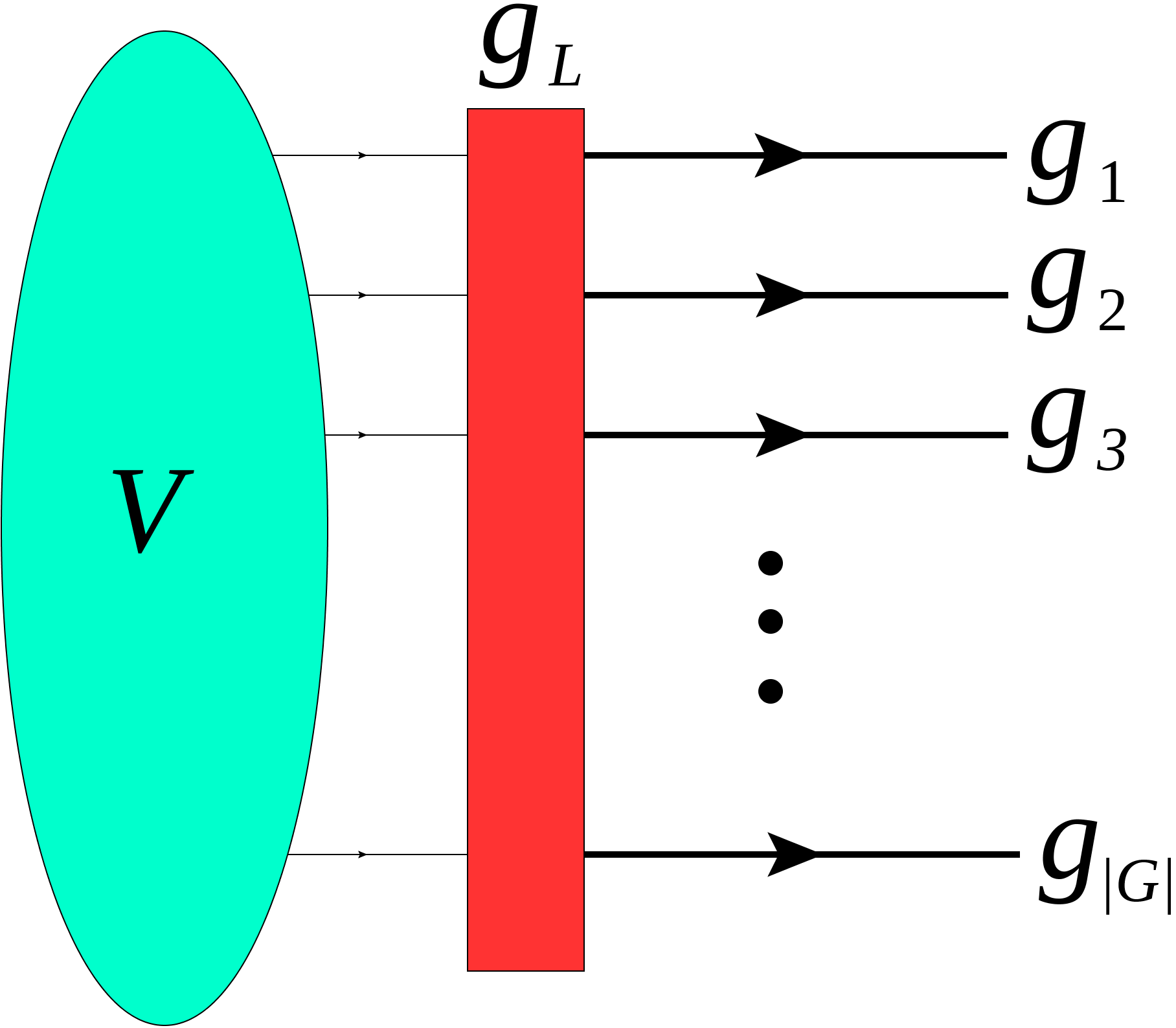}
\caption{Placement of the $G$ gauge variable relative to the supervertex and superlink}
\label{lattice}
\end{center}
\end{figure}
The minimal coupling prescription we use is as follows.  Given any local term in the original ungauged Hamiltonian, for a $G$ gauge field configuration which is gauge equivalent to the trivial configuration in the vicinity of this local term (i.e. has no $G$ fluxes), the form of the corresponding minimally coupled term is completely fixed by $G$ gauge invariance.  For a $G$ gauge field configuration which does contain nonzero $G$ fluxes in the vicinity of this local term, we simply set the corresponding minimally coupled term to $0$.  This actually only occurs for the superplaquette term, when there is a non-zero $G$ flux through it.  Since the Hamiltonian consists of commuting terms which all have negative eigenvalue on the ground state, setting this minimally coupled term to $0$ is actually an energetic penalty for the non-zero $G$ flux.  Lastly, we include `vertex' terms which make the $G$ gauge field fluctuate and thus energetically impose $G$ gauge invariance.  These vertex terms are the only ones which alter the $G$ gauge field configuration.

The above paragraph specifies the minimally coupled Hamiltonian uniquely, but to actually write it out in a compact form it is useful to introduce some additional notation.  It is easiest to start with the vertex terms, given by eq. \ref{eq:Avdef}.  A particular such term involves a specific vertex $v$, which is part of a supervertex $V$.  To minimally couple it, we have to modify each $e_L(g)$ term corresponding to an outgoing superlink $L$ from $V$ (that is, one with $s_V(L)=-1$) to take account of the gauge field $g_L$, and replace it with:
\begin{align}
e_L( g)^R \ket{& \{\eta_L(g')\}_{g'}, g_L } \equiv\notag\\
&\ket{ \{\eta_L(g')e^{-2\pi i\frac{\rho(g_L)}{n}\delta_{g,g'g_L}} \}_{g'}, g_L }
\end{align}
Here $R$ is just a superscript.  To keep the notation compact, we define a superscript-valued function $s'_V(L)=R$ when $s_V(L)=-1$, and $s'_V(L)$ being trivial otherwise, so that we can write the minimally coupled version of $A_V(g)$ simply as:
\begin{align}
\tilde{A}_V( g) &= \prod_{h \in G} e_V(g, h)^{-1}e_V(gh, h^{-1}) \notag\\
&\phantom{{}=}\times\prod_{L\sim V} e_L( g)^{s'_V(L)} \label{eq:A_gauged}
\end{align}

Now let us minimally couple the horizontal plaquette terms $B_{P}(g)$.  When there is a non-trivial $G$ flux through $P$, we simply set ${\tilde B}_{P} (g)=0$.  When this flux is trivial, we first define an auxiliary `book-keeping' operator 
\begin{align}
\al_L \left(\ket{\{\eta_L(g)\}_g,g_L} \otimes \ldots\right) &=\notag\\
 M(g_L)U^{-1}_{g_L}& \left(\ket{\{\eta_L(g)\}_g , g_L}\otimes \ldots \right),
\end{align}
where $M(g_L)$ is the $|G|$ by $|G|$ matrix representing $g_L$ in the regular representation of $G$ (see the discussion around eq. \ref{eq:regular_rep}), $U_{g_L}$ is the global action of $g_L$ (as defined in eq. \ref{eq:def_Uf}), and the ellipses denote the rest of the Hilbert space $\hil$.  Now define
\begin{align}
\tilde{B}_{P}(g) &= \prod_{L\in P} \left(a_L(g)\al_{L}\right)^{s_P(L)} \label{eq:B_gauged}.
\end{align}
Note that even though the operators $\al_L$ are non-local (because they involve the global $U_{g_L}$ operator, when the $G$ flux through $P$ is trivial their combined action in $\tilde{B}_P(g)$ cancels away from the plaquette $P$, and we end up with a local operator.  Also recall that it is the trace of $\tilde{B}_{P}(g)$ over the auxiliary regular representation space which appears in the Hamiltonian.

Next, let us minimally couple the term $C_p$ which involves both a superlink and the adjacent two supervertices.  Any such $p$ is a rectangle, and $C_p$ is a product of $4$ link terms:
\begin{equation}
C_p = (a_L(gh) a_V(g,h))^{-1} a_{V'}(g,h)a_L(g)
\end{equation}
Minimally coupling this inserts the gauge field, in the form of the $\al_L$ operators, along the horizontal links:
\begin{equation}
{\tilde C}_p = (a_L(gh)\al_L a_V(g,h))^{-1} a_{V'}(g,h)a_L(g) \al_L
\end{equation}
Recall that there is never any $G$ flux through such a plaquette $p$.  Finally, the minimal coupling of $C_p$ for a plaquette $p$ entirely within a supervertex $V$ does not involve the gauge field at all, ${\tilde C}_p=C_p$.  

In summary, the minimal coupling is done by starting with the Hamiltonian in eq. \ref{eq:SETham} and making the modifications
\begin{align}\label{eq:gauge_a}
a_L(g) &\rightarrow a_L(g) \al_L\\
s_V(L) &\rightarrow s'_V(L) =
\begin{cases}
\mathbf{1} &\text{ if } L \rightarrow V\\
R &\text{ if } L \leftarrow V,
\end{cases}
\end{align}
resulting in a minimally coupled Hamiltonian $\ham_{\rm{m.c.}}$:
\begin{align}
&\ham_{\rm{m.c.}} = -\sum_{g \in G}\left(\sum_{V}\left( \sum_{m \in \Z_n} \tilde{A}^m_V( g) \right) +\sum_{P} \Tr  \tilde{B}_{P}(g)\right)\notag\\
&\phantom{{}=====}-\sum_{\rm{vertical}\,p} \Tr \tilde{C}_p \label{eq:gauge_ham}\\
&\tilde{A}_V( g) = \prod_{h \in G} e_V( g, h)^{-1}e_V(gh, h^{-1})\notag\\
&\phantom{{}=====}\times\prod_{L\sim V} e_L( g)^{s'_V(L)} \label{eq:A_gauged}\\
&\tilde{B}_{P}(g) = \prod_{L\in P} \left(a_L(g)\al_L\right)^{s_P(L)} \label{eq:B_gauged}\\
&\tilde{C}_p =
\begin{cases}
\omega^{-1}_p a_{V}(f,gh)^{-1}a_{V}(fg,h)a_{V}(f,g)\\ \phantom{{}=====}\rotatebox[origin=c]{180}{$\Lsh$}\text{ if } p\text{ is in $V$}\\
(a_L(gh)\al_L a_{V}(g,h))^{-1}a_{V'}(g,h)a_L(g) \al_L\\ \phantom{{}=====}\rotatebox[origin=c]{180}{$\Lsh$}\text{ if } p\text{ is in $L=\langle V V'\rangle$}
\end{cases} \label{eq:C_gauged}
\end{align}
Here the traces are over both $G$ and $\Z_n$.  The Hamiltonian $\ham_{\rm{m.c.}}$ describes our SET in a fixed background $G$ gauge field configuration, corresponding to some set of extrinsic defects.  We will also want to have a Hamiltonian where the $G$ gauge field is dynamical.  To obtain it, first define operators which change the value of $g_L$ on a single superlink $L$.  We will need two such operators, $\ep_L(g)$ and $\ep_L^R(g)$:
\begin{align}
\ep_L(g)\ket{\{\eta_L(g')\}_{g'}, g_L } &= \ket{\{\eta_L(g^{-1}g')^{\rho(g)}\}_{g'}, gg_L } \\
\ep_L(g)^R \ket{\{\eta_L(g')\}_{g'}, g_L } &=  \ket{\{\eta_L(g')\}_{g'}, g_Lg^{-1} }
\end{align}
Using these we define:
\begin{align}
\mathcal{A}_V(g) =U^{-1}_g(V) \left(\prod_{L\sim V} \ep_L(g)^{s'_V(L)}\right)
\end{align}
where $U_g(V)$ is the action of $g$ on the supervertex $V$.  The Hamiltonian with dynamical $G$ gauge field then becomes
\begin{align}\label{eq:gauge_dyn}
\ham_{\rm{gauged}} =\ham_{\rm{m.c.}} - \sum_{V}\sum_{g \in G} \mathcal{A}_{V}(g)
\end{align}
This is the model whose topological order we have to analyze.

\subsection{Analysis of the topological order}

We claim that the gauged model defined by eq. \ref{eq:gauge_dyn} is equivalent to - i.e. has the same topological order as - an $E$ gauge theory, where the finite group $E$ is a particular extension of $G$ by $\mz_n$.  We derive this equivalence carefully in appendix \ref{ap:derive}, while in this section we just write down the result.  First, let us discuss the extension $E$.  As a set, it is just the product $\mz_n \times G$, so we can label its elements by $(\eta,g)$, with $\eta$ an $n$'th root of unity and $g\in G$.  On the other hand, the group multiplication law is determined by both by $\rho$ and $\omega(g,h)$:

\begin{align}
(\eta_1,g_1)\cdot(\eta_2,g_2)=(\eta_1 (\eta_2)^{\rho(g_1)} \omega(g_1,g_2),g_1 g_2)
\end{align}
Now we write down an $E$ gauge theory on a single copy of the 2d lattice we considered above (as opposed to $|G|$ stacked copies of it).  Since the vertices and links of this 2d lattice are in one to one correspondence with the supervertices and superlinks of the $|G|$ stacked lattice, we will just label them by $V$ and $L$ respectively, and keep calling them supervertices and superlinks to avoid confusion.  Then the degrees of freedom are just pairs $x_L = (\eta_L, g_L)$ (with $\eta_L$ an $n$'th root of unity and $g_L\in G$, and $x_L$ treated as an element of the group $E$) defined on the superlinks $L$.  The Hamiltonian consists of two kinds of terms, vertex terms and plaquette terms.

Intuitively, we define the vertex term at $V$ corresponding to $x\in E$ by multiplying all of the link variables on links $L$ terminating at $V$ by $x$.  However, there is a complication due to the fact that some of the links are outgoing and some are incoming: in one case we want to left multiply by $x$ and in the other we want to right multiply by $x^{-1}$.  We handle both cases by defining the link multiplication operator:

\begin{align}
e_{V,L}(x)|x_L\rangle&=|x x_L \rangle \,\, \text {if } s_V(L)=1 \\
&=|x_L x^{-1}\rangle \,\, \text {if } s_V(L)=-1
\end{align}
With it we then define
\begin{align}
A_V(x) = \prod_{L\sim V} (e_{V,L}(x))
\end{align}

The plaquette term $B_P$ associated to any (super)plaquette $P$ is defined by assigning an energetic penalty for $E$ flux through $P$.  More formally, in the basis of link labellings, we define:

\begin{align}
B_P &= 1 \,\, \text{if } \prod_{L\in P} {x_L}^{s_P(L)} = e \\
  &= 0 \,\, \text{otherwise.}
\end{align}
Using these operators, the Hamiltonian becomes:
\begin{align} \label{eq:Egauge}
\ham_E=-\sum_{V} \sum_{x\in E} A_V(x) - \sum_{P} B_P
\end{align}
The equivalence between the Hamiltonian in eq. \ref{eq:gauge_dyn} and that in eq. \ref{eq:Egauge} is non-trivial, and derived in appendix \ref{ap:derive}.  Here we will just make a couple of comments about how these two Hamiltonians are related.  First of all, the $G$-flux through any plaquette, given by

\begin{align} \label{eq:Gflux}
\prod_{L \in P} {g_L}^{s_P(L)}
\end{align}
in the notation of eq. \ref{eq:gauge_dyn}, is given by the projection from $E$ to $G$ of
\begin{align}
\prod_{L \in P} {x_L}^{s_P(L)}
\end{align}
If we represent each $x_L$ as $x_L=(\eta_L,g_L)$, then this just reduces to the expression in eq. \ref{eq:Gflux}.

Now let us examine configurations with trivial $G$ gauge field, i.e. $g_L=1$ for all $L$.  Then the Hamiltonian in eq. \ref{eq:gauge_dyn} just reduces to the original $\Z_n$ gauge theory on the $|G|$ fold stacked lattice.  On the other hand, the $E$ gauge theory given in eq. \ref{eq:Egauge} reduces to a $\Z_n$ gauge theory on the ordinary 2d lattice.  The identification between these two is non-trivial, but in particular a $\Z_n$ flux in the latter corresponds to that same flux penetrating all $|G|$ layers in the stacked lattice.

Now we will examine some consequences of this equivalence, for a particular example.

\subsection{Example: $\Z_4$ gauge theory}

\begin{figure}[htbp]
\begin{center}
\includegraphics[trim= 3.5cm 9.5cm 6cm 6.5cm, clip,width=0.4\textwidth]{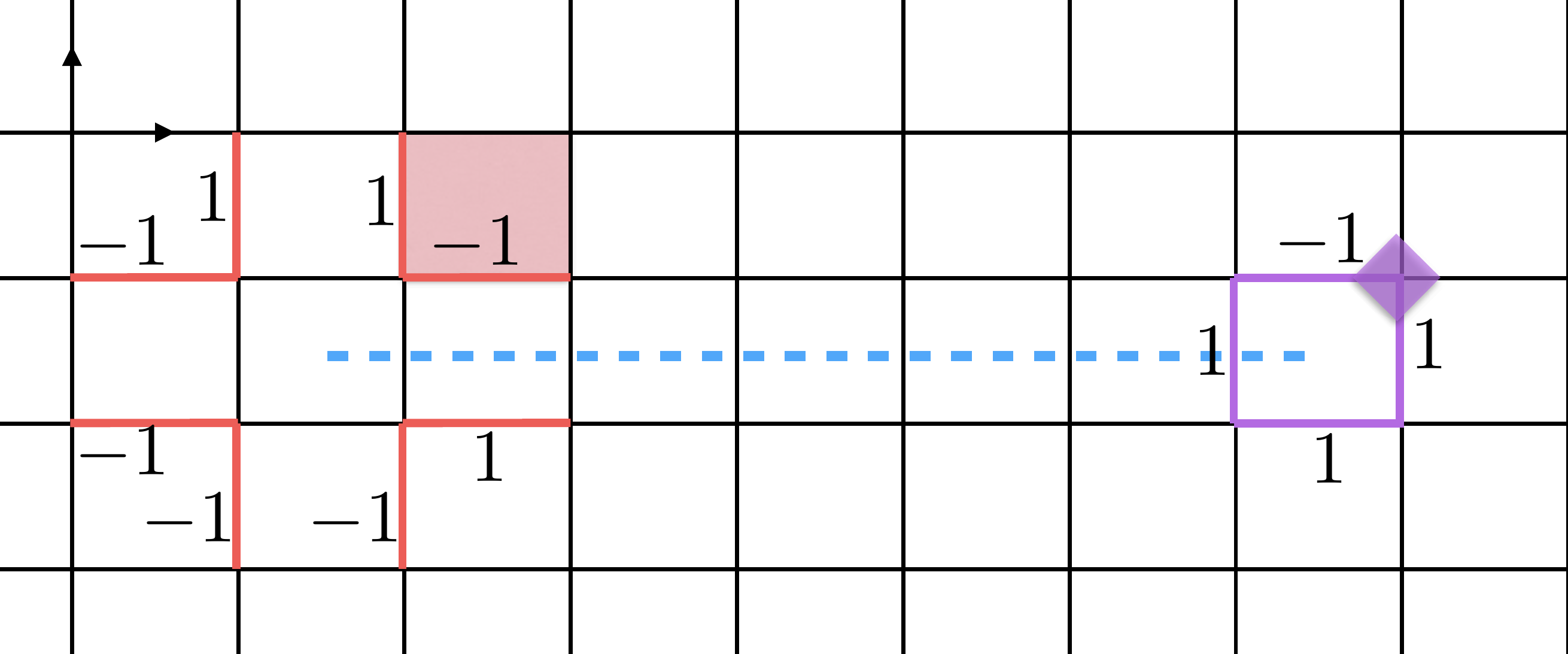}
\caption{$\Z_2$ Gauge field configuration with two widely separated $\Z_2$ defects.  The $\Z_2$ gauge field is non-trivial on the vertical links intersecting with the dashed blue line.  We imagine acting with $e_l$ (the operator which increments $\Z_4$ labels) to the power labeled in the figure ($1$ or $-1$) on the red links.  This is a local operator which commutes with all the terms in the Hamiltonian except the red shaded plaquette, which sees a flux of $2 \in \Z_4=\{0,1,2,3\}$.  Likewise, we imagine acting with $a_l$ to the given power on the purple links.  This produces a charge $2$ in the vertex given by the purple diamond.  Thus we can create even charges and even fluxes locally near a $\Z_2$ defect.}
\label{fig:defect_pair}
\end{center}
\end{figure}

In this subsection we analyze the example of a $\Z_4$ gauge theory with $G=\Z_2=\{1,g\}$ acting by $\rho(g): j\rightarrow-j$, where $j\in \{0,1,2,3\}$.  As we discussed earlier, after gauging $G$ the topological order given by equation \ref{eq:Egauge} is just that of the gauge theory of the extension $E$, where $E$ can be either $\mathbb{D}_8$ or $\mathbb{Q}_8$.

Analyzing extrinsic defects in the ungauged theory is the same as including a non-dynamical background $G$ gauge field, which just amounts to the following modification of the Hamiltonian in equation \ref{eq:Egauge}:

\begin{align} \label{ham_special_case}
\ham_E^{\text{non-dynam.}}=-\sum_{V} \sum_{\eta \in \Z_n} A_V(\eta) - \sum_{P} B_P
\end{align}

Note that we are now working with a purely 2d square lattice.  Let us understand these two terms for a fixed background gauge field configuration, namely that of two widely separated defects, illustrated in figure \ref{fig:defect_pair}.  In this case, the two terms in the Hamiltonian of equation \ref{ham_special_case} reduce to the ordinary $\Z_4$ gauge theory vertex and plaquette term everywhere except for the plaquettes intersected by the dashed blue branch cut line and the vertices directly above them.  For these intersected plaquettes, we have a modified plaquette term where $a_L$ is inverted for the upper horizontal link in the product over plaquette links (recall that $a_L$ is the `phase' operator in the $\Z_4$ gauge theory).  Likewise, just above the dashed blue line we have modified vertex terms, where $e_L$ (the charge operator in the $\Z_4$ gauge theory) is inverted for the lower vertical link (the one bisected by the dashed blue line) in the product over vertex links.

Now, the topological superselection sector of a $\Z_2$ defect can be altered by fusing $\Z_4$ charges and fluxes into it.  However, as shown in figure \ref{fig:defect_pair}, we can create a charge of $2 \in \Z_4=\{0,1,2,3\}$ or a flux of $2 \in \Z_4$ using a local operator near a $\Z_2$ defect.  So the only non-trivial superselection sectors near the $\Z_2$ defect correspond to the parity of charges or fluxes bound to the $\Z_2$ defect, and there are only $4$ such superselection sectors (as opposed to the $16$ anyon types).  This also implies that when we fuse a pair of defects, the fusion product can be changed by an even charge or even flux with a local operator.  This implies that a pair of defects can fuse into any even charge or even flux, so the fusion rules are non-abelian.  This is a general feature of defects in permuting theories.  Indeed, the local operator which creates an even charge in the presence of the defect can be thought of as creating a pair (fundamental charge, fundamental anti-charge), braiding the anti-charge around the defect to turn it into a charge, and re-fusing with the leftover charge to make a charge of $2$, and similarly for the operator that creates an even flux.  In the next section we will generalize this idea to arbitrary permutation actions.

Notice that none of the above discussion depended on whether we chose $E=\mathbb{D}_8$ or $E=\mathbb{Q}_8$.  We will see in the next section that this is a general feature: the superselection sectors of a defect, as well as its fusion rules with anyons, depend only on the permutation $\rho$, and not on $\omega(g,h)$ or any other data.  In order to discriminate between $E=\mathbb{D}_8$ and $E=\mathbb{Q}_8$ we have to perform a more subtle measurement, involving braiding anyons and defects.  We will discuss this in detail in the next section; for now just note that this kind of braiding measurement really comes down to computing commutation relations of the string operator that creates defects with other string operators, and is not something that can be done in a purely fixed background $G$ gauge field configuration.

%%%%%%%%%%%%%%%%%%%%%%%%%%%%%%%%%

\section{Symmetry localization and defect fusion}
\label{general_theory}

In the previous section we constructed a family of lattice models of SETs where the symmetry $G$ permuted the anyons.  For each permutation action $\rho$, we had multiple SETs corresponding to different choices of a function $\omega(g,h) \in H_{\rho}^2(G,\Z_n)$.  We saw that these SETs were distinct by gauging $G$ and examining the full quasiparticle statistics of the gauged theory.  However, we also saw that the distinctions between these SETs (for fixed $\rho$ but different $\omega(g,h)$) were not visible if we examined only non-dynamical background $G$ gauge field configurations (that is, with extrinsic defects, but without fluctuations of the $G$ gauge field).  In this section, we will study the distinctions among such SETs at a formal level, for general anyon theories (not necessarily gauge theories) and general action of finite on-site unitary symmetry $G$.  While we will not fully gauge $G$, we will consider operators which create, annihilate, and move defects.  This algebraic structure turns out to contain the same information as the gauged theory, but is somewhat easier to deal with, because it is smaller, as it does not contain the $G$ charges as topological excitations.

First, let consider ordinary anyons.  The braiding and fusion rules of anyons are encoded in a unitary modular tensor category (UMTC) ${\cal A}$, described concretely by a collection of quasiparticle labels, fusion spaces, notion of antiparticle, and F and R matrices (see appendix E of reference \onlinecite{Kitaev2006} for a review).  Now, the addition of defects expands the list of labels to the set of all defect-anyon composites, and, because defects can also be fused and braided, there must be an extension of the algebraic structure to this larger set of labels.  However, because the defects are extrinsic modifications of the Hamiltonian rather than excitations, and in particular carry branch cuts, this algebraic structure is not a UMTC.  Indeed, braiding around an extrinsic defect can change anyon type and act non-trivially on local operators, which is something that cannot happen when we braid around an ordinary anyon.  The algebraic structure describing anyons and defects is a so-called braided G-crossed category.  See reference \onlinecite{Maissam2014} (sec. VI A.-C.) for a detailed definition of a braided G-crossed category in the context of defect and anyon braiding.  In this paper, we will not discuss all of the properties of braided G-crossed categories all at once; rather, we will only discuss the ones we need, at a physical level, to constrain the possible defect fusion rules.  We build this structure up in stages: rather than imposing all of these properties at once, we first fix the UMTC of the anyons, then look for all consistent solutions for the superselection sectors of a single defect and the F an R matrices defining its braiding with anyons, and only then, having fixed this single-defect structure, study processes involving two defects.\footnote{A final step, which we do not carry out in this paper, is to solve for the 3-defect structure; this gives the entire braided G-crossed category.}  At each stage, we specify the braided G-crossed properties that we use.  This hierarchical approach parallels the mathematical constructions in reference \onlinecite{ENO}.

In particular, we find that the properties of a single defect, including its topological superselection sectors, and braiding and fusion rules with the other anyons, are determined purely by the permutation $\rho$ (more precisely, by the associated action of $G$ by braided auto-equivalences, which we define below), and are independent of the fractionalization class $\omega$.

\subsection{Non-permuting case}

Let us first give a condensed review of symmetry fractionalization in the case that $G$ does not permute the anyons, {\emph{i.e.}} fixes all topological superselection sectors.  When $G$ does not permute the anyons, we assume the existence of an approximate ``local'' action of $G$ on particular anyons - see appendix \ref{long_appendix} for details on how to construct it.  This local action is just an operator $U_g^{\rm{loc}}$ acting on the spins within a distance $r$ of the anyon, whose commutation relations with all local operators acting near the anyon are approximately the same as those of $U_g$.  Here the quality of the approximation is controlled by $e^{-r/\xi}$, where $\xi$ is the correlation length; henceforth we take $r$ sufficiently large and neglect the exponentially small errors.  These properties imply that the $U_g^{\rm{loc}}$ satisfy the group relations up to a possible phase ambiguity $\omega_a(g,h)$ (here $a$ is the anyon acted on):
\begin{equation}\label{eq:ambig}
U_g^{\rm{loc}} U_h^{\rm{loc}} = \omega_a(g,h) U_{gh}^{\rm{loc}}
\end{equation}
Associativity then gives
\begin{equation} \label{eq:basic_assoc}
\omega_a(g,h)\omega_a(f,gh)=\omega_a(fg,h)\omega_a(f,g),
\end{equation}
which, together with the uniqueness of the $U_g^{\rm{loc}}$ up to overall phase, imply that we can extract from them a unique cohomology class $[\omega_a]\in H^2(G,U(1))$.  Furthermore, for every allowed fusion channel $a \times b \rightarrow c$, we must have $\omega_a(g,h) \omega_b(g,h) = \omega_c(g,h)$, since such a fusion is a local operation and carries integral symmetry quantum numbers (see appendix \ref{long_appendix} for a detailed argument).  Now a technical result of category theory (lemma 3.31 of ref. \onlinecite{DGNO}) shows that any $U(1)$-valued function of the anyons which satisfies this property can be represented as
\begin{equation}
\omega_a(g,h)=S_{\omega(g,h),a}
\end{equation}
where $\omega(g,h)$ is an abelian anyon, and $S_{b,a}$ denotes the full braiding phase of the abelian anyon $b$ around a general anyon $a$.  From equation \ref{eq:basic_assoc}, we then have that $[\omega] \in H^2(G, {\cal A})$, where ${\cal A}$ is the subset of abelian anyons viewed as an additive group.  Thus we see that the symmetry fractionalization data is encoded in a cohomology class in $[\omega] \in H^2(G, {\cal A})$.
\vskip 5.0pt
{\emph{Defect fusion rules in the non-permuting case}}
\vskip 5.0pt
Let us now re-interpret this cohomology class $[\omega]$ in terms of defect fusion rules.  As explained in appendix \ref{long_appendix}, a $g$-defect $\tau_g$ is an extrinsic modification of the Hamiltonian which inserts a symmetry flux of $g$ in a certain location, with a $g$-branch cut emanating from that location.  We expect a $g$-defect to have some topological superselection sectors, which can be changed by fusing anyons into the core of the defect (in reference \onlinecite{Maissam2014} these are referred to as topologically distinct types of $g$-defects, and discussed in section V. B.).  Now, we will see below that in the case of permuting symmetries, there exist local processes, involving changing anyon type by braiding around the defect, which seemingly create topological charge in the vicinity of the defect; hence, the structure of defect superselection sectors in this case depends on the permutation being applied.  In the non-permuting case we are considering now, there are no such processes, so we expect that there is no difference between the superselection sectors of $\tau_g$ and those of the trivial defect.  In other words, the superselection sectors of $\tau_g$ are in one to one correspondence with the anyons.

However, a key point is that there is an ambiguity in how to define the Hamiltonian in the core of the $g$-defect.  This means that there is no {\emph{canonical}} way to identify defect superselection sectors with anyons: different choices of the core Hamiltonian yield different ground state superselection sectors.  Now, we can at least demand that the core Hamiltonian be chosen in such a way that a pair of well separated defects $\tau_g$ and $\tau_{g^{-1}}$, connected by a branch cut, has no ground state degeneracy.  This means that the ambiguity in the defect ground state superselection sector is at most by an abelian anyon.  Having no canonical way to resolve this ambiguity, we simply choose a ground state superselection sector $x_g$ arbitrarily for each $g$-defect; this involves making $|G|$ choices, one for each $g\in G$.  

Now consider fusing a $g$-defect and an $h$-defect.  This is accomplished by a $g$ gauge transformation that merges the two branch cuts into a single $gh$ branch cut.  The key point now is that although this fusion product is a $gh$ defect, because of the ambiguity in the Hamiltonian at the core, it may have a different ground state superselection sector than the already chosen $x_{gh}$.  Call this difference $a_{g,h}$; as we just argued, it is an abelian anyon.  In fact, it is the same as the abelian anyon $\omega(g,h)$ that was defined above.  Indeed, the local action $U_g^{\rm{loc}}$ on some anyon $b$ defined above is essentially given by braiding $b$ with a $g$-defect.  Then the difference between $U_g^{\rm{loc}} U_h^{\rm{loc}}$ and $U_{gh}^{\rm{loc}}$ is just the difference between braiding $b$ around a pair consisting of a $g$ defect and an $h$ defect versus braiding $b$ around a single $gh$ defect.  According to eq. \ref{eq:ambig} above, this is the braiding phase of $b$ with $\omega(g,h)$, whereas in the present discussion, it is the braiding phase of $b$ with $a_{g,h}$.  Since these two must be equal for all $b$, modularity implies that $a_{g,h}=\omega(g,h)$.

Of course, one has to be a little more precise in defining the local action of the symmetry: $U_g^{\rm{loc}}$ must be an operator that acts on just the anyon $b$, and any $g$-defects involved in the process must be created from the ground state and likewise disappear at the end of the process.  A careful treatment, given in appendix \ref{long_appendix}, involves a dual picture of holding the anyon $b$ fixed and braiding the $g$-defect around it instead.  Nevertheless, the above argument applies equally well to this definition of $U_g^{\rm{loc}}$, and the conclusion is the same: $a_{g,h}=\omega(g,h)$.

Thus we can extract the symmetry fractionalization data from the defect fusion rules in the non-permuting case.

\subsection{The permuting case: single fixed defect}

The case where $G$ permutes the topological superselection sectors is significantly more complicated.  For one thing, there is now no canonical `trivial' SET, as opposed to the untwisted case, where we can construct a trivial SET simply by defining $G$ to act trivially on the microscopic degrees of freedom for any Hamiltonian realizing the desired intrinsic topological order.  Thus, we should not expect to be able to assign group cohomology classes to permuting SETs, since there is now no preferred choice of a trivial SET corresponding to the trivial cohomology class.  Instead, given a particular choice of intrinsic topological order and twisted action of $G$, our approach will be to start with an arbitrary SET realizing this twisted action, and look for all possible gauge inequivalent ways of deforming the defect fusion rules.  This approach is rooted in the general idea of classifying SETs via braided G-crossed categories \cite{Fidkowski2014, Teo_Hughes, Maissam2014}, where constructing defect fusion rules - formally, a tensor product of bimodule categories - is one step in a systematic construction of the braided G-crossed category.  For now, however, let us examine a single $g$-defect in the twisted case where $\rho(g)$ non-trivially permutes the anyons.

To appreciate the complexity of this twisted case, note that the $g$-defects are now generically non-abelian.  For example, consider a defect of a $\mz_2$ symmetry which exchanges $e$ and $m$ in a model with toric code topological order (for an example of such a model see \cite{Kitaev2006, Kitaev_Kong}).  By nucleating a pair of $e$'s, braiding one around the defect to turn it into an $m$, and fusing with the remaining $e$, we perform a local process that nucleates the $e \times m = f$ particle in the vicinity of the defect.  This is reminiscent of an Ising defect, and indeed there is a two-fold degeneracy for widely separated defects associated with absorbing and emitting the $f$ fermion (also, the defects actually become Ising anyons upon gauging the $\mz_2$ symmetry \cite{Lu2013, Teo_Hughes, Maissam2014}).

\begin{figure}[htbp]
\begin{center}
\includegraphics[width=0.4\textwidth]{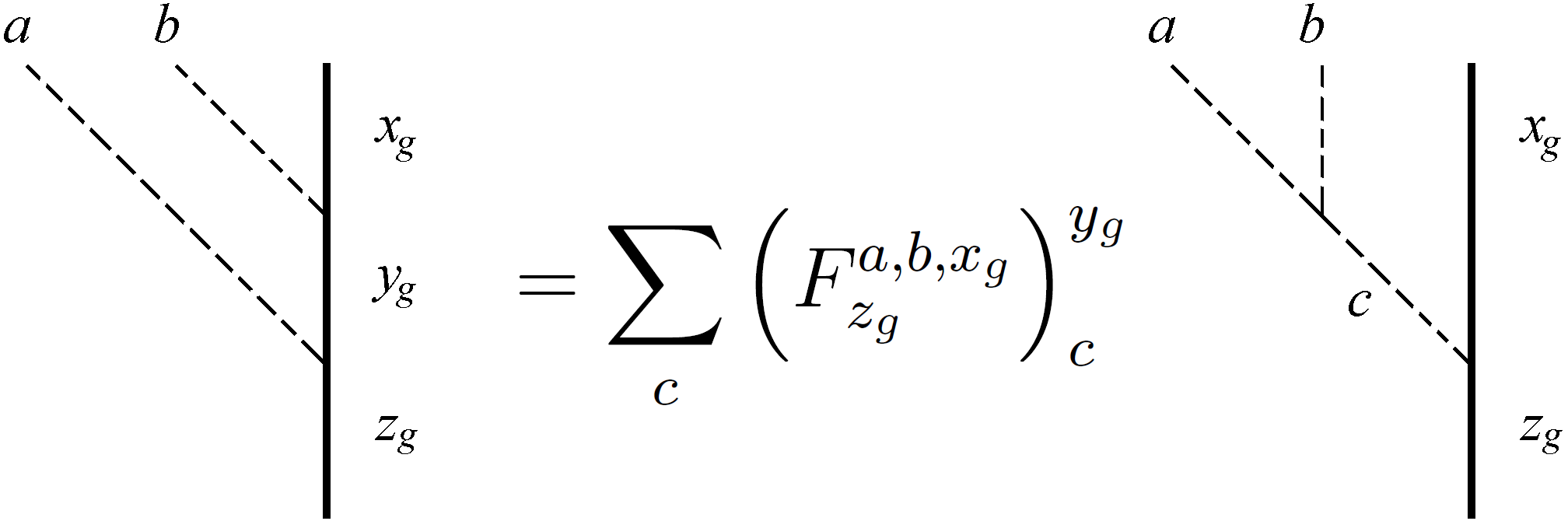}
\caption{Basic associativity relation for a left module category structure of a $g$-defect.  The $g$-defect superselection sectors $x_g$,$y_g$, $z_g$ are the simple objects.  For simplicity we assume in this figure that all fusion spaces are one dimensional.  The sum on the right hand side is over all anyons $c$ which satisfy the fusion rules, {\emph{i.e.}} correspond to non-trivial splitting spaces $V^{a,b}_c$ and $V^{c, x_g}_{z_g}$.}
\label{fig1}
\end{center}
\end{figure}

To get a handle on this complexity, suppose first that we are given a Hamiltonian corresponding to a single fixed $g$-defect, and suppose further that we know nothing about the location of the branch cut or the action of the symmetry on our microscopic degrees of freedom.  What kind of data can we extract about the symmetry enriched phase, using only braiding and fusion operations of anyons in the background of this fixed defect configuration?  Certainly we can perform experiments that braid anyons around the defect and measure the topological charge before and after braiding: this is just measuring the permutation action of $g$:

\begin{equation}
a\rightarrow g\cdot a.
\end{equation}
Furthermore, we can also measure the action of $g$ on anyon fusion spaces:
\begin{align}
g: V^{a,b}_c &\rightarrow V^{g\cdot a,g\cdot b}_{g\cdot c} \\
X\in V^{a,b}_c &\rightarrow g\cdot X \in V^{g\cdot a, g\cdot b}_{g\cdot c},
\end{align}
A physical process that can measure this action on anyon fusion spaces is as follows.  Given an operator $X$ which splits $c$ into the pair $a,b$, i.e. $X \in V^{a,b}_c$, define $g \cdot X$ be the operator that acts as follows: first braid $g \cdot a$ and $g \cdot b$ around the defect to turn them into $a,b$ respectively, apply $X$, and then braid the fusion product $c$ in the opposite direction to get $g \cdot c$.  Note that because braiding an anyon $a$ around the defect is only defined up to a phase $\alpha_a$, $g \cdot X$ is only well defined up to a phase $\alpha_a \alpha_b \alpha_c^{-1}$.

A permutation action on anyons together with a compatible action by unitary linear transformations on anyon fusion spaces is called a braided tensor autoequivalence of ${\cal A}$ (in reference \onlinecite{Maissam2014}, this also goes under the name of topological symmetry, discussed in section III. A.).  There is also a notion of two braided autoequivalences being the same: this occurs when the two autoequivalences have the same permutation action, and their actions on the fusion space $V^{a,b}_c$ differ by $\alpha_a \alpha_b \alpha_c^{-1}$, where the $\{\alpha_a\}$ is some fixed set of $U(1)$ phases.  Hence, by performing braiding experiments around a fixed $g$ defect configuration we uniquely recover precisely a braided tensor autoequivalence corresponding to $g$, up to this notion of braided tensor autoequivalences being the same.\footnote{Note that in an actual SET, the action of $g$ on fusion spaces, given by $X\rightarrow U_g^{-1} X U_g$, is well defined with no phase ambiguity.  Some actions of $G$ by braided autoequivalences cannot be realized by any SET, because there is no way to fix the phase in the action on fusion spaces, in a way compatible with the group law.  The corresponding obstruction class is valued in $H^3(G,{\cal A}_{\rm{abelian}})$ - see references \onlinecite{ENO, Maissam2014, Teo_Hughes}.}

Now, the goal of the classification program of SETs is to construct a larger algebraic structure - the braided $G$-crossed category - which describes the fusion and braiding rules of both anyons and defects.  How much of this larger structure can one recover from knowing the braided tensor autoequivalence corresponding to each $g \in G$?  The answer is given by theorem 5.2 of reference \onlinecite{ENO}: the braided tensor autoequivalence uniquely determines an `invertible bimodule category' corresponding to the $g$-defect.  Roughly, the `invertible bimodule category' corresponding to the $g$-defect is the subset of the braiding and fusion data of the theory that involves a single $g$-defect and an arbitrary number of anyons, modulo gauge equivalences.  The intuition is that, while it may be possible to find a gauge equivalence between such a subset of braiding and fusion data for two different theories, it might be impossible to extend this gauge equivalence to a gauge equivalence of the full theories, involving braiding and fusion data of an arbitrary number of defects and anyons.  So the invertible bimodule categories associated to the various defects constitute coarse data that only partially classifies different SETs.  In particular, in the context of the lattice models from the previous section, this data is sensitive only to the permutation $\rho$, and not to the co-cycle $\omega$.

Let us now define invertible bimodule categories in a little more detail.

\vskip 5.0pt
{\emph{Invertible bimodule categories}}
\vskip 5.0pt

First of all, let us put all of our defects and anyons on a line.  Then we can hold the defect at a fixed position, and imagine fusing anyons into and out of it.  We can do this either from the left or the right; the corresponding structures, consisting of the superselection sectors of the $g$-defect, together with the F-move isomorphisms for these fusion processes, are called left and right module categories respectively.  More precisely, a (say) left module category involves this fusion and F-move data for anyons fusing in from the left, modulo gauge equivalence corresponding to basis redefinitions within the anyon-defect fusion spaces, and satisfying pentagon equation constraints involving 3 anyons and one defect.  The data defining a left module category is shown in figure \ref{fig1}. 

Now, physically we can extract the data of both a left and right module category structure for a single fixed defect.  However, in a physical system these data are not independent: there are additional data and compatibility constraints between fusion from the left and right, beyond those encoded in the left and right module structures alone.  This is the associativity of a process where one anyon fuses in from the left and another fuses in from the right, constrained by the appropriate pentagon equations.  A left and right module category together with such additional data and constraints is called a bimodule category.  Thus a single defect gives rise to a bimodule category over the anyons.

\begin{figure}[htbp]
\begin{center}
\includegraphics[width=0.4\textwidth]{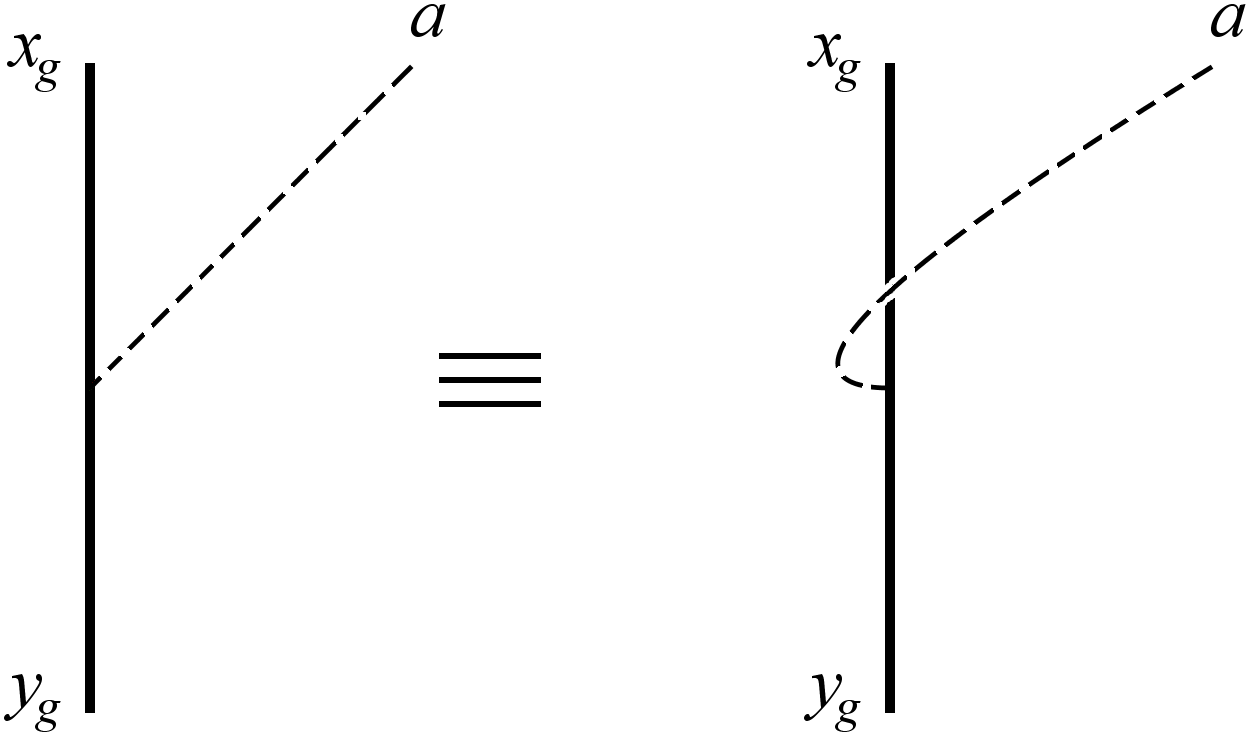}
\caption{In the algebraic fusion and braiding theory of anyons and defects, one considers two separate splitting spaces $V^{a,x_g}_{y_g}$ and $V^{x_g,a}_{y_g}$, corresponding to the anyon $a$ splitting off from the defect either to the left or to the right.  Because of the existence of braiding, these spaces are actually isomorphic: $V^{a,x_g}_{y_g} \cong V^{x_g,a}_{y_g}$.  In our graphical calculus, we fix this isomorphism to be the one given by braiding the anyon $a$ in front of the defect, so as to avoid the defect branch cut, which extends into the page.  Having chosen a basis for the same anyon splitting from the left, this fixes the basis for the fusion space of the anyon splitting from the right, one in which the half-braid R-move is equal to the identity: $R^{a,x_g}_{y_g}=1$. }
\label{fig2}
\end{center}
\end{figure}

Additionally though, all of the bimodule categories that we encounter have further properties due to the existence of a braiding structure.  For now this braiding structure involves leaving the $g$-defect fixed, and just braiding anyons around it.  Let us imagine that the branch cut emanating from the defect goes towards the back of the page (away from the line where all the quasiparticles sit).  Then braiding an anyon $a$ in front of the defect, so as to avoid its branch cut, gives an isomorphism of anyon defect fusion spaces:

\begin{equation} \label{eq:defect_braid}
R^{a,x_g}_{y_g}: V^{a,x_g}_{y_g} \rightarrow V^{x_g,a}_{y_g}
\end{equation}
Here $x_g$ and $y_g$ are defect superselection sectors and $a$ is an anyon.  We can identify the two fusion spaces $V^{a,x_g}_{y_g}$ and $V^{x_g,a}_{y_g}$ using the isomorphism in eq. \ref{eq:defect_braid}, as illustrated in figure \ref{fig2}.  Then the compatibility of braiding and fusion, formally expressed in terms of the hexagon equations, uniquely determines all the F-move isomorphisms involving two anyons and one defect in terms of those where the anyons only fuse in from the left, illustrated in figure \ref{fig1}.  Mathematically one says that in this case the bimodule category structure is induced from a left module category via the braiding in ${\cal A}$.  Thus a single defect gives rise to a bimodule category over the anyons, where the bimodule structure is induced form a left module structure by using the anyon braiding. 

A second property of these bimodule categories comes from the fact that anyons can braid behind the defect as well.  Various hexagon equation constraints\footnote{Some of these hexagon equations have also been called `heptagon' equations in reference \onlinecite{Maissam2014}, because passing an fusion vertex behind the defect incurs an extra phase due to the defect branch cut, and is hence considered an extra step.} turn out to determine the half-braiding of an anyon behind a defect up to a $U(1)$ phase which depends on the anyon being braided, but not on the superselection sector of the defect \cite{ENO}.  These are just the phases $\alpha_a$ mentioned earlier.  Mathematically, these extra constraints imply that the bimodule category is `invertible' - see reference \onlinecite{ENO} for a precise definition of invertibility, which we will not state here.  Physically, however, invertibility turns out to mean that for any defect, there exists an anti-defect that can annihilate with it, under the defect fusion product defined in the next section.

Thus, in summary, just from braiding experiments involving a single static $g$-defect one can determine the action of $g$ by braided tensor autoequivalences, which is equivalent to all fusion and braiding rules involving a single $g$-defect, with the exception of braiding behind the defect, which is determined only up to a phase that depends on the anyon being braided (but not the superselection sectors of the defect).  The uniqueness here is up to a gauge freedom corresponding to a change of orthonormal bases in the anyon-defect and defect-anyon fusion spaces.  Mathematically, this collection of data is an invertible bimodule category with bimodule structure induced from braiding in the anyons ${\cal A}$.

%To get a handle on this complexity, let us see what kind of gauge invariant data is encoded in a defect.  Recall that with ordinary anyons, we can put them on a line and consider associativity of various splitting or fusion processes.  This associativity is described by isomorphisms of certain tensor products of fusion spaces (see Appendix E of reference \onlinecite{Kitaev2006} for a detailed exposition), which for a specific choice of basis can be written as unitary matrices - so-called F-matrices.  Viewing basis redefinitions as gauge transformations, one obtains some gauge invariant data encoded in these F-matrices, which is highly constrained by coherence conditions like the pentagon equation.  Now, having fixed this structure for the anyons, we can proceed with defects in essentially the same fashion, but it is useful to organize the list of F-matrices by the number of defects involved.  In particular, we can first take just one fixed $g$-defect, and consider various anyons splitting off and fusing to it from the left - see Fig. \ref{fig1}.  The defect superselection sectors, fusion rules of anyons into the defect, and associativity isomorphisms relating these fusion spaces together constitute the structure of a so-called `module category' over the anyons.  

\vskip 8pt
{\noindent\bf{Example: $\Z_4$ gauge theory with $\Z_2$ symmetry}}
\vskip 3pt
Let us give a non-trivial example of the above structure.  Let $G=\Z_2$, and consider $\Z_4$ gauge theory, whose anyon content is $\Z_4 \times \Z_4$.  This is the theory we discussed in the previous section.  We will think of the first $\Z_4$ as charge and the second $\Z_4$ as flux, and denote anyons by $(i,j)$, $0 \leq i,j <4$.  We will take the non-trivial generator $g\in \Z_2$ to act on the anyons by $(i,j) \rightarrow (4-i,4-j)$.  What is the invertible bimodule category structure associated to a $g$-defect?

First of all, by the argument above, we can nucleate anyons of the form $(2i,2j)$ in the presence of the $g$-defect.  Thus the $g$-defect has only $4$ superselection sectors, namely $(\alpha,\beta)_g$, $0 \leq \alpha,\beta \leq 1$.  The fusion rules with the anyons are simply given by addition mod $2$
\begin{equation}
(i,j) \times (\alpha,\beta)_g = (\alpha +i\,\rm{mod}\,2, \beta+j\,\rm{mod}\,2)_g.
\end{equation}
These are the same fusion rules that we found directly in the previous section.

The $F$ symbols involving one defect and two anyons are non-trivial.  There is a choice of basis in which they can all be set equal to $\pm 1$, and they can be determined explicitly using a general equivalence between such invertible bimodule categories and Lagrangian subgroups of $H \otimes H^*$ for any abelian $H$ gauge theory - see the discussion in section 10.3 of reference \onlinecite{ENO}.  More explicitly, in a particular gauge they can be taken as follows.  Given even anyons $a_1=(2i_1,2j_1), a_2=(2i_2,2j_2)$, let
\begin{equation}
F^{a_1,a_2,x}_y=(-1)^{i_1 j_2 - i_2 j_1}.
\end{equation}
Anyons not of this form can always be uniquely written as $a_1=(2i_1+\mu_1,2j_1+\nu_1), a_2=(2i_2+\mu_1,2j_2+\nu_1)$, where $\mu_{1,2}, \nu_{1,2}=0,1$.  For these anyons we define the F-symbol the same way, $F^{a_1,a_2,x}_y=(-1)^{i_1 j_2 - i_2 j_1}$.  F-matrices for fusion of anyons from the right are then uniquely determined using the rule in figure \ref{fig2} and ordinary anyon fusion rules.  The R-move corresponding to braiding an anyon $a$ behind a defect is also determined uniquely up to a phase that depends on $a$ but not on the superselection sector of the defect, as is the action on the anyon fusion spaces, which turns out to be trivial.

Let us emphasize that the F-matrices we have written down here, involving one defect and two anyons, will be valid for both of the SETs with this permutation action on the anyons.  The data which distinguishes between the two SETs, namely the cohomology classes $\omega_{\pm}(g,h)$ discussed in the previous section, will show up only in quantities involving two defects.  Below we will discuss this quantity $\omega$ in our more general context.

\subsection{Defect fusion in the twisted case}\label{ssec:twisted}

Above we saw that the gauge invariant data contained in a single $g$-defect is entirely determined by the action of $g$ by braided autoequivalences on the anyons.  In particular, we cannot distinguish among SETs with the same braided autoequivalence action of $G$ by considering only single defects.  In this section, we will see that such SETs can be distinguished by considering pairs of defects and their splitting and fusion rules.

\begin{figure}[htbp]
\begin{center}
 \includegraphics[width=0.4\textwidth]{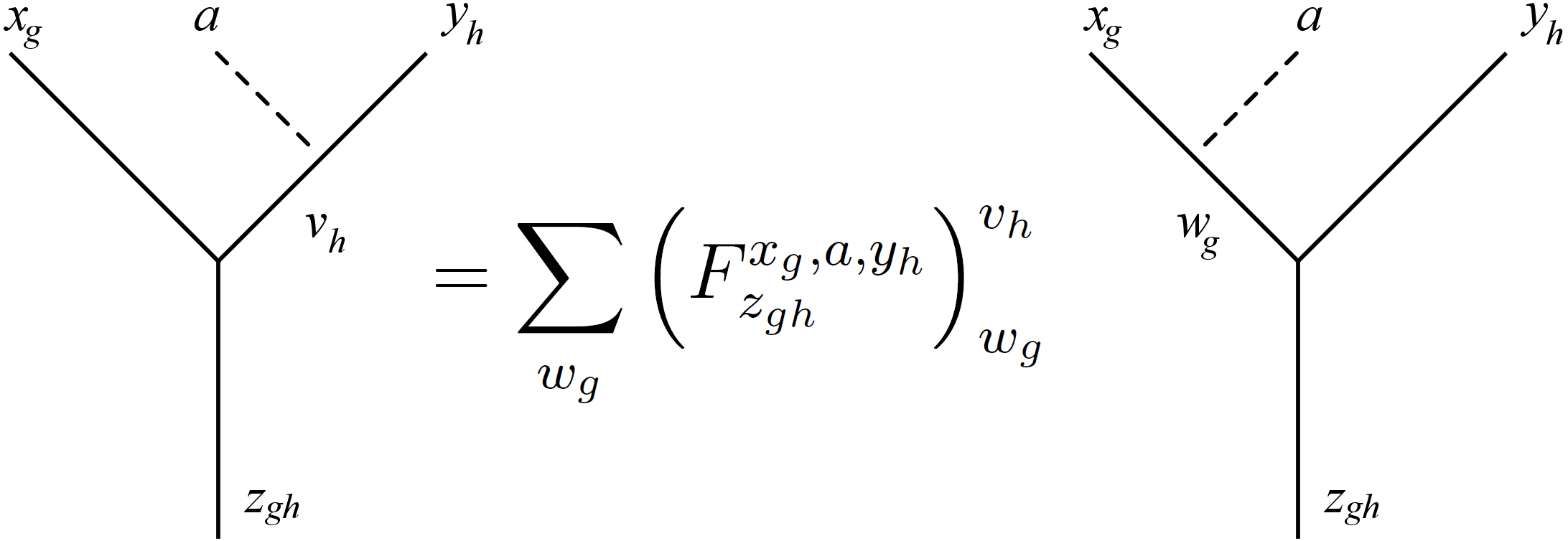}\vspace{3mm}
    \includegraphics[width=0.4\textwidth]{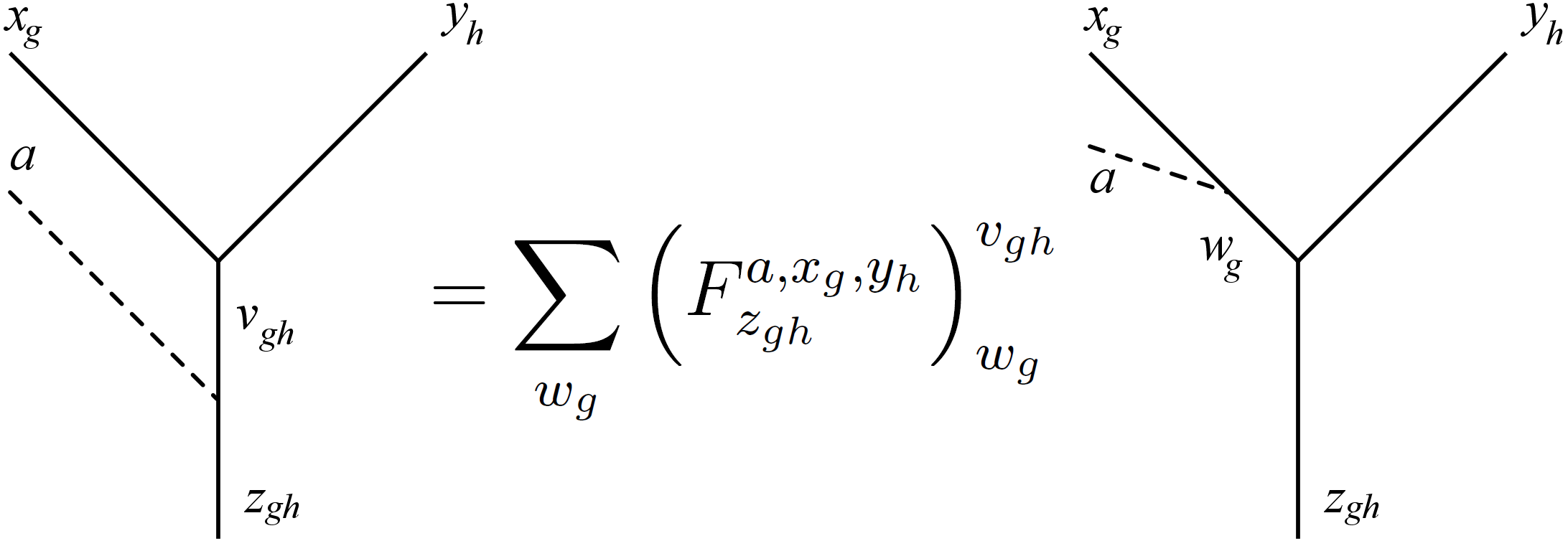}\vspace{3mm}
    \includegraphics[width=0.4\textwidth]{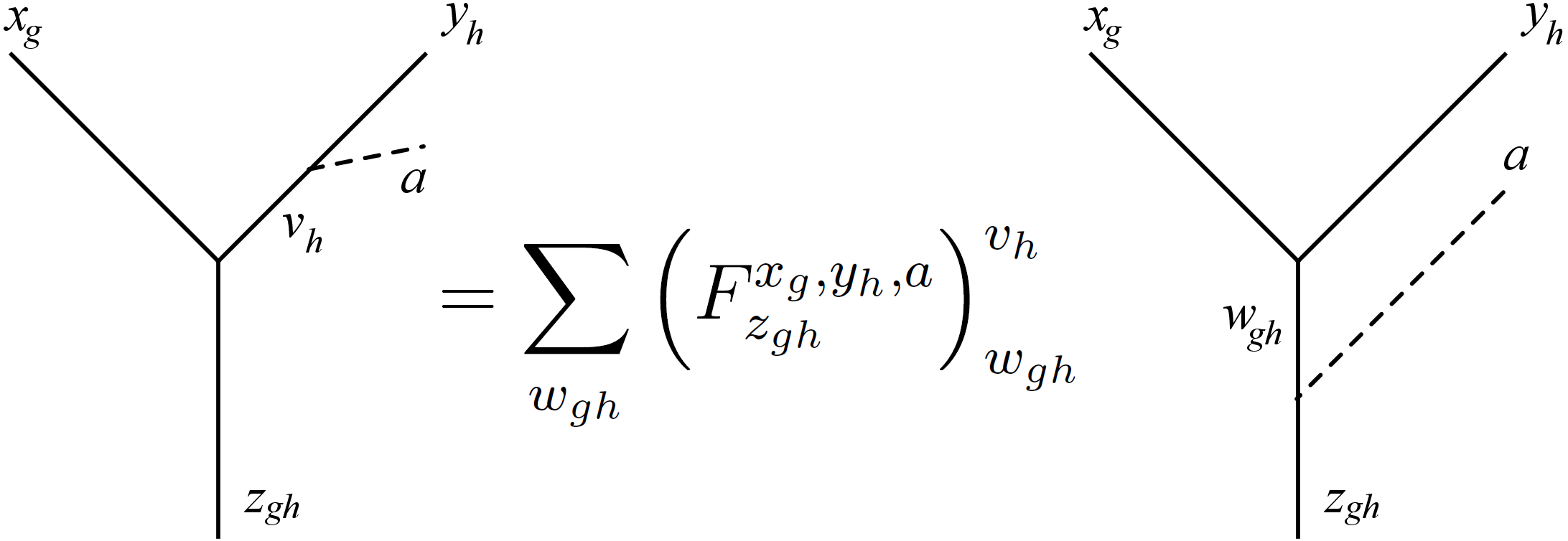}
\caption{Data defining a defect fusion product of a $g$ defect and an $h$ defect into a $gh$ defect.  Dotted lines represent anyons, solid lines represent defects.}
\label{fig3}
\end{center}
\end{figure}

First let us fix a pair $g,h \in G$, and consider a $gh$ defect splitting into a $g$ defect and an $h$ defect.  One can again look at associativity relations for processes which start with the $gh$ defect and end in a $g$ defect, an $h$ defect, and an additional anyon $a$.  There are three types of processes, depending on whether $a$ appears to the left, to the right, or in between the defects in our chosen line ordering; they are illustrated in figure \ref{fig3}.

\begin{figure}[htbp]
\begin{center}
\includegraphics[width=0.35\textwidth]{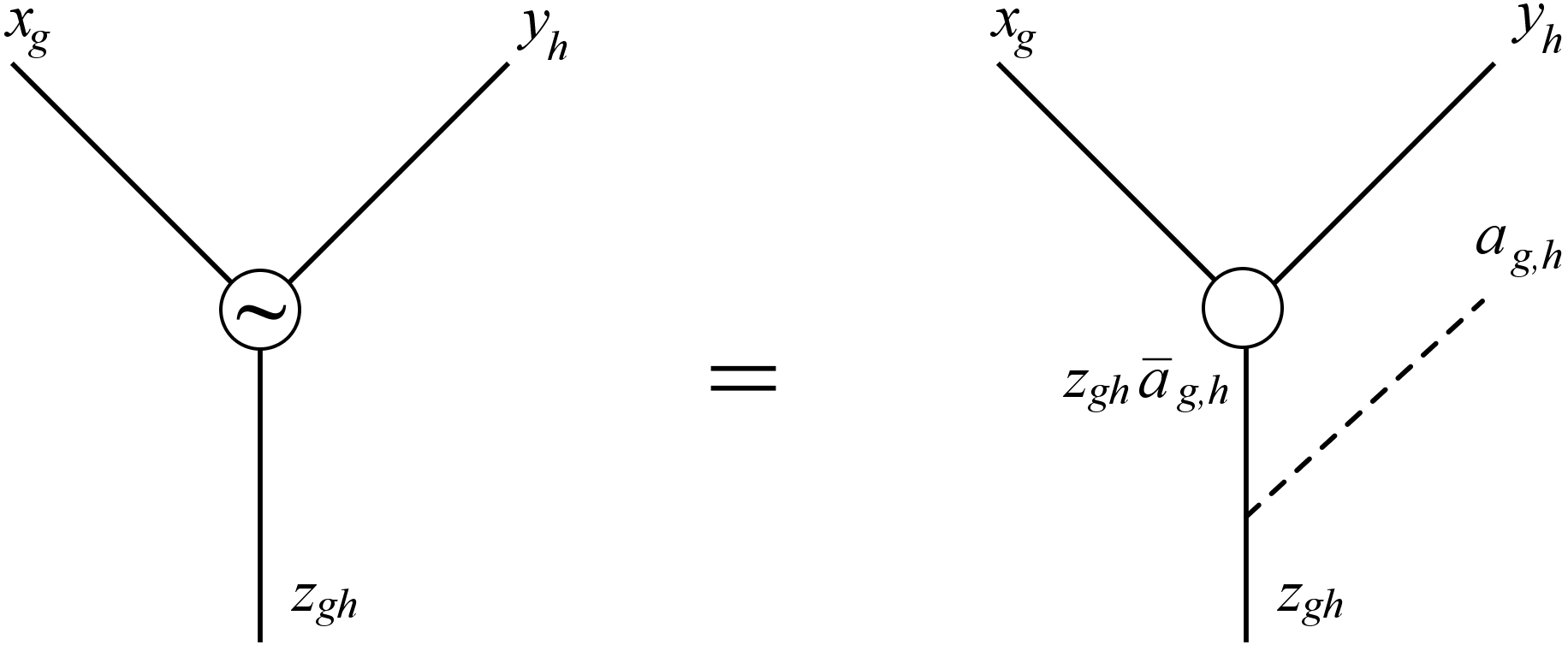}
\caption{Given a valid defect fusion product, one can construct a deformed defect fusion product, shown here with a tilde, as shown in the above figure.  The equation in the above figure amounts to a construction of new $F$-symbols involving the $g$ and $h$ defects and arbitrary anyons.  For details on how these new $F$-symbols are defined, see appendix \ref{ap:deform}.} 
\label{fig4}
\end{center}
\end{figure}

The coherence conditions on the F-matrices illustrated in figure \ref{fig3} are highly constraining.  Indeed, these F-matrices must satisfy compatibility conditions with each other, with the invertible module category structure, and with the braiding and fusion structure of the original anyons.  These all take the form of pentagon equations and variants of the hexagon equations involving $0$, $1$, or $2$ defects.  Despite the complicated nature of these equations, reference \onlinecite{ENO} again proves a theorem that constrains the possible solutions.  Specifically, proposition 7.3(ii) of \onlinecite{ENO} shows that, for a fixed pair ($g$,$h$), and having fixed all of the one defect data (namely, the map from $G$ into the braided autoequivalences of ${\cal A}$), the set of gauge equivalence classes of solutions for the fusion rules of $g$ and $h$ defects into a $gh$ defect together with the associated $F$-matrices is in one to one correspondence with the set of abelian anyons ${\cal A}$.  Indeed, given one solution, reference \onlinecite{ENO} shows how to `deform' it by any abelian anyon $a_{g,h}$.  Intuitively, one can think of this deformation as splitting off $a_{g,h}$ from the $gh$ defect before it splits into $g$ and $h$ - see figure \ref{fig4}.  We stress that this intuitive picture is just graphical shorthand for a precise formula, written down in appendix \ref{ap:deform}, that expresses the new, deformed fusion rules and F-matrices in terms of the original ones.

Now, suppose we have a complete solution for all of the data in our theory, including F-matrices with any number of anyons or defects.  We can try to find a new solution by leaving alone all data involving at most one defect (in particular the fusion and braiding data of the anyons and the invertible bimodule category structure), deforming, for each pair $(g,h)$, the fusion and $F$-matrix data involving two defects $g$ and $h$ by some anyon $a_{g,h}$, as in figure \ref{fig4} and attempting to solve for the rest of the structure (including associativity conditions with 3 defects, and the defect braiding matrices).  That is, we perform the deformation discussed above for every pair of defects $g$ and $h$.  Now, there is no guarantee that if we do this then we will be able to solve all of the remaining coherence conditions and obtain a consistent theory.  A basic requirement on $a_{g,h}$ is
\begin{equation} \label{eq:gen_cocycle}
\rho_f(a_{g,h})a_{f,gh}=a_{fg,h} a_{f,g}
\end{equation}
where $\rho_f$ represents the permutation action of $f$ on the anyon labels.  In the non-permuting case this condition is simply the requirement that we obtain the same topological superselection sector regardless of the order in which we fuse three defects $f,g,h$.  In the twisted case, equation \ref{eq:gen_cocycle} is not as obvious, but rather follows from imposing all pentagon equations involving three defects and one anyon: a solution for the three defect F-matrix can be found only if eq. \ref{eq:gen_cocycle} is satisfied.  In the language of reference \onlinecite{ENO}, these pentagon equations translate to the statement that the tensor product of bimodule categories referred to previously is associative, and the condition follows from their Theorem 8.5.  We note that even when this is so, this associativity isomorphism might not satisfy the $4$-defect pentagon equation and fail to give a consistent theory \cite{ProjS} - however, we will not be concerned with this issue here.

Because deformations of the form $a_{g,h} = \rho_g (b_h) b_{gh}^{-1} b_g$, where the $b_g$ are abelian, can be undone by a re-labeling of defect superselection sectors, eq. \ref{eq:gen_cocycle} shows that the potentially non-trivial deformations correspond to nonzero twisted cohomology classes in $[\omega] \in H_{\rm{twisted}}^2(G,{\cal{A}})$, represented by the functions $\omega(g,h)=a_{g,h}$.  In section IV of reference \onlinecite{Maissam2014}, there is defined a collection of $U(1)$-valued 2-cocycles $\tau_a(g,h)$, where $a$ ranges over all of the anyons in the theory (not just abelian ones).  This $\tau_a(g,h)$ is related to our $\omega(g,h)$ by:
\begin{align}
\tau_a(g,h)=S_{a,\omega(g,h)}
\end{align}
where $S_{a,\omega(g,h)}$ is the full $U(1)$ braiding phase of the two anyons.

Of course, given such a nonzero cohomology class, it is not a priori clear how many consistent theories it leads to, if any.  Nevertheless, assuming there does exist at least one consistent deformed solution for a given $[\omega]$, one can ask what gauge invariant quantity distinguishes this deformed solution from the un-deformed one.  In principle, one can simply compute all of the relevant (two defect, one anyon) F-matrices for both solutions and check if there is a gauge equivalence between them.  However, in practice this is difficult, both because of the large number of F-matrices involved, and also because extracting universal data associated to F-matrices is significantly more subtle than extracting braiding data for a given physical system.

Fortunately, it turns out that once we specify the deformation $[\omega]$, various coherence conditions - namely pentagon and analogues of the hexagon equations - force much of the remaining fusion and braiding structure.  Indeed, according to the general classification of references \onlinecite{Fidkowski2014, Maissam2014}, the SET is then actually determined almost uniquely: the only ambiguity is the stacking of an extra $G$-SPT (symmetry protected topological phase) on top of the system.  In particular, the braiding of anyons around defects is completely determined.  As we will see in the example below, the gauge invariants that can be constructed from this braiding data contain a much more convenient signature of the deformation.

\subsection{Example: $\Z_4$ gauge theory with $\Z_2$ symmetry}

Let us determine the defect fusion rules in the example discussed above.  The only non-trivial case is a $g$-defect fusing with itself, and in this case, since we know that braiding around two such $g$-defect induces the trivial permutation on anyon labels and the trivial map on anyon fusion spaces, from theorem 5.2 of reference \onlinecite{ENO} it follows that the product of two $g$-defects is a trivial defect.  Recall that we are representing the anyons by pairs $(i,j)$, where $i$ denotes charge $\Z_4$ charge and $j$ denotes $\Z_4$ flux, that the superselection sectors of the non-trivial $g$-defect are labeled by pairs $(\alpha,\beta)_g$, with $\alpha, \beta$ defined modulo $2$, and that the fusion rules are:

\begin{equation} \label{eq:fusion_rules_example}
(\alpha,\beta)_g \times (\alpha',\beta')_g = \oplus_{i,j} (i,j),
\end{equation}
where the sum on the right is over $0 \leq i,j <4$ with $\alpha+\alpha'= i\,\rm{mod}\,2$ and $\beta+\beta'=j\,\rm{mod}\,2$.  Notice that these fusion rules are non-abelian.

Now, according to the above discussion, the potential deformations of this defect product are parametrized by $H^2_{\rm{twisted}} (\Z_2, \Z_4 \times \Z_4) = \Z_2 \times \Z_2$.  Let us now focus on the deformations where the anyons $a_{g,h}$ are fluxes of the $\Z_4$ gauge theory, which restricts us to a $\Z_2$ subgroup of $\Z_2 \times \Z_2$; the non-trivial co-cycle is then $a_{g,g}=(0,2)$.  We do this because this $\Z_2$ is the one that can be accessed in the particular class of exactly solved models given in the previous section.  Indeed, these models, like those of reference \onlinecite{Hermele_string_flux} on which they are based, were constructed to allow for fractional symmetry charge assignments to gauge theory charges, since it is the gauge charges that are sensitive to the Berry flux in the vertical plaquette terms.  The other $\Z_2$ would correspond to a phase where instead the gauge fluxes carry fractional symmetry charge, and would require a different lattice construction \footnote{Of course, different choices of $\omega$ can lead to the same SET phase after a relabeling of anyons.}.

Note that deforming by this co-cycle does not change the fusion rules in equation \ref{eq:fusion_rules_example}.  Thus it is now much more difficult to distinguish the deformed and undeformed products of bimodule categories.  Indeed, as opposed to the untwisted case, where the distinction can be ascertained by just examining the fusion rules for defect superselection sectors, we now have to examine subtle gauge invariant F-matrix data for processes involving two defects and one anyon.  Fortunately, as discussed above, this distinction appears also in the braiding data of anyons with the defect in the full solutions (`braided $G$-crossed categories') corresponding to the deformed and undeformed data.  As shown in appendix \ref{ap:deform}, imposing hexagon coherence conditions shows that the deformed and undeformed cases differ in the sign of the gauge invariant braiding quantity

\begin{equation}\label{eq:criterion}
R_{x_g, (3,0)} R_{(3,0),x_g}  R_{x_g,(1,0)} R_{(1,0),x_g},
\end{equation}
where $x_g$ is an arbitrary superselection sector of the $g$-defect, and $R_{a,b}$ is the R-matrix corresponding to the exchange of $a$ and $b$.  Recall that $(i,j)$ represents an anyon which is a composite of a charge $i$ and flux $j$ $i,j=0,\ldots,3$.  Thus equation \ref{eq:criterion} simply gives the braiding phase for the process of braiding a fundamental $\Z_4$ charge $(1,0)$ twice around the $g$-defect.  Notice that braiding it only once turns it into the opposite charge $(3,0)$ and hence does not lead to a gauge invariant phase.

Now recall that in the previous section we had an exactly solved lattice construction for precisely this topological order and action of symmetry, and found two distinct SETs.  We claim that these are lattice realizations of the two cases - deformed and un-deformed - that we are discussing right now.  In order to argue this, we will just show that the two distinct lattice SETs we constructed in the previous section have a different sign of the gauge invariant quantity defined in eq. \ref{eq:criterion}.  Recall that upon gauging $G$, the two SETs we constructed in the previous section resulted in ${\mathbb D}_4$ and ${\mathbb Q}_8$ topological orders, respectively.

First, let us establish some more notation.  The second group cohomology $H^2_{\rho}(\Z_2, \Z_4)=\Z_2$, where $\Z_2=\{1,g\}$ acts as $\rho: x\rightarrow x^{-1}$ on $\Z_4 = \{1,i,-1,-i\}$.  We found two representative co-cycles, $\omega_{+}$ and $\omega_{-}$.  $\omega_{+}$ was equal to $1$ for all values of its arguments, and $\omega_{-}(g,g)=-1$.  According to Appendix \ref{ap:extend}, $\omega_{+}$ corresponds to the dihedral group ${\mathbb{D}}_4$, and $\omega_{-}$ to the quaternion group ${\mathbb{Q}}_8$.  The first of these just consists of the symmetries of the square, with a $\Z_4$ rotation subgroup which we refer to as $\Z_4 = \{1,i,-1,-i\}$, while the second consists of the elements $\{\pm1,\pm i, \pm j, \pm k\}$ which follow the multiplication rules of the quaternion group.  Here we again take the $\Z_4$ subgroup to be ${\pm 1, \pm i}$.  The key difference between these two extensions, captured by the respective co-cycles, is that in ${\mathbb{D}}_4$, any lift of $g$ squares to $+1$, whereas in ${\mathbb{Q}}_8$, any such lift squares to $-1$.  

We now perform the double braid computation in eq. \ref{eq:criterion} in the fully gauged theories.  Here, the charges $(1,0)$ and $(3,0)$ (obtained from it by acting with $g$) pair up into the unique two dimensional irreducible representation of either ${\mathbb{D}}_4$ or ${\mathbb{Q}}_8$ depending on whether we are in the $\omega_{+}$ or $\omega_{-}$ case, respectively.  Now, the gauge theories (or `quantum doubles') of these two groups can be constructed using a general prescription outlined {\emph e.g.} in section 3.2 of reference \onlinecite{lect_mtc}.  The key formula for computing braiding is the general formula for the universal $R$ matrix given there.  Here we are interested in moving a pure charge - our two dimensional irreducible representation of ${\mathbb{D}}_4$ or ${\mathbb{Q}}_8$ - around a $g$-flux twice.  Using the formula for the R matrix given in section 3.2 of reference \onlinecite{lect_mtc}, the action of such a double braid operation is simply given by $g^2$ acting within the two dimensional irreducible representation of $E$ that contains the fundamental $\Z_4$ charge.  This is just the identity in ${\mathbb{D}}_4$, and $-1$ in ${\mathbb{Q}}_8$, which acts as $-1$ in the two dimensional irreducible representation, so that the value of the double braid in eq. \ref{eq:criterion} differs by a sign for the two theories, as desired.

\section{Summary and new directions}
\label{summary}

In this paper we presented a framework for studying 2d bosonic symmetry enriched topological phases with a finite, unitary symmetry group $G$.  We demonstrated that even when $G$ acts non-trivially on the topological superselection sectors, we can still extract fractionalization data and distinguish different symmetry enriched phases using the fusion rules of extrinsic twist defects.  Moreover, some non-trivial features such fractionalization data are visible only at the level of associativity rules - i.e. F-move isomorphisms.  We illustrated these results in a class of exactly solved lattice models of $\Z_n$ gauge theory, for which the fractionalization data, in the form of a twisted group cohomology class $[\omega]\in H_{\rho}^2(G,\Z_n)$, is one piece of input defining the Hamiltonian.  By gauging $G$ we found that the intrinsic topological order of the resulting model is that of an $E$ gauge theory, where $E$ is the extension of $G$ by $H$ determined by the permutation action $\rho$ and $[\omega]$.  In particular, whenever these gauge theories are distinct, so are the underlying SETs. 

In the future we would like to extend our results to the case of time reversal symmetry.  Its anti-unitary nature makes it difficult to define the notion of a twist defect and of gauging, but there are indications \cite{Chen_TR} that some notion of symmetry fractionalization should still hold in this case.  In particular, we believe that the exactly-solved models of reference \onlinecite{Hermele_string_flux} can be generalized to this case.  Also, we would like to study $3$ dimensional generalizations of these models.  Indeed, topological order and symmetry enrichment are not as well understood in $3$ dimensions as they are in $2$ dimensions, and it would be nice to have a class of exactly solved 3d models to benchmark potential classification schemes.  Another set of systems we would like to understand are the Walker-Wang models of 3d SPTs - perhaps the ideas presented in this paper can be used to give a more geometric construction of the symmetry decorated semion model of reference \onlinecite{ProjS}.  Finally, it would be good to see whether these ideas generalize to fermionic systems, and in particular, to what extent it is useful to think of gauging the global fermion parity symmetry.

\section*{Acknowledgments}

We would like to thank A. Kitaev for many insightful discussions, and especially for pointing out the mathematical work of \onlinecite{ENO} and \onlinecite{DGNO}.

\appendix

\section{Derivation of $E$ gauge theory Hamiltonian} \label{ap:derive}

In this sub-section we will determine the low energy physics of the gauged Hamiltonian given in eq. \ref{eq:gauge_dyn}.  Again, this sub-section is rather technical, but the final result is quite simple: the low energy physics of this gauged Hamiltonian is equivalent to that of the non-abelian lattice gauge theory based on the extension $E$ of $G$ by $\Z_n$ determined by $\rho$, given in equation \ref{eq:ham_final_mod}.

Our analysis here is similar to that of reference \onlinecite{Hermele_string_flux} in the non-anyon-permuting case.  The general strategy is to simplify the problem by imposing certain terms in the gauged Hamiltonian in eq.~\ref{eq:gauge_dyn} as constraints.  This gets rid of states in the Hilbert space which violate these constraints, leaving us with a smaller Hilbert space that is hopefully easier to understand.  However, we have to be careful and ensure that by doing this we have not removed any topologically non-trivial excitations from the spectrum, because if we have, we no longer have a model with the same topological order.  A second issue is that the resulting Hilbert space, although smaller and in principle easier to understand than the original, is a constrained Hilbert space.  That is, it is not of the form of a tensor product of site Hilbert spaces.  Because we would like to work with unconstrained Hilbert spaces, we resolve this issue by rewriting the constrained Hilbert space as an unconstrained Hilbert space together with a reparametrization redundancy.  Again, we show that violating this reparametrization redundancy does not introduce additional topological superselection sectors, so that that the reparametrization redundancy condition can be imposed energetically.  Thus, at the end we have another generalized spin model with the same topological order and low energy physics; in effect we have carried out a duality transformation.  This new generalized spin model, after some trivial manipulations, turns out to be the gauge theory of the extension $E$.

\subsection{Establishing the Hilbert space constraints}
\subsubsection{Supervertex constraints \label{sssect:supervertex}}

We first impose as constraints the plaquette terms $\tilde{C}_p$ for plaquettes $p$ contained entirely within a single supervertex $V$: that is, we restrict to the states $|\Psi\rangle$ for which $\tilde{C}_p |\Psi\rangle = |\Psi\rangle$.  For each plaquette triple $(f,g,h)$, this constraint reads
\begin{align} \label{eq:supervertex}
\eta_{V}(f,g)\eta_{V}(fg,h)  \eta_{V}(f,gh)^{-1} =\omega(g,h)^{\rho(f)}
\end{align}
Using eq. \ref{eq:cocycle1}, we see that one solution to this equation is $\bar{\eta}(g,h) = \omega(g,h)$.  Now, any other solution to eq. \ref{eq:supervertex} must be related to $\bar{\eta}(g,h)$ by a gauge transformation:\footnote{This is because any two solutions to eq. \ref{eq:supervertex} must differ by a flat $\Z_n$ gauge field configuration on the Cayley graph of the supervertex $V$: that is, a $\Z_n$ gauge field with no flux.  Because the complex consisting of the Cayley graph and all its plaquettes has no non-trivial cycles, such a flat $\Z_n$ gauge field configuration must be gauge equivalent to the trivial one.}
\begin{align}
\label{eq:background}
\eta_V(g,h) = \theta^{-1}_V(gh)\bar{\eta}(g,h)\theta_V(g)
\end{align}
We now define an unconstrained Hilbert space by trading in the $\Z_n$ link variables $\eta_V(g,h)$, i.e. those with links entirely within supervertex $V$, for $\Z_n$ valued `clock' variables $\theta_V(g)$.  We introduce the usual operator algebra associated to each $\Z_n$ clock: $\Theta_V(g)$ is the operator that has $\theta_V(g)$ as an eigenvalue, and $\Phi_V( g)$ acts as $\ket{\theta_V(g')} \rightarrow \ket{e^{ \frac{2\pi i}{n}\delta_{g,g'}}\theta_V(g')}$.  One advantage of representing $\eta_V(g,h)$ in terms of the $\theta_V(g)$ as in eq. \ref{eq:background} is that it simplifies the vertex term $\tilde{A}_{V}(g)$.  Indeed, the contributions to $\tilde{A}_V(g)$ coming from links within the supervertex $V$ amount to just $\Phi^{-1}_{V}( g)$, so the vertex term becomes:
\begin{align}
\tilde{A}_{V}(g) \rightarrow  \Phi^{-1}_{V}( g) \prod_{L\sim V} e_L( g)^{s'_{V}(L)}
\end{align}
%An application of eq.\ref{eq:Ufl_operators} shows that the $\Theta_v(g)$ transform projectively under G, satisfying
%\begin{align}
%U_f^{-1} \Theta_v(g) U_f =  \omega(f,f^{-1}g) \Theta_v(f^{-1}g)^{\rho(f)}\\
%\therefore U_g^{-1} U_f^{-1} \Theta_v(h) U_f U_g = \omega(f,g) U_{fg}^{-1} \Theta_v(h) U_{fg}
%\end{align}
Note that $\{ \theta_V(g) \}$ form a redundant parametrization of the original link degrees of freedom, in that shifting all $\theta_V(g)$ simultaneously by the same $g$-independent amount does not change any $\eta_V(g,h)$ defined in eq. \ref{eq:background}.  Thus the Hilbert space of the gauged Hamiltonian in eq. \ref{eq:gauge_dyn} with the supervertex constraints $\tilde{C}_p |\Psi\rangle = |\Psi\rangle$, $p\in V$ imposed, is recovered by projecting onto the subspace where
\begin{align}
\label{eq:gauge_condition}
\prod_{g\in G} \Phi^m_V( g)\ket{\Psi} = \ket{\Psi},~\forall~m\in \Z_n
\end{align}
This is the reparametrization redundancy condition we mentioned earlier.

What about the Hamiltonian?  We can certainly write down a Hamiltonian which reduces to that in eq. \ref{eq:gauge_dyn} when acting on the subspace satisfying the condition in eq. \ref{eq:gauge_condition}, and which also includes projectors onto this subspace, such that these projectors commute with the rest of the Hamiltonian.  However, before we do this we will impose the second set of constraints in the next subsection.  For now, however, we just note that such a Hamiltonian will necessarily have the same topological order as the original one in eq. \ref{eq:gauge_ham}.  To see this, we argue as follows.  First of all, one may worry that there are non-trivial topological excitations in the Hamiltonian of eq. \ref{eq:gauge_ham} which are removed from the spectrum once the constraint $\tilde{C}_p=1$ for $p\in V$ is imposed.  However, it is fairly easy to see that this is not the case.  Indeed, any violation of $\tilde{C}_p$ associated to $V$ can be removed with a local operator supported only on $V$.  This operator may create other excitations which are violations of $\tilde{C}_{p'}$ for $p'$ containing some horizontal links, but crucially, it will not create violations of any $\tilde{C}_{p''}$ for $p''$ contained entirely in some supervertex.  Now, the violations of $\tilde{C}_{p'}$ for $p'$ containing horizontal links can again be removed by acting on the appropriate horizontal link variables.  These will not create excitations at any other $\tilde{C}_p$.  Thus, any local excitation can be pushed off to violating only the $B_p(g)$ and remaining terms in the Hamiltonian using a local process, so that its topological superselection sector remains in the spectrum even after the constraint is imposed.  A similar argument shows that imposing the reparametrization redundancy condition in eq. \ref{eq:gauge_condition} does not introduce new topological superselection sectors.

\subsubsection{Superlink constraints \label{sssect:superlink}}

Let us now also impose the constraints $\tilde{C}_p|\Psi\rangle=|\Psi\rangle$ for $p$ involving a superlink $L=\langle V V' \rangle$.  The definition of $\tilde{C}_p$ in eq. \ref{eq:C_gauged} shows that $\tilde{C}_p$ just measures the $\Z_n$ flux in a certain rectangular plaquette involving the superlink $L$.  A slight rearrangement of eq. \ref{eq:C_gauged} shows that setting this $\Z_n$ flux to $1$ is equivalent to:
\begin{align}
\eta_L(gh)\eta_V(g_L ^{-1} g , h )^{\rho(g_L)} &= \eta_{V'}(g,h) \eta_L(g)
\end{align}
Now inserting the representation of $\eta_V(g,h)$ in terms of the $\theta_V(g)$ variables introduced in the previous section (see eq. \ref{eq:background}) we obtain:
\begin{align}
&\eta'_L(gh,g_L) \bar{\eta}(g_L^{-1} g , h )^{\rho(g_L)} = \bar{\eta}(g,h) \eta'_L(g,g_L) \label{eq:slink}\\
&\text{where } \eta'_L(g,g_L) \equiv \theta_{V'}(g) \eta_L(g) \theta_V(g_L ^{-1}g)^{-\rho(g_L)}
\end{align}
This equation shows that once we specify $\eta'_L(g,g_L)$ for a single $g$, it is then determined uniquely for all $g$. A useful one to fix is $\eta'_L(g_L,g_L)$, simplifying eq. \ref{eq:slink} to
\begin{align}
\eta'_L(g_L h,g_L) = \omega(g_L,h) \eta'_L(g_L,g_L)
\end{align}
 From this point forward, we will refer to $\eta'_L(g_L,g_L)$ as $\eta'_L$, {\emph{i.e.}} drop the two arguments. Imposing the superlink plaquette constraint thus leads to a Hilbert space which can be parametrized by introducing a single $\Z_n$ clock variable $\eta'_L$ for each superlink $L$, in place of the link variables $\eta_L(g)$.  Note that there is no additional gauge redundancy introduced by this parametrization.  We let $a'_L$ and $e'_L$ be the evaluation and shift operators corresponding to the $\eta'_L$ variable.
\begin{figure}[htbp]
\begin{center}
\includegraphics[width=0.3\textwidth]{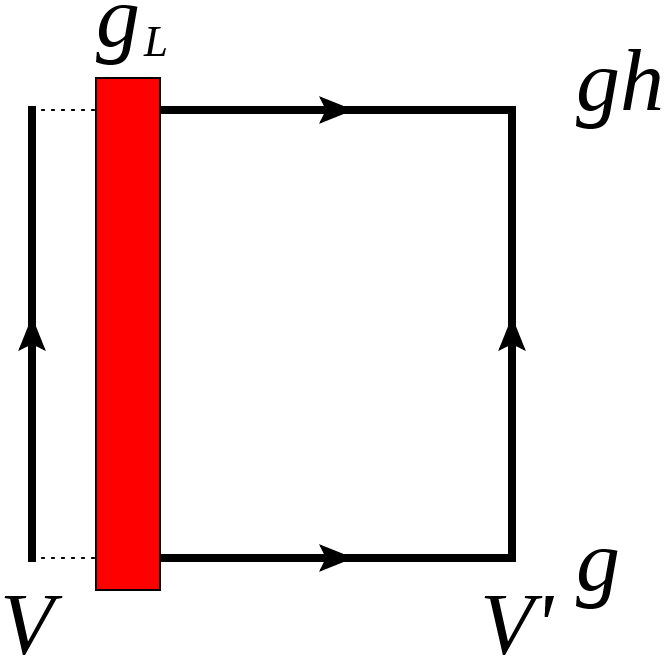}
\caption{Links acted upon by the superlink constraint}
\label{lattice}
\end{center}
\end{figure}
\subsection{Hamiltonian after imposing constraints on the gauged theory}

We have thus constructed a Hilbert space parametrized by $\{ \theta_{V}(g) \}_V$, $\{\eta'_L, g_L \}_L$.  Our goal here is to write out the Hamiltonian in this Hilbert space. We begin with the plaquette term $\tilde{B}_P(g)$ in eq. \ref{eq:B_gauged}, assuming, without loss of generality, that all links are oriented positively relative to the plaquette $p$, which we take to be $j$-sided.  Order the $j$ links $1, \ldots, j$ starting from an arbitrary point, and for convenience let $g_i = g_{L_i}$ denote the $G$ gauge field variable on the $i$th link, $1\leq i \leq j$.  In appendix \ref{ap:gauge_plaq} we show that the following term reproduces $\tilde{B}_P(g)$ in the new Hilbert space:

\begin{align}
\label{eq:gauge_plaq}
&\text{Tr}_{G,\Z_n} \tilde{B}_p(g) \rightarrow \text{Tr}_{\Z_n} \Omega_p {B}'_p  \text{ Tr}_G \mathcal{B}_p \\
 &\Omega_p = \prod_{i=1}^j \omega\left(\prod_{r=1}^{i-1}g_r, g_i\right)\\
 &B'_p= \prod_{i=1}^j \left(a'_l\right)^{\rho(\prod_{r=1}^{i-1}g_r)}\\
 &\mathcal{B}_p =  \prod_{i=1}^j \al_{i}
\end{align}
where $\text{Tr}_G$ refers to tracing over the regular representation of $G$.

Now, for each superlink $L$, the pair $(\eta'_L, g_L)$ can naturally be thought of as an element of the extension $E$ of $G$ by $\Z_n$ determined by $\omega(g,h)$, since the extension is just a twisted product of $\Z_n$ and $G$ (see appendix \ref{ap:extend} for some background on group extensions).  The twisting in this product is manifested only in the non-trivial nature of the group multiplication law in $E$:

\begin{equation}
(\eta'_1,g_1)\cdot(\eta'_2,g_2) = (\eta'_1 {\eta'_2}^{\rho(g_1)}\omega(g_1,g_2), g_1g_2)
\end{equation}
This way of thinking about the degrees of freedom on superlinks is especially convenient because the plaquette term in eq. \ref{eq:gauge_plaq} precisely just enforces that the pairs $(\eta'_L, g_L)$ around a plaquette multiply to the identity in the extension $E$.  Indeed, repeated applications of the above group multiplication law in $E$ give
\begin{align}
\prod_{i}(\eta'_i,g_i) = \left(\prod_i (\eta'_i)^{\prod_{j=1}^{i-1}\rho(g_j)}  \prod_{i} \omega\left(\prod_{r=1}^{i-1}g_r, g_i\right),  \prod_i g_i \right)
\end{align}
which is precisely the quantity that the term in eq. \ref{eq:gauge_plaq} energetically prefers to be trivial.  Thus we just have the plaquette term of a non-abelian $E$ gauge theory.

As for the vertex term, a quick calculation shows the action of $\tilde{A}_V( g)$ within the constrained Hilbert space is the same as that of the operator $\Phi'_V( g)$ defined by:
\begin{align}
 \Phi'_V( g) &\left( \bigotimes_{g'\in G}\ket{\theta_V(g')} \otimes \bigotimes_{L \sim V}\ket{\eta'_L, g_L}\right)=\notag\\
 & \bigotimes_{g'\in G }\ket{e^{ \frac{2\pi i}{n}\delta_{g,g'}}\theta_V(g')} \otimes \bigotimes_{L \sim V}\ket{\eta'_L, g_L} 
\end{align}
This is a great simplification, because $\Phi'_V( g)$ does not act at all on the $\eta'_L$ degrees of freedom.  On the other hand, the constraint \ref{eq:gauge_condition}, which now also has to be imposed energetically, is less trivial.  We have:
\begin{align}
\prod_{g\in G} & \Phi_V( g) \left(\bigotimes_{g'\in G }\ket{\theta_V(g')} \otimes \bigotimes_{L \sim V}\ket{\eta'_L, g_L} \right)\notag\\
=&\begin{cases}
\bigotimes \ket{ \theta_L(g)e^{\frac{2\pi i}{n}}} \otimes \bigotimes \ket{e^{\frac{2\pi i}{n}} \eta'_L, g_L}, L \rightarrow V\\
\bigotimes \ket{ \theta_L(g)e^{\frac{2\pi i}{n}}} \otimes \bigotimes \ket{ \eta'_L e^{-2\pi i\frac{\rho(g_L)}{n}}  , g_L}, L \leftarrow V
\end{cases}
\end{align}
This is equal to the action of the operator
\begin{align}
A'_V \prod_{g\in G} \Phi'_V( g)
\end{align}
with
\begin{align}\label{avprime}
A'_V &\equiv \prod_{L\sim V} \left(e'_L\right)^{s'_V(L)},
\end{align}
where
\begin{align}
e'_L\ket{\eta'_L, g_L} &= \ket{e^{\frac{2\pi i}{n}} \eta'_L, g_L}\\
\left(e'_L\right)^R\ket{\eta'_L, g_L} &= \ket{\eta'_L e^{-2\pi i\frac{\rho(g_L)}{n}}, g_L}
\end{align}
Now, since $\prod_g \Phi'_V(g)$ already appears in the Hamiltonian, and since all the terms in the Hamiltonian will commute, we can replace it with $1$ in the expression $A'_V\prod_{g\in G} \Phi'_V( g)$, so ultimately the constraint \ref{eq:gauge_condition} is imposed by including the terms $-\sum_m A'^m_V$ in the Hamiltonian.  Now, according to eq. \ref{avprime}, $A'^m_V$ is just the gauge transformation by $m \in \Z_n \subset E$ at supervertex $V$, when we interpret the degrees of freedom as those of an $E$ gauge theory.  Thus so far we have recovered the plaquette term of the $E$ gauge theory, and a term that imposes gauge invariance with respect to the $\Z_n$ subgroup of $E$.   

Finally, we must write down terms corresponding to $\mathcal{A}_V(g)$, which make the $G$ gauge field fluctuate.  There is some ambiguity in the choice of these terms because the action on the portion of the Hilbert space violating the constraint \ref{eq:gauge_condition} is not uniquely determined.  However, one valid choice, which commutes with this constraint and the rest of the terms in the Hamiltonian, is to simply define $\mathcal{A}_V(g)$ to act as the gauge transformation by $(1,g)\in E$; the arbitrariness is due to the fact that we could have equally chosen $(e^{2\pi i m/n},g)\in E$.  

So, our Hamiltonian, when written in terms of operators that act on  $\{ \theta_V(g),\eta'_L, g_L \}$, becomes
\begin{align} 
\ham'_{E} =&- \sum_V \sum_{m\in \Z_n}\Phi'^m_V(g)- \sum_P  \text{Tr}_H \Omega_P {B}'_P  \text{ Tr}_G \mathcal{B}_P\notag\\& - \sum_V \sum_m A'^m_V - \sum_V \sum_{g\in G}\mathcal{A}_V(g) \label{eq:ham_final}
\end{align}

This is almost the Hamiltonian of an $E$ gauge theory.  Indeed, the first term simply enforces that the ground state is symmetrized over the $\theta$'s, and the $\theta$ degrees of freedom do not appear anywhere else in the Hamiltonian.  Thus we can remove them from the Hamiltonian without changing the low energy physics or the topological order.  As far as the rest of the Hamiltonian, we can make it exactly equal to that of an $E$ gauge theory by adding in terms of the form $A'^m_V\mathcal{A}_V(g)$; these do not introduce any new conditions:

\begin{align}
\ham_{E} &=- \sum_V \sum_{\substack{g\in G \\ m \in \Z_n}} A'^m_V\mathcal{A}_V(g)\notag\\&\phantom{{}=} - \sum_P  \text{Tr}_{\Z_n} \Omega_P B'_P  \text{ Tr}_G \mathcal{B}_P \label{eq:ham_final_mod}
\end{align}
This is simply the Hamiltonian of the discrete lattice gauge theory based on the group $E$, the extension of $G$ by $\Z_n$ defined by our group cohomology class $\omega(g,h)$.  In particular, $A'_V$ and $\mathcal{A}_V(g)$ satisfy
\begin{align}
\mathcal{A}_V(g_1)\mathcal{A}_V(g_2)&= A'^{q(g_1, g_2)}_V\mathcal{A}_V(g_1 g_2)\\&\text{ where }e^{\frac{2\pi i}{n} q(g_1,g_2)} = \omega(g_1,g_2) \notag\\
\mathcal{A}_V(g)A'^m_V &= A'^{\rho(g)m}_V\mathcal{A}_V(g)
\end{align}
The Hamiltonian $\ham_E$ is equivalent to that in eq. \ref{eq:Egauge}.  While the arguments we have given were written down only for $\Z_n$ gauge theory, it is quite easy - though notationally cumbersome - to generalize to the gauge theory of any finite abelian group.

\section{Symmetry fractionalization, local action of $G$, and defect fusion rules} \label{long_appendix}

\subsection{Local action of $G$ on anyons \label{sub:Trivial-permutation-action}}

\begin{figure}[htp]
\begin{center}
\includegraphics[scale=0.3]{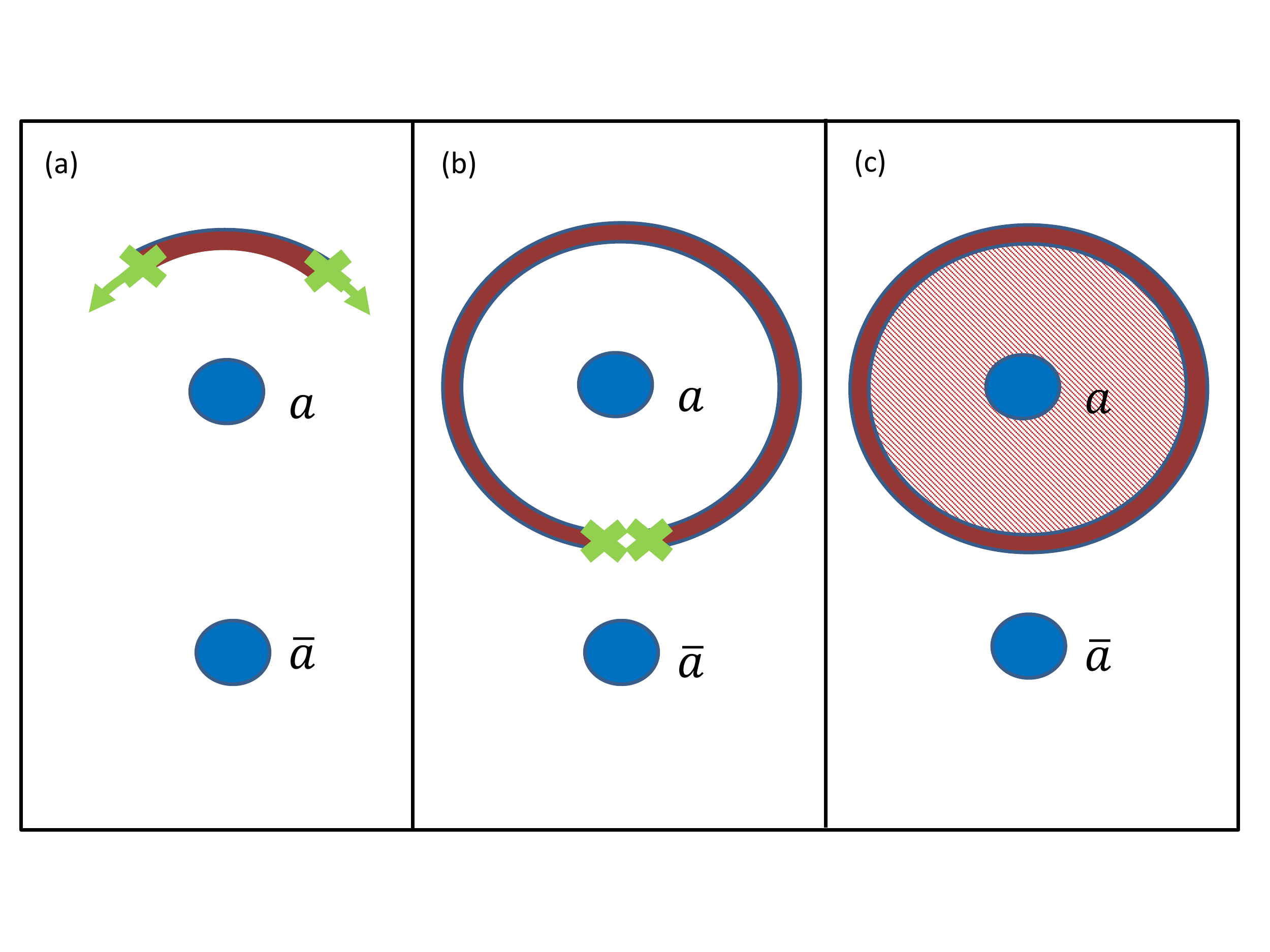}
\caption{Local application of $g\in G$ for a trivial permutation of the anyon
types. The initial system contains two anyons of types $a$ and $\bar{a}$.
A $g$-defect and $g^{-1}$ defect are nucleated, braided around the
$a$ anyon and annihilated, leaving a branch cut surrounding the anyon.
The symmetry $g\in G$ is then applied to all degrees of freedom inside
defect's path, thereby resulting in a low energy state of the original Hamiltonian.
\label{fig:Local-application-of-symmetry}}
\end{center}
\end{figure}

Throughout this subsection, we assume that $G$ fixes the topological superselection sectors.  Suppose we have a state with anyon $a$ at position $x$, and its anti-particle somewhere far away at position $y$. In order to extract the fractional quantum numbers, we would like to define a local action of $g\in G$ in a neighborhood of $x$ - not including $y$ - so that we are sensitive to the Berry phases associated with acting on $a$, without also picking up the opposite phases from $\bar{a}$.  More precisely, for each $g\in G$ we would like to have an operator $U^x_g$ that acts only on spins near $x$, that commutes with the Hamiltonian, and that has the same commutation relations with operators localized near $x$ as the global action $U_g$.  The naive proposal would be to act only on some subregion $R_{x}$ of spins near $x$, but this is clearly unsatisfactory: such an action does not commute with the Hamiltonian at the boundary of this subregion, and so creates spurious excitations at this boundary. Also, it clearly has no projective character, since the underlying spins are honest linear representations of $G$.  Instead, we define $U^x_g$ as follows.

Our operator $U_{g}^{x}$ will be designed to act non-trivially only when the state being acted on looks like the ground state at the boundary of $R_x$ - {\emph{i.e.}} its reduced density matrix is the same as the ground state's; if not, we define $U_g^x$ to act as $0$.  When the state does look like the ground state at the boundary of $R_x$, we act by braiding a twist defect of $g$ around this boundary and acting with $U^g$ inside $R_x$.  To explain this, let us first explain what we mean by a twist defect.

\begin{figure}[htb]
\begin{center}
\includegraphics[trim= 3cm 2cm 13cm 2.5cm, clip,scale=0.3]{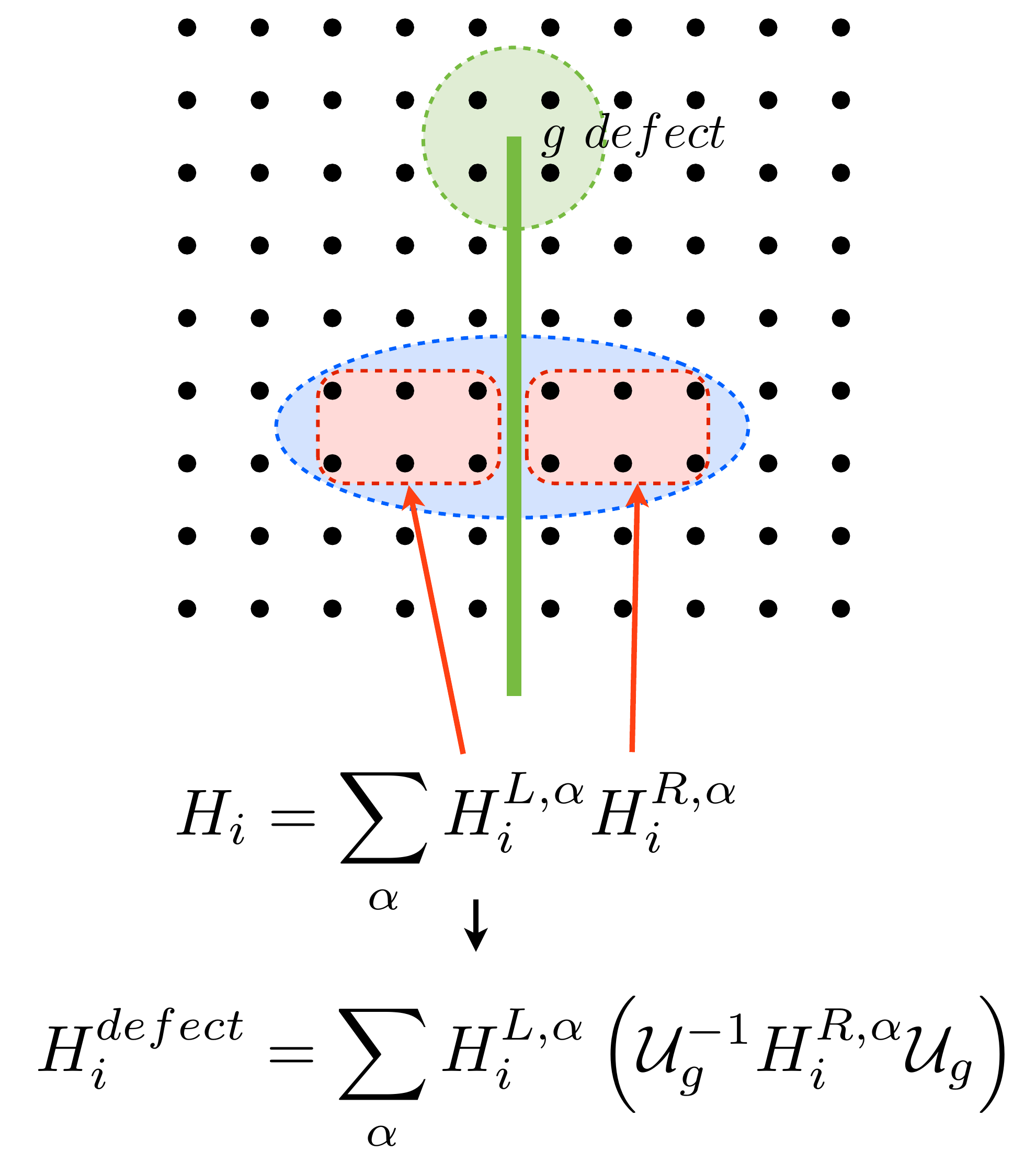}
\caption{The Hamiltonian $H=\sum_{i}H_{i}$, where the $H_{i}$ are local terms.
The introduction of the $g$-defect modifies any $H_i$ (blue) which straddles the branch cut (green).
In such a case, $H_{i}$ is written as a sum of terms which
are products of operators acting on the spins to the
left and right of the branch cut, and then conjugates only the
right ones by the action of $g\in G$. Note that this definition is
unambiguous only far away from the $g$-defect itself.\label{fig:branch_cut_definition}}
\end{center}
\end{figure}

A twist defect is a modification of the Hamiltonian along a branch cut terminating at the defect.  The local terms in the Hamiltonian which straddle the branch cut then get twisted by $g$, as illustrated in Fig \ref{fig:branch_cut_definition}.  For example, if $G=\mathbb{Z}_{2}$ is the Ising symmetry, a term $\sigma_{i}^{z}\sigma_{j}^{z}$ coupling two sites $i$ and $j$ on opposite sides of the branch cut flips sign: $\sigma_{i}^{z}\sigma_{j}^{z}\rightarrow-\sigma_{i}^{z}\sigma_{j}^{z}$.  Note that this procedure is ambiguous near the endpoints of the cut, where the ``sides'' of the cut are not well defined - such microscopic choices turn out to lead to discrete gauge ambiguities in the definitions below.

Now to define $U_g^x$, we simply nucleate a pair consisting of a $g$-defect and a $g^{-1}$ defect, connected by a branch cut, and then grow the branch cut to encircle the region $R_{x}$ (Fig. \ref{fig:Local-application-of-symmetry}).  This process defines a continuous path in the space of gapped Hamiltonians, and $U_{g}^{x}$ acting on a state $|\psi\rangle$ is then defined by adiabatically following $|\psi\rangle$ through the branch cut nucleation and growth process, and then acting by $U_g$ on all of the spins inside the region $R_x$.  We fix the overall phase of $U_{g}^{x}$ by demanding that $U_{g}^{x}|0\rangle=|0\rangle$, where $|0\rangle$ is a state containing no anyons. Because the Hamiltonian is gapped, we expect that the adiabatic continuation can be approximated exponentially well by an operator acting in a thin ribbon (several correlation lengths thick) around the boundary of $R_{x}$. Hence $U_{g}^{x}$ acts like $U_{g}$ in the interior of $R_{x}$, like the identity outside $R_{x}$, and interpolates smoothly between these options within a few correlation lengths of the boundary of $R_{x}$.

Because we made some arbitrary choices - {\emph e.g.} the nature of the Hamiltonian at the defect core - we want to see to what extent $U_{g}^x$ is well defined.  A naive argument would say that two different such constructions of $U_g^x$ have the same commutation relations with all local operators near $x$, and their overall phases are fixed to be the same, so they must be equal.  This is not quite correct, however, because there exist excitations near $x$ - namely anyons - which cannot be created with local operators, and are sensitive, via braiding, to possible topological charge bound to the encircling $g$-defect.  That is, if we had chosen a different nucleation process with different defect cores, it is possible that it would differ from the first by an additional abelian anyon stuck on the defect, which could measure topological charge at $x$ via braiding (a non-abelian anyon stuck to the defect would be more problematic, and may not lead to a well-defined $U_g^x$; we evade this possibility by insisting on a fully gapped defect).  Now, there is no way to unambiguously measure the abelian anyon charge on a single defect, so we cannot specify the abelian anyon charge carried by the defect in our procedure - the best we can do is to specify the local $G$-action up to the gauge equivalence:

\begin{equation}
U_{g}^{x}|a\rangle\rightarrow e^{2\pi i\langle a,\alpha(g)\rangle}U_{g}^{x}|a\rangle\label{eq:gauge_ambiguity}
\end{equation}

\noindent Here $|a\rangle$ is any state with total anyon charge $a$ inside the branch cut loop, $\alpha(g)$ is an arbitrary abelian anyon, and the angular brackets denote the full braiding phase of $\alpha(g)$ around $a$ (i.e. $e^{2\pi i\langle\alpha(g),a\rangle}=S_{\alpha(g),a}$ is their $S$-matrix element).

\subsection{Projective nature of local action of G on anyons\label{sub:Projective-nature-trivial-eta}}

While the global action of $G$ satisfies the group relation $U_g U_h = U_{gh}$, the corresponding local operators $U_{g}^{x}, U_h^x, U_{gh}^x$ might only satisfy it up to a Berry's phase, which depends on $g,h$, and the anyon charge $a$ near $x$.  To demonstrate this, let us form the operator that would measure this Berry's phase:

\begin{equation}
V_{g,h}^{x}\equiv U_{gh}^{x}(U_{h}^{x})^{-1}(U_{g}^{x})^{-1}\label{eq:def_of_V}
\end{equation}

\noindent Now, consider any single site operator $S_{i}$ acting on
a spin $i$ well inside the local neighborhood $R_{x}$ of $x$ -
i.e. inside the encircling branch cut. Its commutation relations with
$U_{g}^{x}$ are the same as its commutation relations with $U_g$
- this is just the statement that $U_{g}^{x}$ acts just like $U_g$
inside the branch cut region.
\begin{figure}[htbp]
\begin{center}
\includegraphics[trim= 7cm 7cm 16cm 6.5cm,clip,scale=0.35]{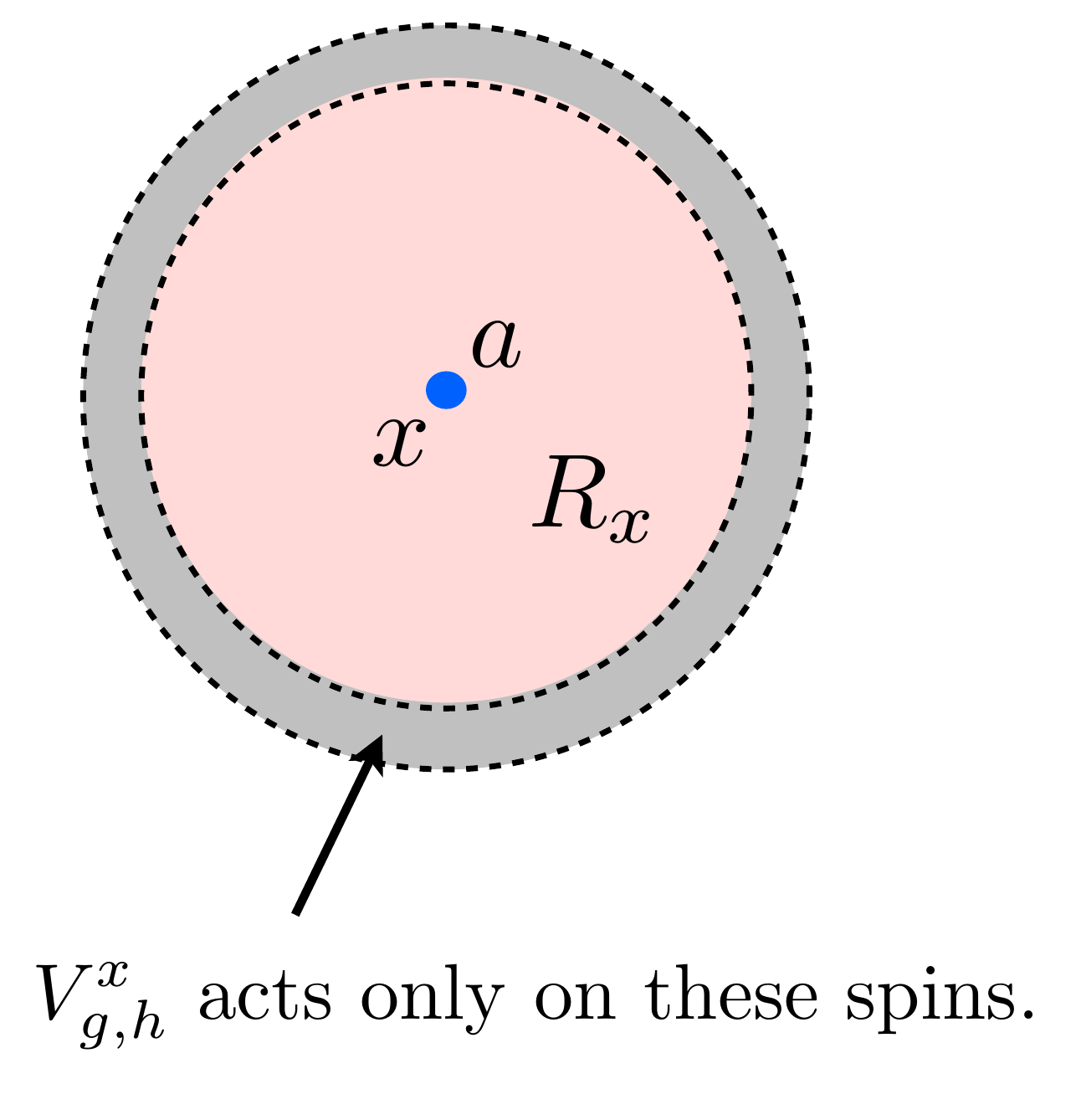}
\caption{Local action of $G$ on anyon $a$ at position $x$. While the operators
$U_{g}^{x}$, $U_{h}^{x}$, and $U_{gh}^{x}$ act non-trivially in
the entire disk, $V_{g,h}^{x}\equiv U_{gh}^{x}(U_{h}^{x})^{-1}(U_{g}^{x})^{-1}$
can be approximated by an operator that acts only on the spins within
the narrow surrounding grey ribbon region.\label{fig:ribbon}}
\end{center}
\end{figure}

\noindent $S_{i}$ must commute with $V_{g,h}^{x}$, since the latter acts as $U_{gh} (U_h)^{-1}(U_g)^{-1}=1$
well inside $R_{x}$. Thus, since $V_{g,h}^{x}$ commutes with all
observables inside the cut, we can approximate it by an operator that
acts only on spins in a thin ribbon region around the cut itself,
insofar as its action on excitations well inside the cut is concerned
- see Fig. \ref{fig:ribbon} (the approximation error is exponentially
small in the thickness of the ribbon divided by the correlation length).

When an interior excitation is created by a local operator acting
on the vacuum, i.e. is of the form $|loc\rangle=Y|0\rangle$, with
Y a local operator acting only on spins inside the cut, we have

\begin{center}
$V_{g,h}^{x}|loc\rangle=V_{g,h}^{x}Y|0\rangle=YV_{g,h}^{x}|0\rangle=|loc\rangle$;
\par\end{center}

\noindent $V_{g,h}^{x}$ and $Y$ commute because they are spatially
separated. Thus there are no Berry's phases for excitations that can
be made locally, as expected. However, consider now a state $|a\rangle$
which contains an anyon $a$ at $x$ (and no other quasiparticles
within the branch cut region). $|a\rangle$ cannot be nucleated locally
from the ground state, but rather must be created using a string operator
that intersects the branch cut, and hence might fail to commute with
$V_{g,h}^{x}$, resulting in a Berry's phase,

\begin{flushleft}
\begin{equation}
V_{g,h}^{x}|a\rangle=e^{i\xi(g,h;a)}|a\rangle,\label{eq:def_xi}
\end{equation}

\par\end{flushleft}

To compute this action of $V_{g,h}^{x}$, note first that it must be by an $a$-dependent phase, since $V_{g,h}^{x}$ must be proportional to the identity within each topological superselection sector, by the previous argument.  Furthermore, if $c$ is in the fusion product of $a$ and $b$, the phase for $c$ must be the sum of the phases for $a$ and $b$, because there exists a local operator that fuses $a$ and $b$ into $c$.  Mathematically, any such function from the anyons to $U(1)$ must be equivalent to the braiding of an abelian anyon - which we call $\omega(g,h)$ - around the anyon in question.  Note that $V_{g,h}^x$ has support on a thin ribbon surrounding the anyon in question, so this is certainly an allowed possibility.  We thus have
\begin{flushleft}
\begin{equation}
V_{g,h}^{x}|a\rangle=e^{2\pi i\left\langle a,\omega(g,h)\right\rangle }|a\rangle=S_{\omega(g,h),a}|a\rangle\label{eq:VS}
\end{equation}

\par\end{flushleft}

\noindent where $S_{\omega(g,h),a}$ is the full braid $S$-matrix element.

Note that there are constraints among the $\omega(g,h)$.  Specifically, consider the operator $W=U_{fgh}^{x}(U_{h}^{x})^{-1}(U_{g}^{x})^{-1}(U_{f}^{x})^{-1}$.  We can write it in the following two ways:

\begin{center}
$W=V_{fg,h}^{x}\, V_{f,g}^{x}$
\par\end{center}

\noindent and

\begin{center}
$W=V_{f,gh}^{x}\, U_{f}^{x}V_{g,h}^{x}(U_{f}^{x})^{-1}$.
\par\end{center}

\noindent The first tells us that $W$ acts by braiding $\omega(f,g)$
and then braiding $\omega(fg,h)$, while the second says that it is
equal to the composition of braiding operations of $\omega(g,h)$
and $\omega(f,gh)$ (the conjugation by $U_{f}^{x}$ does not affect
anything, because $U_{f}^{x}$ does not change the topological charge
$a$ that is being acted on). Thus, we have

\begin{center}
$\omega(fg,h)\omega(f,g)=\omega(f,gh)\omega(g,h)$ 
\par\end{center}

\noindent or

\begin{center} 
$\omega(g,h)\omega(fg,h)^{-1}\omega(f,gh) \omega(f,g)^{-1}=1$
\par\end{center}

\noindent Those acquainted with group cohomology
will recognize the above as the cocycle equation

\begin{flushleft}
\begin{equation}
d\omega(f,g,h)=1\label{eq:cocycle}
\end{equation}

\par\end{flushleft}

Furthermore, we can use the definition (\ref{eq:def_of_V}) of $V_{g,h}^{x}$
to see that under the inherent gauge ambiguity (\ref{eq:gauge_ambiguity}),
$\omega(g,h)$ must transform as

\begin{center}
$\omega(g,h)\rightarrow\omega(g,h)\alpha(g)\alpha(h)\alpha(gh)^{-1}$ 
\par\end{center}

\noindent \begin{flushleft}
which can also be re-written in group cohomology terms as
\par\end{flushleft}

\begin{flushleft}
\begin{equation}
\omega(g,h)\rightarrow\omega(g,h) \cdot d\alpha(g,h)\label{eq:gauge_transform}
\end{equation}

\par\end{flushleft}

\noindent Functions $\omega(g,h)$ satisfying eq. (\ref{eq:cocycle}),
modulo the gauge equivalences in eq. (\ref{eq:gauge_transform}),
form a finite abelian group called the second group
cohomology $H^{2}(G,{\cal A})$.  Here ${\cal A}$ is the abelian group formed by anyons of quantum dimension $1$.  The equivalence class of $\omega$ in this set is denoted $[\omega]\in H^{2}(G,{\cal A})$, and is an invariant which encapsulates the projective properties of the anyons in our symmetry enriched topological phase.

\section{Deformations of the defect fusion product} \label{ap:deform}

Here we give a precise description of the deformed defect product defined in figure \ref{fig4}.  Recall that the data defining the original defect product of two defects of type $g$ and $h$ is a collection of fusion rules for $g$ and $h$ superselection sectors, and F-matrices involving the triple consisting of a sector of $g$, a sector of $h$, and an anyon, in any order.  In the following we assume no fusion multiplicity, which makes the formulae simpler; the general case is not any harder but the formulae are more cumbersome to write out.

Suppose we deform the defect product by $a_{g,h}$, which we denote $a$ for short.  Then the deformed fusion rules are simply of the form $x\times y=z$ where $x$, $y$, and $z$ are sectors of the $g$, $h$, and $gh$ defects respectively and $x\times y = \overline{a} z$ is an allowed fusion rule of the undeformed defect product.  Notice that sometimes, as in the case of our $\Z_4$ gauge theory with twisted $\Z_2$ symmetry example with $a=(0,2)$, the deformed and undeformed fusion rules are identical.  To see the difference between the two fusion products, we then have to examine the F-matrices.

Figure \ref{fig4} is a shorthand for describing the deformed $F$ matrices, which we denote $\tF$ in contrast to the original ones, which we call $F$.  The precise formulae for $\tF$ in terms of $F$ are

\begin{equation} \label{bxx}
\left[\tF^{b,x,y}_z\right]^{x'}_{z'a}= \left[ F^{b,x,y}_{z\overline{a}} \right]^{x'}_{z'} \left[ F^{b,z',a}_z \right]^{z\overline{a}}_{z'a},
\end{equation}

\begin{equation} \label{xbx}
\left[\tF^{x,b,y}_z \right]^{x'}_{y'}= \left[F^{x,b,y}_{z\overline{a}}\right]^{x'}_{y'}
\end{equation}
\begin{align}
&\left[\left(\tF^{x,y,b}_z\right)^{-1}\right]^{y'}_{z'a}\notag=\\
&\phantom{=}\left[ \left(F^{x,y,b}_{z\overline{a}}\right)^{-1}\right]^{y'}_{z'} \left[F^{z',b,a}_z\right]^{z\overline{a}}_{ba} R^{b,a} \left[ \left(F^{z',a,b}_z\right)^{-1}\right]^{ba}_{z'a} \label{xxb}
\end{align}

\begin{figure}[htbp]
\begin{center}
\includegraphics[width=0.4\textwidth]{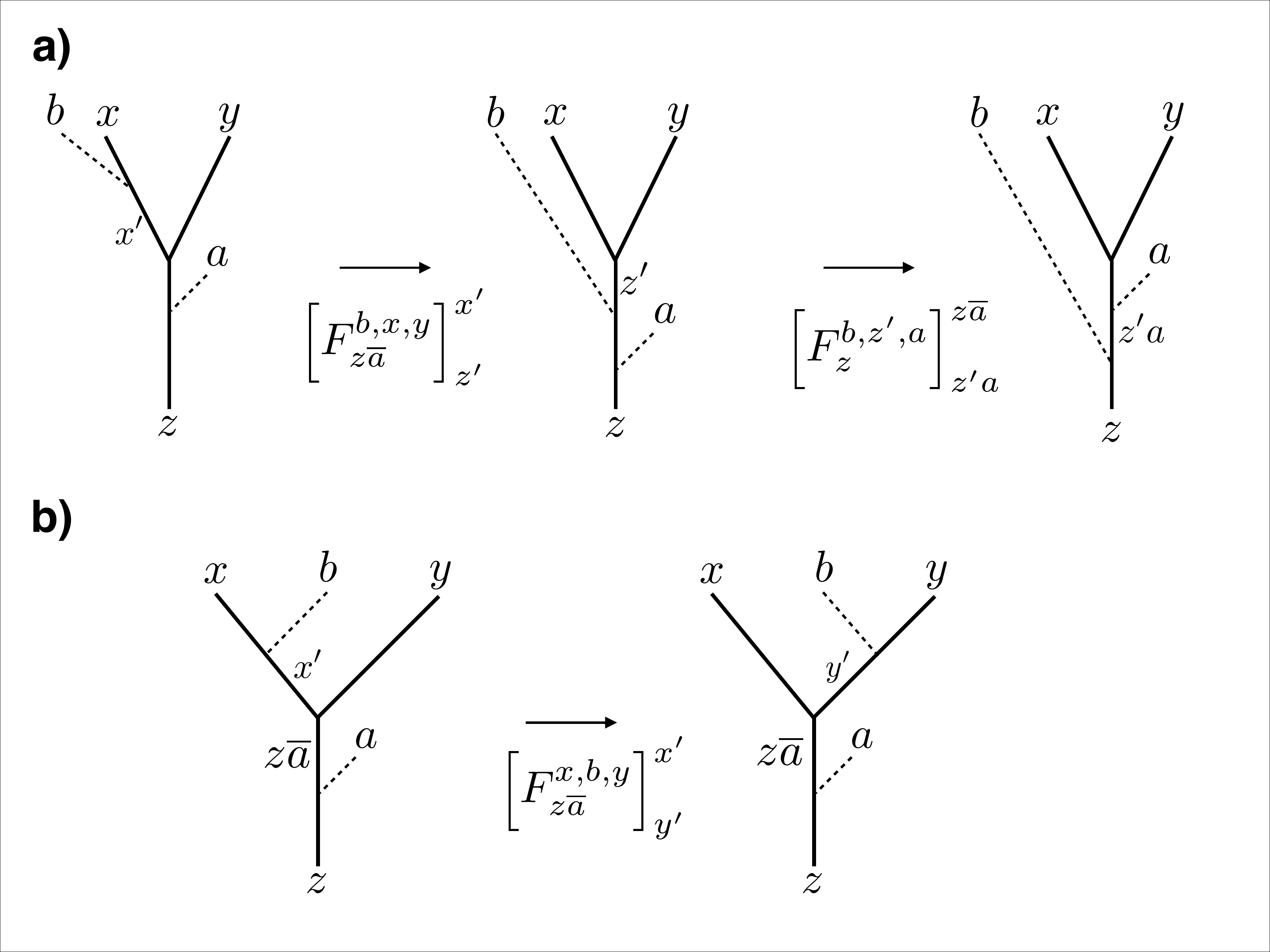}
\caption{Definition of deformed F-matrix.}
\label{deform_fig1}
\end{center}
\end{figure}

\begin{figure}[htbp]
\begin{center}
\includegraphics[width=0.4\textwidth]{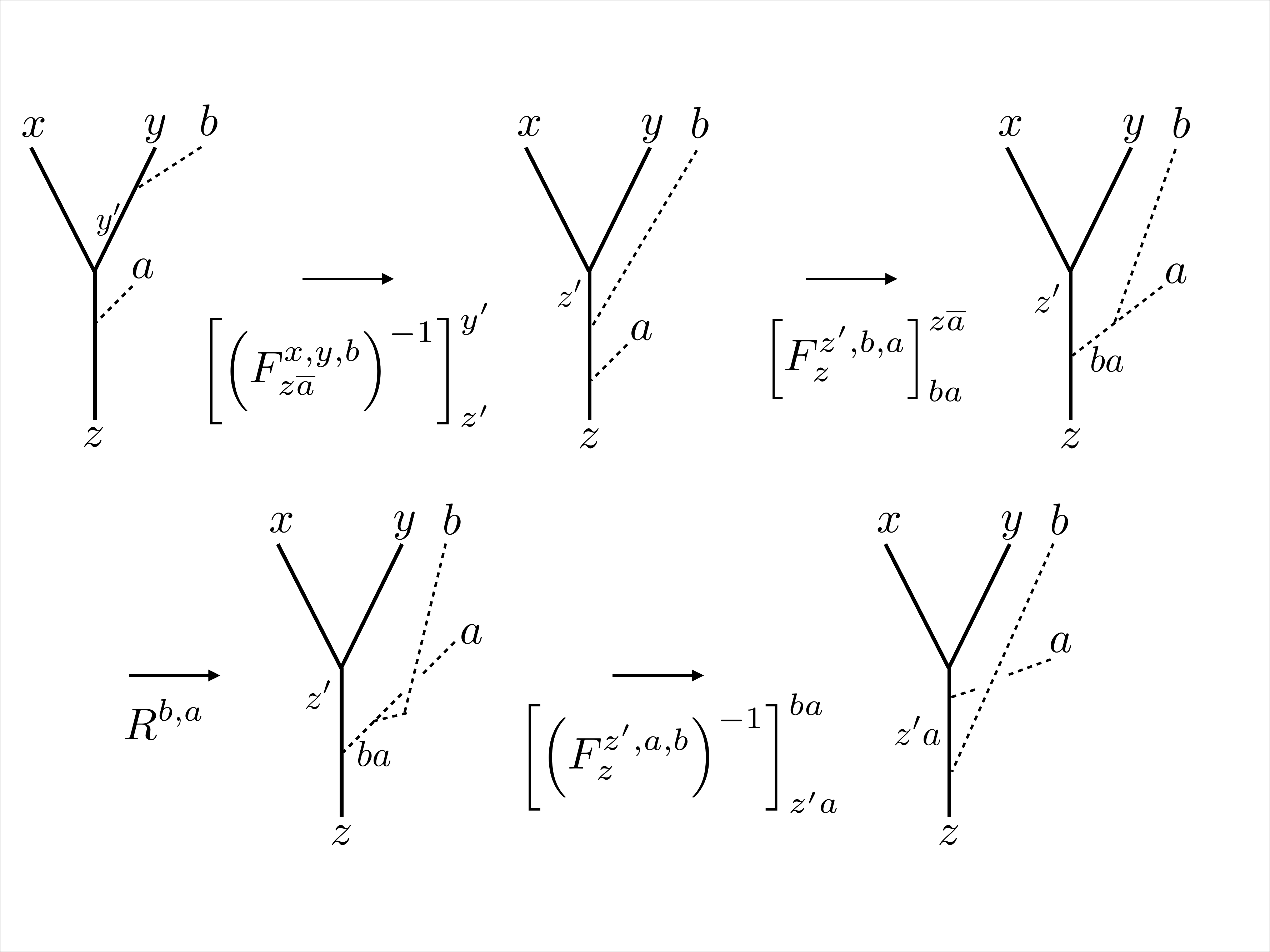}
\caption{Definition of deformed F-matrix.}
\label{deform_fig2}
\end{center}
\end{figure}

The way that these three equations were obtained from the prescription in figure \ref{fig4} is illustrated graphically in figures \ref{deform_fig1}(a), \ref{deform_fig1}(b), and \ref{deform_fig2} respectively.

\begin{figure}[htbp]
\begin{center}
\includegraphics[width=0.4\textwidth]{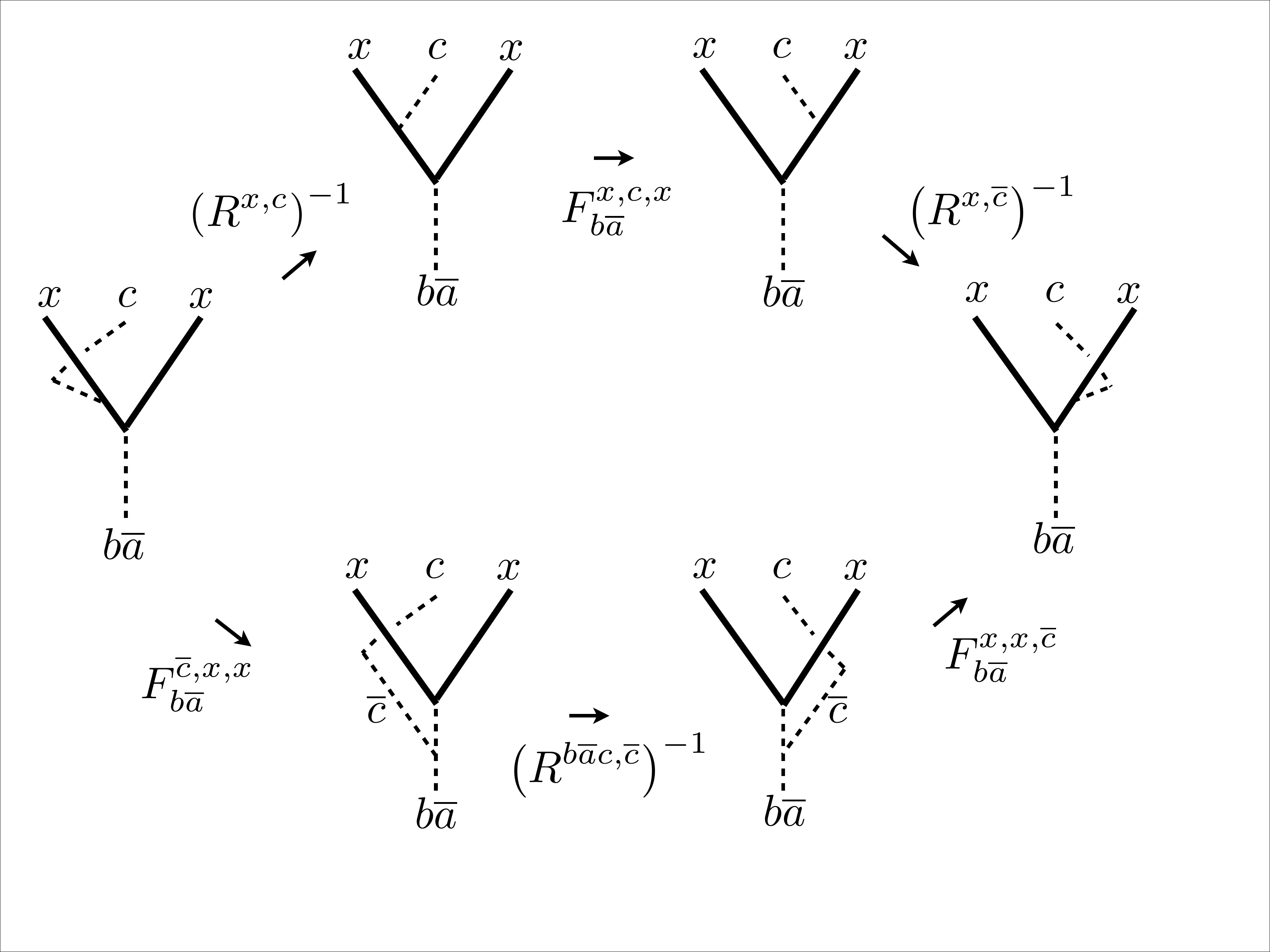}
\caption{Instance of hexagon equation for the original defect fusion product.}
\label{deform_hex3}
\end{center}
\end{figure}

\begin{figure}[htbp]
\begin{center}
\includegraphics[width=0.4\textwidth]{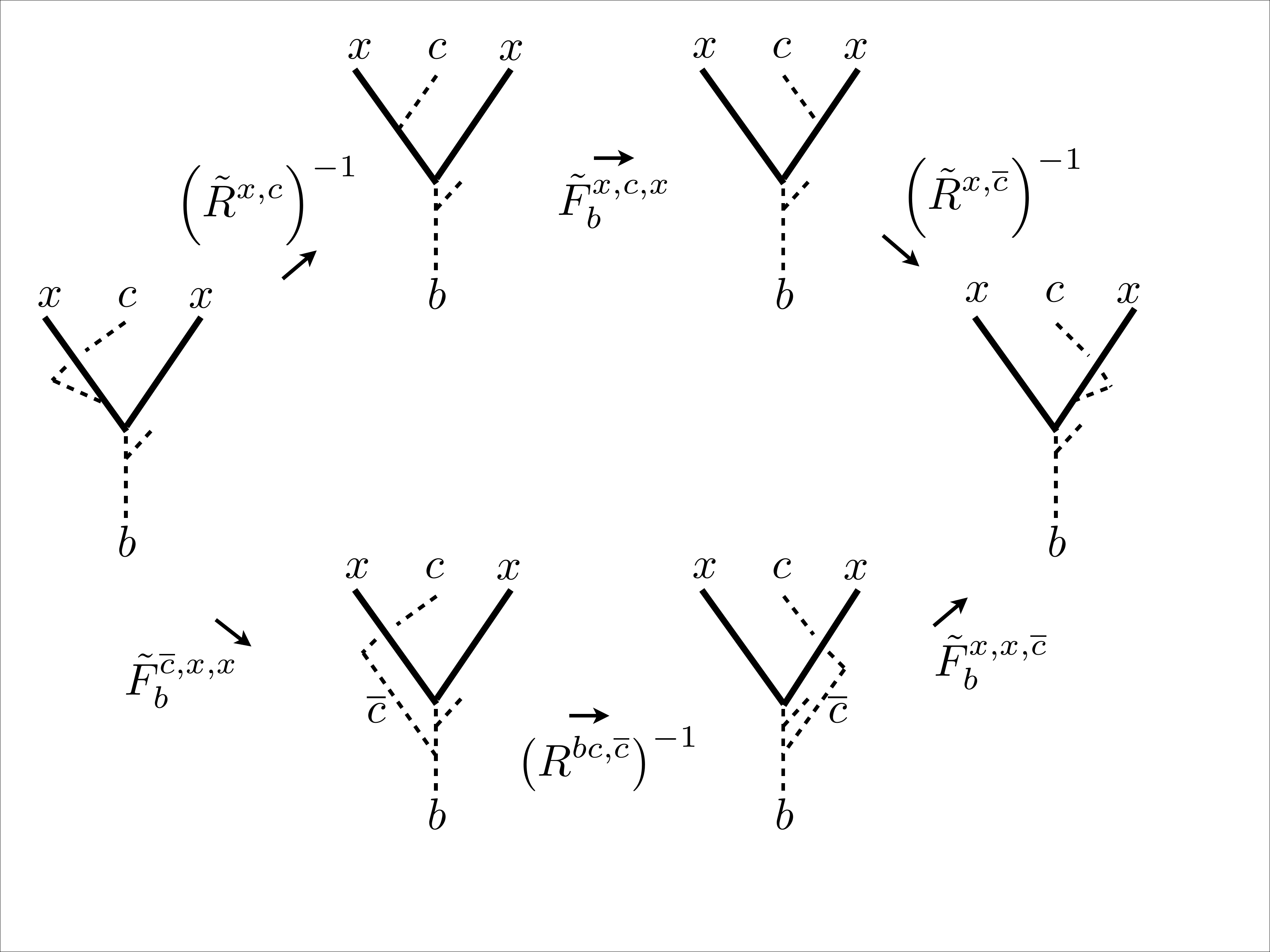}
\caption{Instance of hexagon equation for the deformed defect fusion product.}
\label{deform_hex2}
\end{center}
\end{figure}

Now let us analyze the specific example of the $\Z_4$ gauge theory with twisted $\Z_2={1,g}$ symmetry.  Our claim is that deforming the $g\times g=1$ fusion rule by the anyon $a=a_{g,g}=(0,2)$ ({\emph{i.e.}} the strength $2$ flux of $\Z_4$) forces the braiding process of taking a $c=(1,0)$ charge around the $g$ defect twice to acquire a minus sign.  Note that braiding this charge around the $g$ defect only once changes it into a $\overline{c}=(3,0)$ charge, and there is no gauge invariant amplitude associated with this process, at the level of extrinsic defects (once the theory is gauged, $c$ and $\overline{c}$ combine into a single non-abelian anyon and a single braid maps the appropriate fusion space to itself).

This relation between deforming F-matrices and deforming braiding phases of an anyon around a defect follows from the hexagon equation illustrated in figures \ref{deform_hex3} and \ref{deform_hex2}.  To see this, first note that all of the F and R matrices are phases, as the anyons $a,b,c$ are all abelian.  The equations in figures \ref{deform_hex3} and \ref{deform_hex2} read:

\begin{align}
R^{x,c} R^{x,\overline{c}}&=F^{x,c,x}_{b\overline{a}} \left(F^{\overline{c},x,x}_{b\overline{a}}\right)^{-1} \left( F^{x,x,\overline{c}}_{b\overline{a}} \right)^{-1} R^{b\overline{a}c,\overline{c}}\\
{\tilde{R}}^{x,c} {\tilde{R}}^{x,\overline{c}}&=\tF^{x,c,x}_b \left(\tF^{\overline{c},x,x}_{b}\right)^{-1} \left( \tF^{x,x,\overline{c}}_b \right)^{-1} R^{bc,\overline{c}}
\end{align}

Now using equations \ref{bxx},\ref{xbx}, and \ref{xxb} and the fact that the F-matrices for the $\Z_4$ gauge theory are all equal to $1$ we obtain

\begin{align}
\tF^{\overline{c},x,x}_b &= F^{\overline{c},x,x}_{b\overline{a}} \\
\tF^{x,x,\overline{c}}_b &= F^{x,x,\overline{c}}_{b\overline{a}} \left( R^{\overline{c},a} \right)^{-1} \\
\tF^{x,c,x}_b&=F^{x,c,x}_{b\overline{a}}
\end{align}
so that

\begin{align}
\frac{{\tilde R}^{x,c} {\tilde R}^{x,\overline{c}}}{R^{x,c} R^{x,\overline{c}}} &=R^{\overline{c},a} R^{bc,\overline{c}} \left(R^{b\overline{a}c,\overline{c}}\right)^{-1}
\end{align}

This is equal to $R^{\overline{c},a} R^{a,\overline{c}}=-1$.  Using the fact that in our convention, braiding in front of the defect gives a trivial phase of $1$, this shows that the quantity in equation \ref{eq:criterion} acquires a $-1$ when the defect fusion rules are deformed by $a=(0,2)$, as desired.

\section{$H^2_{\rho}(G,H)$ and group extensions \label{ap:extend}}

Symmetry enriched phases are classified by the group extensions of $H$ by $G$. Such a group extension satisfies
\begin{align}
E\big/H \cong G
\end{align}
which means that $H$ appears as a normal subgroup of $E$. Each element of $E$ can be labeled by a pair of elements $h \in H$ and $g \in G$. However, we note that $(e,g_1)\cdot(e,g_2) \neq (e, g_1g_2)$ in general, as only the cosets $E\big/H$ multiply as $G$. Thus, a function $\omega: G^2 \rightarrow H$ should be introduced such that
\begin{align}
(e,g_1)\cdot(e,g_2) = (\omega(g_1,g_2), g_1g_2)
\end{align}
$G$ also has a nontrivial effect on $H$, permuting the elements inside the normal subgroup. This is an automorphism of $H$, and so we can summarize this effect by
\begin{align}
(e,  g)\cdot (h, e) = (h^{\rho(g)}, g)\\
\rho \in \text{Hom}(G, \text{Aut}(H))
\end{align}
Combining these data, the total multiplication rule is
\begin{align}
(h_1,g_1)\cdot(h_2,g_2) = (h_1h_2^{\rho(g_1)}\omega(g_1,g_2), g_1g_2) \label{eq:ext_mult}
\end{align}
Associativity imposes a constraint on $\omega$, namely
\begin{align}
\omega(g_2,g_3)^{\rho(g_1)}\omega(g_1, g_2g_3) = \omega(g_1, g_2) \omega(g_1 g_2, g_3)
\end{align}
This is precisely the 2-cocycle condition from a group cohomology of $G$, $H^2_{\rho}(G,H)$. This information will be encoded into the plaquette terms that live entirely within vertical links in our Hamiltonian. Different cocycles correspond to different group extensions that we can construct, producing distinct SET phases.

\section{Explicit derivation of plaquette terms after gauging \label{ap:gauge_plaq}}
As it appears often in this derivation, let $\bar{g}_i = \prod_{r=1}^{i}g_r$. Introducing our $\al_L$ operator following our prescription in eq. \ref{eq:gauge_a}, we find the plaquette terms are modified as
\begin{align*}
\text{Tr}_H  \prod_{L\in P} a_l(g) & \rightarrow \text{Tr}_{H,G} \prod_{L\in P} a_L(g)\al_L\\
 &= \text{Tr}_H \prod_{i=1}^j a_i\left(g_i \left(\bar{g}_{i}\right)^{-1} g \right)^{\rho(\bar{g}_{i-1})} \text{ Tr}_G  \prod_{i=1}^j \al_{i}\\
&= \text{Tr}_H  \Omega_P(g) \prod_{i=1}^j \left( a'_i\right)^{\rho(\bar{g}_{i-1})} \text{ Tr}_G  \prod_{i=1}^j \al_{i}\\
\Omega_P(g)& = \prod_{i=1}^j \omega\left(g_i, \left(\bar{g}_{i} \right)^{-1} g \right)^{\rho(\bar{g}_{i-1})}
\end{align*}
where we recall that $a'_L \ket{\eta'_L} = M(\eta'_L) \ket{\eta'_L}$. Focusing on $\Omega_P(g)$ we observe that we can simplify it as follows
\begin{align*}
\Omega_P(g)& = \prod_{i=1}^j \omega\left(g_i, \left(\bar{g}_{i}\right)^{-1} g \right)^{\rho(\bar{g}_{i-1})}\\
& = \prod_{i=1}^j \frac{\omega\left(\bar{g}_{i}, \left(\bar{g}_{i} \right)^{-1} g \right)}{\omega\left(\bar{g}_{i}, \left(\bar{g}_{i} \right)^{-1} g \right)}  \omega\left(\bar{g}_{i-1}, g_i\right)\\
&= \prod_{i=1}^j \omega\left(\bar{g}_{i-1}, g_i\right) \equiv \Omega_P
\end{align*}
which is manifestly independent of $g$, and thus greatly simplifies our plaquette term.
\bibliographystyle{unsrt}
\bibliography{References}

\end{document}